\documentclass[twocolumn]{aastex62}

\usepackage{graphicx}
\usepackage{amsfonts,amsmath,amssymb}
\usepackage{multirow,booktabs}
\usepackage{subfloat}
\usepackage[caption=false]{subfig}
\usepackage{txfonts}
\usepackage{afterpage}
\usepackage{xtab}
\usepackage{fp}
\usepackage{enumerate}

\newcommand{\nh} {{$N_{\rm{H}}$}}
\newcommand{\es}  {{erg s$^{-1}$}}

\newcommand{\be}  {{B\emph{e}}}

\newcommand{\chandra}{{\em Chandra}}

\newcommand{\einstein}{{\em Einstein}}
\newcommand{\rxte}{{\em Rossi X-ray Timing Explorer}}
\newcommand{\rxtet}{{\em RXTE}}
\newcommand{\xmmn}{{\em XMM-Newton}}
\newcommand{\rosat}{{\em ROSAT}}

\newcommand{\nustarf}{{\em Nuclear Spectroscopic Telescope Array}}
\newcommand{\nustar}{{\em NuSTAR}}
\newcommand{\nustars}{{\em NuSTAR's}}
\newcommand{\swift}{{\em Swift}}
\newcommand{\intg}{{\em INTEGRAL}}

\newcommand{\galex}{{\em GALEX}}
\newcommand{\spitzer}{{\em Spitzer}}
\newcommand{\bepposax}{{\em BeppoSAX}}
\newcommand{\uhuru}{{\em Uhuru}}
\newcommand{\twomass}{{2MASS}}

\newcommand{\ligo}{Laser Interferometer Gravitational Wave Observatory}

\newcommand{\irast}{{\em IRAS}}
\newcommand{\suzaku}{{\em Suzaku}}

\newcommand{\msun}{M$_{\odot}$}

\newcommand{\lx}  {{$L_{\rm{X}}$}}

\newcommand{\ks}  {{$K_{\rm{s}}$}}
\newcommand{\sfr}  {{M$_{\odot}$ yr$^{-1}$}}

\newcommand{\numclass}  {{90}}	

\newcommand{\numstate}  {{87}}	

\newcommand{\numsrc}  {{128}}	

\newcommand{\numbh}  {{47}}	

\newcommand{\numns}  {{43}}	

\newcommand{\numap}  {{21}}	

\newcommand{\numbhh}  {{27}}	

\newcommand{\numnsh}  {{14}}	

\newcommand{\full}  {{$4-25$~}}
\newcommand{\hard}  {{$12-25$~}}

\newcommand{\fitaf}  {{39.49}}

\newcommand{\fitbf}  {{0.83}}

\newcommand{\fiterraf}  {{0.15}}

\newcommand{\fiterrbf}  {{0.24}}
\newcommand{\fitah}  {{38.80}}

\newcommand{\fitbh}  {{0.79}}

\newcommand{\fiterrah}  {{0.21}}

\newcommand{\fiterrbh}  {{0.32}}

\newcommand{\fitalphaf}  {{3.560}}

\newcommand{\fitbetaf}  {{1.902}}

\newcommand{\fiterralphaf}  {{1.163}}

\newcommand{\fiterrbetaf}  {{0.837}}
\newcommand{\fitalphah}  {{3.426}}

\newcommand{\fitbetah}  {{5.150}}

\newcommand{\fiterralphah}  {{2.205}}

\newcommand{\fiterrbetah}  {{1.878}}

\FPeval\fitalphahten{round(\fitalphah/10:3)}	

\FPeval\fitbetahten{round(\fitbetah/10:3)}	

\FPeval\fiterralphahten{round(\fiterralphah/10:3)}	

\FPeval\fiterrbetahten{round(\fiterrbetah/10:3)}	

\received{}
\revised{}
\accepted{}
\submitjournal{ApJ}
\shorttitle{Black Holes \& Neutron Stars from \nustar}
\shortauthors{N. Vulic et al.}

\begin{document}

\title{Black Holes and Neutron Stars in Nearby Galaxies: Insights from \nustar}

\correspondingauthor{N. Vulic}
\email{nvulic@uwo.ca}

\author[0000-0001-7855-8336]{N. Vulic}
\affil{Laboratory for X-ray Astrophysics, Code 662, NASA Goddard Space Flight Center, Greenbelt, MD 20771, USA}
\affil{Department of Astronomy and Center for Space Science and Technology (CRESST), University of Maryland, College Park, MD 20742-2421, USA}

\author[0000-0001-8667-2681]{A. E. Hornschemeier}
\affil{Laboratory for X-ray Astrophysics, Code 662, NASA Goddard Space Flight Center, Greenbelt, MD 20771, USA}
\affil{Department of Physics \& Astronomy, Johns Hopkins University, 3400 North Charles Street, Baltimore, MD 21218, USA}

\author[0000-0001-9110-2245]{D. R. Wik}
\affil{Department of Physics \& Astronomy, University of Utah, Salt Lake City, UT 84112-0830, USA}
\affil{Laboratory for X-ray Astrophysics, Code 662, NASA Goddard Space Flight Center, Greenbelt, MD 20771, USA}

\author[0000-0001-6366-3459]{M. Yukita}
\affil{Department of Physics \& Astronomy, Johns Hopkins University, 3400 North Charles Street, Baltimore, MD 21218, USA}
\affil{Laboratory for X-ray Astrophysics, Code 662, NASA Goddard Space Flight Center, Greenbelt, MD 20771, USA}

\author{A. Zezas}
\affil{Physics Department \& Institute of Theoretical \& Computational Physics, University of Crete, 71003 Heraklion, Crete, Greece}
\affil{Harvard-Smithsonian Center for Astrophysics, 60 Garden Street, Cambridge, MA 02138, USA}

\author[0000-0001-5655-1440]{A. F. Ptak}
\affil{Department of Physics \& Astronomy, Johns Hopkins University, 3400 North Charles Street, Baltimore, MD 21218, USA}
\affil{Laboratory for X-ray Astrophysics, Code 662, NASA Goddard Space Flight Center, Greenbelt, MD 20771, USA}

\author[0000-0003-2192-3296]{B. D. Lehmer}
\affil{Department of Physics, University of Arkansas, 825 West Dickson Street, Fayetteville, AR 72701, USA}

\author[0000-0001-7539-1593]{V. Antoniou}
\affil{Harvard-Smithsonian Center for Astrophysics, 60 Garden Street, Cambridge, MA 02138, USA}

\author[0000-0003-0976-4755]{T. J. Maccarone}
\affil{Department of Physics \& Astronomy, Box 41051, Science Building, Texas Tech University, Lubbock, TX 79409-1051, USA}

\author[0000-0002-7502-0597]{B. F. Williams}
\affil{Department of Astronomy, Box 351580, University of Washington, Seattle, WA 98195, USA}

\author[0000-0002-9286-9963]{F. M. Fornasini}
\affil{Harvard-Smithsonian Center for Astrophysics, 60 Garden Street, Cambridge, MA 02138, USA}

\begin{abstract}

Nearby galaxy surveys have long classified X-ray binaries (XRBs) by the mass category of their donor stars (high-mass and low-mass). The \nustar\ observatory, which provides imaging data at E $>10$ keV, has enabled the classification of extragalactic XRBs by their compact object type: neutron star (NS) or black hole (BH). We analyzed \nustar/\chandra/\xmmn\ observations from a \nustar-selected sample of 12 galaxies within 5 Mpc having stellar masses ($M_{\star}$) $10^{7-11}$ \msun\ and star formation rates (SFR) $\approx0.01-15$ \sfr. We detect \numsrc\ \nustar\ sources to a sensitivity of $\approx10^{38}$ \es. Using \nustar\ color-intensity and color-color diagrams we classify \numns\ of these sources as candidate NS and \numbh\ as candidate BH. We further subdivide BH by accretion states (soft, intermediate, and hard) and NS by weak (Z/Atoll) and strong (accreting pulsar) magnetic field. Using 8 normal (Milky Way-type) galaxies in the sample, we confirm the relation between SFR and galaxy X-ray point source luminosity in the \full and \hard keV energy bands. We also constrain galaxy X-ray point source luminosity using the relation $L_{\rm{X}}=\alpha M_{\star}+\beta\text{SFR}$, finding agreement with previous work. The XLF of all sources in the \full and \hard keV energy bands matches with the $\alpha=1.6$ slope for high-mass XRBs. We find that NS XLFs suggest a decline beginning at the Eddington limit for a 1.4 \msun\ NS, whereas the BH fraction shows an approximate monotonic increase in the \full and \hard keV energy bands. 
We calculate the overall ratio of BH to NS to be $\approx1$ for \full keV and $\approx2$ for \hard keV.

\end{abstract}

\section{Introduction}	\label{sec:intro}

Until the launch of the first focusing telescope to operate at E $>10$ keV, the \nustarf\ \citep[\nustar;][]{harrison06-13}, we knew very little about the behaviour and nature of extragalactic black hole (BH) and neutron star (NS) populations at harder energies. In the absence of an X-ray bright supermassive BH, the total X-ray emission of a galaxy above 2 keV is dominated by X-ray binaries (XRBs), classified as low-mass (LMXB) or high-mass (HMXB) based on their donor star. Previous studies of nearby galaxies in the soft X-ray band ($0.5-10$ keV) by, e.g.\ \chandra\ and \xmmn\ \citep[e.g.][]{stiele10-11,mineo01-12,mineo01-14,long06-14,haberl02-16,peacock02-16} have revealed important new information on compact object populations, such as strong correlations between properties of XRBs and galaxy star formation rate (SFR), stellar mass, and metallicity \citep[e.g.][]{basu-zych02-16}. Extrapolation of these local-Universe measurements as well as supporting measurements at high-redshift \citep{lehmer07-16} have indicated a possible significant role of XRBs in heating the Intergalactic Medium (IGM) of the early Universe \citep[e.g.][]{fragos10-13, mesinger04-14, pacucci09-14, madau05-17, sazonov06-17, das07-17}.

However, there are questions about the extragalactic XRB population that are difficult to answer at E $<10$ keV, including whether compact objects are BH or NS. The rich suite of thousands of \rxte\ (\rxtet) PCA spectra of BH/NS XRBs in the Milky Way galaxy provide critical diagnostics in the \full keV band of both compact object type (BH vs. NS) and accretion state \citep[e.g.][]{maccarone01-03, mcclintock04-06, done12-07}. With \nustar, for the first time, we are able to leverage the knowledge gained from compact objects in our own galaxy by applying these harder X-ray diagnostics to extragalactic populations.

The hard X-ray coverage with \nustar\ is crucial for distinguishing different types of accreting binaries, such as BH/NS XRBs and accreting pulsars.
Compact object diagnostics have already been successfully applied to characterize XRBs in several nearby galaxies observed by \nustar. These studies include simultaneous {\em NuSTAR/Chandra/XMM-Newton/Swift} studies of the nearby star-forming galaxies NGC 253 \citep{lehmer07-13, wik12-14} and M83 \citep{yukita06-16}, as well as Local Group galaxy M31 \citep{maccarone06-16, yukita03-17, lazzarini06-18}; for a description of the \nustar\ galaxy program please see \citet{hornschemeier-16}. Using \full keV color-color and color-intensity diagnostics, these studies have shown that the starburst galaxies are dominated by luminous BH-XRB systems, mostly in intermediate accretion states. Specifically, ultraluminous X-ray sources (ULXs) with $3-30$ keV spectra indicative of super-Eddington accretion (e.g.\ \citealt{gladstone08-09}) appear to dominate the hard X-ray emission of starburst galaxies \citep{walton12-13, bachetti10-14, rana02-15, lehmer06-15}. 
Meanwhile, M31 has a significant contribution from NS accretors (pulsars and low-magnetic field Z-type sources; \citealt{maccarone06-16, yukita03-17}). As expected, the pulsars trace the young stellar population in the spiral arms and the Z-type sources are concentrated in globular clusters and the bulge/field of the galaxy. \nustar\ data were crucial to the reclassification of previously identified BH candidates in M31 globular clusters as NS, based on their hard X-ray spectra \citep{maccarone06-16}.

\nustar\ has previously resolved the XRB population in 3 galaxies. Thus, it is now time for a broader investigation of the relationship between the properties of a galaxy and the X-ray source types and accretion states as determined from hard X-ray observations.
Specifically, what is the relationship between galaxy properties such as the stellar mass and recent star formation rate/history and compact object type/accretion state as determined from hard X-ray diagnostics?
To estimate the number of BH and NS that will be formed in a galaxy requires binary population synthesis, and a detailed understanding of concepts such as supernova explosions, which is not well understood \citep[e.g.][]{pejcha03-15}. Alternatively, we can use observational data and methods to determine the BH fraction and its dependence on X-ray luminosity and specific star formation rate (sSFR).

With \nustar\ we can measure local-galaxy SEDs over $0.5-30$ keV that are applicable to high-$z$ galaxies detected by \chandra. One of our goals is to determine what sources are contributing to the $0.5-30$ keV emission. Furthermore, we would like to be able to predict, based on galaxy properties such as star formation rate/history and stellar mass, what the distribution of binaries and their emitting properties are. Achieving this goal is rather complicated, as there are parameters such as the duty cycle that result in a broad range of population properties for different stellar ages, etc. One approach to this complicated problem is to make direct measurements over a variety of galaxy properties. Each snapshot view of an individual galaxy measures the state of the overall population, giving us a constraint on duty cycles \citep{binder01-17}. Hard X-ray diagnostics allow us to determine the distribution of BH spectral states, similar to Galactic BH studies \citep[e.g.][]{tetarenko02-16}. Using this approach, we can obtain baseline estimates of XRB formation rate, duty cycles, spectral states, and galaxy SEDs. 
Understanding these properties at E $>10$ keV is critical to compare to the results of XRB evolution in the $0.5-10$ keV bandpass. \nustar\ is well-matched to the rest-frame energies of high-$z$ galaxies at $z=3-4$ probed by \chandra\ and is thus a new window into XRB evolution.

The X-ray luminosity function (XLF) represents the distribution of sources in a galaxy based on their luminosity. 
Seminal studies of LMXBs in elliptical galaxies \citep[e.g.][]{gilfanov03-04, zhang10-12} and HMXBs in spiral galaxies \citep[e.g.][]{grimm03-03, mineo01-12} found that their XLFs were (approximately) universal when normalizing by the stellar mass and SFR of a galaxy, respectively (see \citealt{gilfanov-04} for a summary). Small variations in the power law slope and cutoff are dependent on factors such as metallicity \citep{basu-zych02-16} and star formation history \citep{lehmer12-17}. 
We will investigate how scaling \nustar\ XLFs by SFR compares with results from \chandra/\xmmn\ studies. 

To date, studies of the XLFs of nearby galaxies have mostly focused on LMXB or HMXB populations. However, certain XLF characteristics can be attributed to compact object types \citep{lutovinov05-13}, such as the break at $\sim$few$\times10^{38}$ \es\ corresponding to the Eddington limit for NS. This break is often argued to reflect the transition from a population of NS to BH XRBs \citep{sarazin12-00, kim08-04, wang09-16}. 
\nustar\ is well-suited to distinguish between BH and NS accretors, therefore allowing a first-look at BH-only and NS-only XLFs.
In addition, this can elucidate how the $0.5-30$ keV SED of galaxies depends on the compact object type and accretion states of BH and NS. 

Our goals are to study the hard X-ray properties of the XRB population of 12 nearby galaxies ($<5$ Mpc) using joint \nustar\ and \chandra/\xmmn\ data. We will use knowledge of galaxy parameters such as SFR and stellar mass to investigate the connection between XRB populations and host galaxy properties. 
In Section \ref{sec:sample} we describe the sample selection and calculation of SFR and stellar mass for galaxies in the sample. In Section \ref{sec:obs} we summarize the \nustar, \chandra, and \xmmn\ observations. In Section \ref{sec:data} we outline our analysis methods, which focus on the PSF fitting procedure for \nustar\ data. In Section \ref{sec:resdis} we present \nustar\ diagnostic diagrams, XLFs, and scaling relations, and discuss their implications. We summarize our conclusions in Section \ref{sec:con}.

\section{Sample Selection}	\label{sec:sample}

Using the HyperLeda Database\footnote{\url{http://leda.univ-lyon1.fr/}} \citep{makarov10-14} and the Updated Nearby Galaxy catalog \citep{karachentsev04-13} we searched for all galaxies within 10 Mpc that have been observed by \nustar\ as of 1 July 2017. We created the sample based on reaching an X-ray point source sensitivity limit of $\approx10^{38}$ \es\ (\full keV), corresponding to the expected approximate luminosity of luminous non-magnetized NS XRBs\footnote{e.g.\ Sco X-1; \lx\ ($2-20$ keV) $\approx2\times10^{38}$ \es, \citealt{bradshaw02-99}}, for each observed galaxy. We excluded M51, NGC 4258, and NGC 4395 because they did not reach this limit. We also excluded Centaurus A due to the presence of a luminous AGN, whose emission contaminated the field of view and prevented the detection of faint point sources. The nearby dwarf galaxy IC 10 was excluded due to contamination from stray light. 

In Table \ref{tab:gals} we list the 12 galaxies in the sample and	include their coordinates, morphological type, dimensions, inclination, distance, Galactic column density, stellar masses, and SFR (see Section \ref{sec:smsfr}). Several of these galaxies are part of either the \nustar\ nearby galaxies program \citep{hornschemeier-16} or were targeted because they contained individual ULX sources that are likely high accretion rate XRBs \citep{bachetti01-14, kaaret08-17}. 

\begin{figure*}[ht!]
\includegraphics[width=1.0\textwidth]{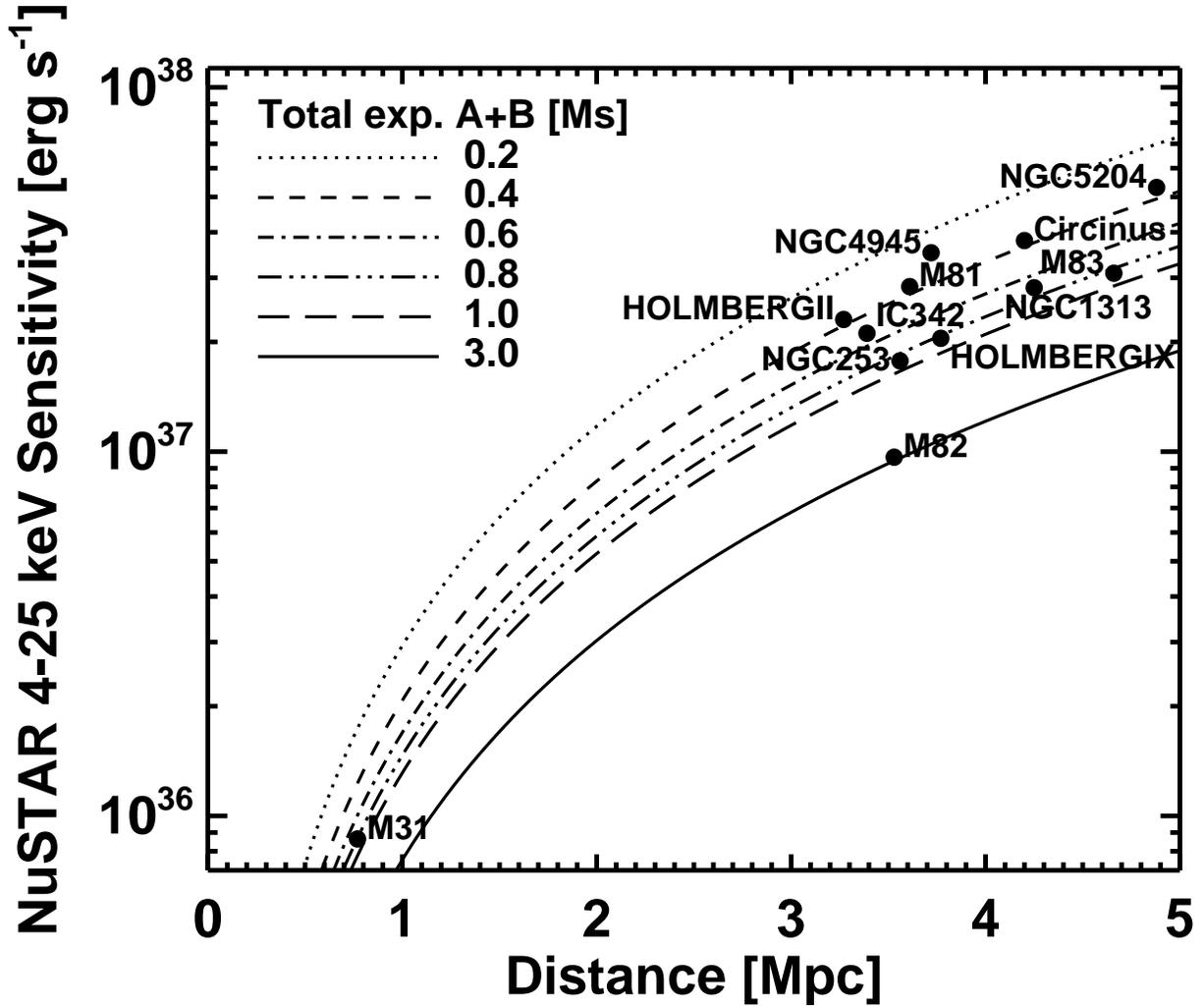}
\caption{Shown is the X-ray point source sensitivity of each \nustar-observed galaxy, which is affected most directly by the exposure time and the distance to the galaxy. \nustar\ point source (absorbed) sensitivity limits were calculated for each galaxy based on a $3\sigma$ detection (30\arcsec\ extraction region for telescopes A and B) using the distance and $N_{\rm{H}}$ values from Table \ref{tab:gals} and a spectral index of $\Gamma=1.7$. Exposure times were taken from Table \ref{tab:obsid-info} (telescopes A+B) and we assumed a constant background based on the value from the 30\arcsec\ \nustar\ background spectrum file. Lines of constant exposure time for telescopes A+B in Ms are shown based on the same assumptions with a constant $N_{\rm{H}}$ value of $10^{20}$ cm$^{-2}$.}
\label{fig:sensgals}
\end{figure*}

There is sufficient \nustar\ exposure for the entire galaxy sample for detection of all point sources with \lx\ above $\sim10^{38}$ \es\ (\full keV). However, there is spatial variation of sensitivity within the galaxies due to source confusion in regions with higher relative source density. In Figure \ref{fig:sensgals} we plot the \full keV point source sensitivity limit against the distance of each galaxy and show sensitivity curves for total exposure times ranging from 200 ks to 3 Ms. 

To date, while there have been studies of individual sources or galaxies, there has not been a systematic analysis of the hard X-ray point source populations for an ensemble of these galaxies. The X-ray point source populations of these galaxies have been well-studied in the $0.5-10$ keV bandpass in the past by various X-ray observatories such as \chandra, \xmmn, and \rosat, with the exception of NGC 1313 and NGC 5204, where the focus has been on ULX sources as opposed to the point source population. In Appendix \ref{sec:sampsum} we summarize individual galaxy properties and previous X-ray studies for each galaxy in the sample. A detailed study of M31 will be presented by D. Wik et al.\ 2018 (in prep.), thus we exclude M31 from our analysis (except for total galaxy X-ray luminosity fitting in Section \ref{sec:corr}).

\begin{table*}
\scriptsize
\caption{Galaxy Properties\label{tab:gals}}
\begin{tabular}{@{}  c c  c  c  c  c  c  c  c  c  c  c  c  @{}}
\hline\hline
Galaxy	&	R.A.	&	Decl.		&	Type	&	D$_{25}$	&	d$_{25}$	&	Inclination	&	Linear Scale	&	Distance	&	Uncertainty	&	$N_{\rm{H}}$	&	Stellar Mass	&	SFR	\\
&	\multicolumn{2}{c}{(J2000.0)}	&	&	(\arcmin)	&	(\arcmin)	&	(degrees)		&	(pc/\arcsec)	&	(Mpc)	&	(Mpc)	&	($10^{20}$ cm$^{-2}$)	&	($10^{9}$ \msun)	& (\sfr)		\\
(1)	&	(2)	&	(3)	&	(4)	&	(5)	&	(6)	&	(7)	&	(8)	&	(9)	&	(10)	&	(11)	&	(12)	&	(13)	\\	\hline
            M31\tablenotemark{a} &        10.685 &        41.269 &    Sb   &      177.8 &       69.2 &    72 &       3.73 &          0.77 &          0.04 &     6.7 &      93.88 &    0.32
\\      HOLMBERGII &       124.768 &        70.722 &    I    &        7.9 &        5.6 &    51 &      15.85 &          3.27 &          0.18 &     3.4 &       0.11 &    0.06
\\           IC342 &        56.705 &        68.101 &    SABc &       20.0 &       19.1 &    18 &      16.44 &          3.39 &          0.22 &    28.7 &      22.64 &    3.90
\\             M82 &       148.968 &        69.680 &    Scd  &       11.0 &        5.1 &    76 &      17.11 &          3.53 &          0.26 &     4.0 &      32.45 &   12.52
\\          NGC253 &        11.888 &       -25.288 &    SABc &       26.9 &        4.6 &    90 &      17.26 &          3.56 &          0.26 &     1.4 &      71.63 &    5.82
\\             M81 &       148.888 &        69.065 &    Sab  &       21.4 &       11.2 &    62 &      17.50 &          3.61 &          0.20 &     4.2 &      88.22 &    0.47
\\         NGC4945 &       196.364 &       -49.468 &    SBc  &       23.4 &        4.1 &    90 &      18.04 &          3.72 &          0.27 &    14.9 &      38.15 &    4.35
\\      HOLMBERGIX &       149.383 &        69.046 &    I    &        2.5 &        2.1 &    34 &      18.28 &          3.77 &          0.28 &     4.3 &       0.02 &    0.01
\\        Circinus &       213.291 &       -65.339 &    Sb   &        8.7 &        4.3 &    64 &      20.36 &          4.20 &          0.78 &    59.9 &      53.70 &    3.23
\\         NGC1313 &        49.565 &       -66.498 &    SBcd &       11.0 &        9.1 &    34 &      20.60 &          4.25 &          0.31 &     4.0 &       1.17 &    0.58
\\             M83 &       204.254 &       -29.866 &    Sc   &       13.5 &       13.2 &    14 &      22.59 &          4.66 &          0.30 &     3.7 &      44.06 &    3.41
\\         NGC5204 &       202.402 &        58.419 &    Sm   &        4.5 &        2.8 &    58 &      23.66 &          4.88 &          0.38 &     1.4 &       0.21 &    0.08
\\
\hline
\end{tabular}
\tablecomments{Galaxy properties. Unless indicated, values have been taken from the HyperLeda Database \citep[][\url{http://leda.univ-lyon1.fr/}]{makarov10-14}. Columns (5) and (6): major and minor isophotal diameters D$_{25}$ and d$_{25}$, respectively, for $\mu_{B}=25$ mag arcsec$^{-2}$. Column (7): inclination in degrees. Column (8): linear scale in pc representing 1\arcsec\ at the adopted distance. Distances are from \citet{tully10-13} except for Circinus, which is from \citet{tully08-09}. Column (10): $1\sigma$ distance uncertainty. Column (11): Galactic column density from \citet{kalberla09-05}. Columns (12) and (13): stellar mass and SFR as determined using the methods from Section \ref{sec:smsfr}.}
\tablenotetext{a}{Results for M31 can be found in D. Wik et al.\ 2018 (in prep.)}
\end{table*}

\subsection{Stellar Masses and Star Formation Rates} \label{sec:smsfr}

The LMXB and HMXB populations in a galaxy trace the host galaxy stellar mass and SFR, respectively. Therefore, to investigate this relationship, we need to determine accurate values of galaxy stellar mass and SFR. To calculate stellar masses we used the results from \citet{into04-13} that were corrected for self-consistency by \citet{mcgaugh11-14}. This relation was derived assuming a \citet{kroupa-98} initial mass function (IMF). We chose the parameterization reproduced in equation \ref{eq:mass} because $(B-V)$ colors were readily available from the HyperLeda Database \citep{makarov10-14} and the \ks-band luminosity is a robust indicator of stellar mass. Extinction-corrected \twomass\ \ks-band magnitudes were taken from \citet{tully08-16} and converted to luminosities using the distances from Table \ref{tab:gals} and $K_{\rm{s},\odot} = 3.302$ \citep{pecaut09-13, casagrande12-12}. The revised relation from \citet{mcgaugh11-14} was based on \spitzer\ 3.6 \micron\ data and required conversion of $K$-band magnitudes using their prescription $K_{\rm{s}} - [3.6] = 0.31$. The resulting stellar masses were each multiplied by 1.29 to convert from 3.6 \micron\ to $K$-band. 
Typical $M/L$ uncertainties were estimated to be $\sim0.1$ dex in the near-IR as a result of dust and complex star-formation histories. 
\begin{equation}
\text{log}(M_{\star}/\rm{M}_{\odot}) = \text{log}(L_{K_{\rm{s}},gal}/L_{K_{\rm{s}},\odot}) + 0.849(B-V) - 0.861	\label{eq:mass}
\end{equation}
The \ks-band magnitude and ($B-V$) color for M31 were adjusted using the values from Table 3 of \citet{kormendy08-13}, corrected for angular extent. The stellar mass agrees with the recently determined value from \citet{sick04-15}. NGC 4945 suffers from large internal extinction that affects the stellar mass estimate, so we used the ($B-V$) value from \citet{mccall05-14} that was corrected for internal extinction.

We determined SFRs from the relations presented in \citet{calzetti10-13} by adding the contribution from the UV and IR luminosity (\citealt{calzetti10-13} equation 1.11). These relations all assumed a \citet{kroupa04-01} IMF. Equation \ref{eq:sfruv} (\citealt{calzetti10-13} equation 1.2) was used to estimate the UV (dust-obscured) component of SFR: 
\begin{equation}
\text{SFR(UV) } [\rm{M}_{\odot}/\text{yr}^{-1}] = 3\times10^{-47}\lambda [\text{\AA}] L(\lambda) [\text{erg s}^{-1}]	\label{eq:sfruv}
\end{equation}
We used \galex\ far-UV (1539 \AA, FWHM of 269 \AA) asymptotic (total) magnitudes from \citet{lee01-11} and \citet[][only for M31]{gil-de-paz12-07} to calculate SFR(UV). 
We calculated the IR (dust-unobscured) component of SFR with equation \ref{eq:sfrir} (\citealt{calzetti10-13} equations 1.5-1.7) using 24 \micron\ fluxes from \citet{dale09-09}. 
\begin{equation}
\text{SFR(IR) } [\rm{M}_{\odot}/\text{yr}^{-1}] = 2.04\times10^{-43} L(\lambda)[\text{erg s}^{-1}]		\label{eq:sfrir}
\end{equation}
M31 and IC 342 did not have 24 \micron\ fluxes in \citet{dale09-09} and were instead taken from \citet{tempel01-10} and \citet{jarrett01-13}, respectively. For Circinus, we used the 25 \micron\ \irast\ flux from the NASA/IPAC Extragalactic Database (NED) and adjusted the coefficient in equation \ref{eq:sfrir} to $1.789\times10^{-43}$. NGC 4945 did not have a UV flux estimate and so we used the SFR from Atacama Large Millimeter Array results \citep{bendo11-16}, which were not affected by dust attenuation in the nuclear starburst. Circinus also had no UV flux estimate and thus we only used the SFR(25 \micron) value as it agrees well with other studies \citep{grimm03-03, for09-12}. 

\begin{figure*}[ht!]
\includegraphics[width=1.0\textwidth]{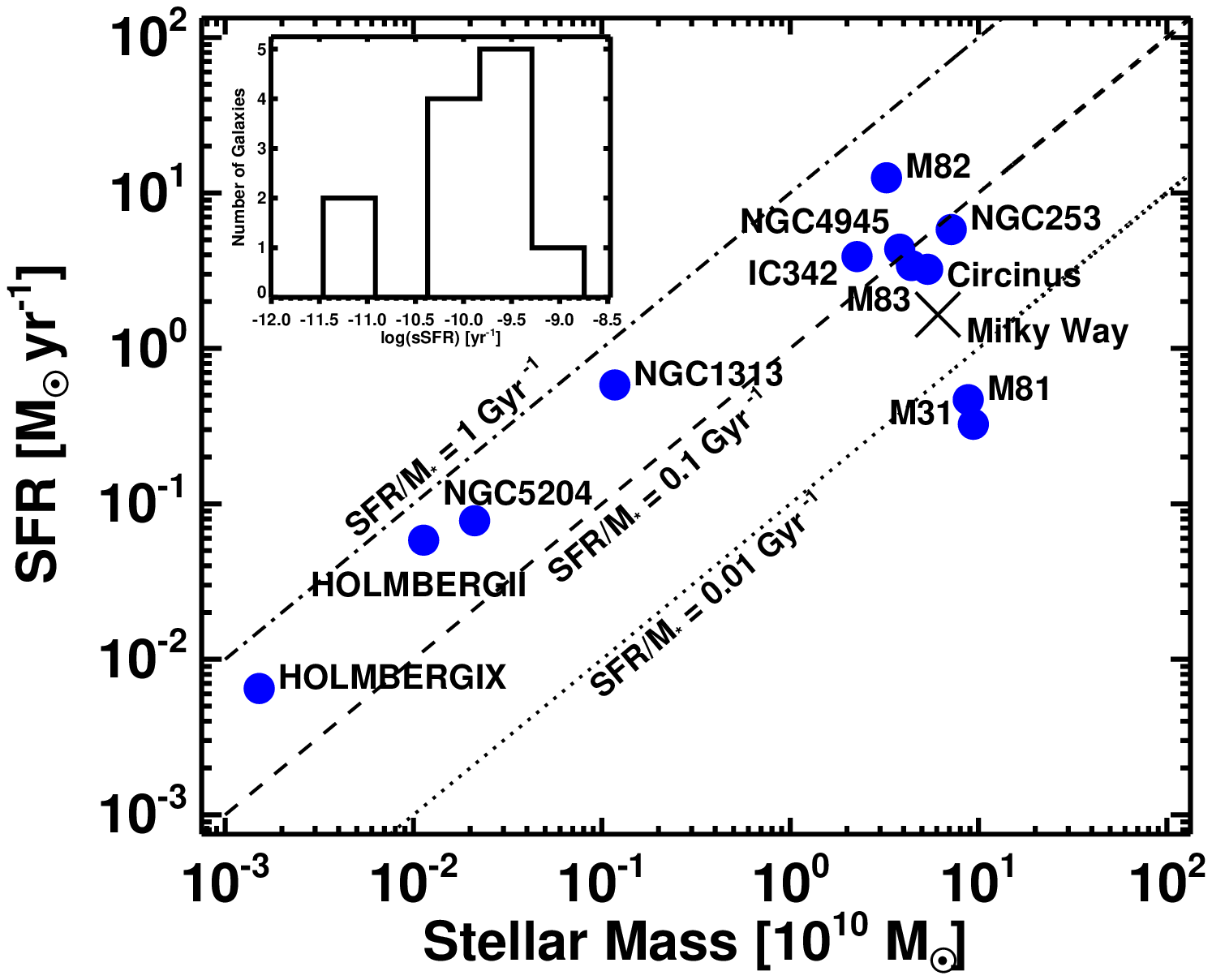}	
\caption{Shown are the star formation rates (SFR) and stellar masses $M_{\star}$ for all 12 galaxies in the sample (see Table \ref{tab:gals} and Section \ref{sec:smsfr} for SFR and stellar mass calculations). Lines of constant specific SFR (SFR/$M_{\star}$) are indicated to gauge the contribution from HMXBs. The Milky Way has been included as a comparison to the sample. The inset shows a histogram of the sSFR for all galaxies in the sample. The calculation of stellar mass and SFR assumed a \citet{kroupa-98} and \citet{kroupa04-01} IMF, respectively.}
\label{fig:sfrmass}
\end{figure*}

In Figure \ref{fig:sfrmass} we plot the SFR vs.\ the stellar mass for each galaxy in the sample. We also included the Milky Way for reference. The stellar mass and SFR for the Milky Way, $6.08\pm1.14\times10^{10}$ \msun\ and $1.65\pm0.19$ \sfr, respectively, were taken from \citet{licquia06-15}. Lines of constant specific SFR (sSFR) are indicated to help compare the relative amount of star formation per galaxy across a variety of stellar masses. 
One expects fractionally more HMXBs in galaxies with higher values of sSFR. Most of the galaxies in the sample (8 of 12) have stellar masses comparable to the Milky Way galaxy. There is a range of sSFR values with a peak around the value for the Milky Way. The \nustar\ archive represents a biased nearby galaxy sample that tends towards intermediate sSFR as seen in Figure \ref{fig:sfrmass}. This results from the relative lack of nearby massive elliptical galaxies that have low sSFR (e.g.\ Cen A, Maffei 1) and few nearby starbursts with large sSFR (e.g.\ NGC 253, M82).

\section{Observations}\label{sec:obs}

Table \ref{tab:obsid-info} summarizes the simultaneous/archival \nustar\ and \chandra/\xmmn\ observations for galaxies in the sample that were analyzed in this work (Table \ref{tab:gals}). We have excluded all \nustar\ observations of galaxies that are shorter than $\sim10$ ks (before data reduction) because it is not possible to robustly constrain the background. Each observatory's FOV covers the D$_{25}$ ellipse of all galaxies except the \nustar\ observations of M83 ($\sim90$\% of D$_{25}$, see \citealt{yukita06-16}), IC 342 ($\sim75$\% of D$_{25}$), and M31 ($\sim40$\% of D$_{25}$). There was non-contemporaneous archival X-ray data available for these galaxies, however, simultaneous data are particularly important for study of the highly variable XRB population. Thus, we prioritized simultaneous \chandra/\xmmn\ data for our analyses. Such data were available for 11 of the 12 galaxies; for M81 we relied solely upon archival data. Due to the high frequency of observations for M31, M82, and Holmberg IX, only some observations were simultaneous with \chandra/\xmmn. 
If simultaneous observations were not present, we used observations as close in time as possible. We used \chandra\ ACIS observations and \xmmn\ PN for 11 of the 12 galaxies because of the higher signal-to-noise ratio compared to \xmmn\ MOS. The \xmmn\ PN observation of M81 was taken in small window mode and so we used the MOS detector data in order to cover the entire FOV. 
This combination of \chandra/\xmmn\ observations ensures we have high spatial resolution X-ray data to create point source lists used in \nustar\ data processing (see Section \ref{sec:psd}).

\afterpage{
\clearpage
\startlongtable
\begin{deluxetable*}{c c  c  c  c  c  c  c  c  c  c  c }
\tablecaption{\nustar, \chandra, and \xmmn\ Observations of the \nustar\ Galaxy Sample \label{tab:obsid-info}}
\tabletypesize{\scriptsize}
\tablecolumns{12}
\tablehead{
\colhead{Galaxy} & \colhead{\nustar\ ObsID} & \colhead{Date} & \colhead{R.A.} & \colhead{Decl.} & \colhead{Livetime} & \colhead{Observatory} & \colhead{ObsID} &  \colhead{Date} & \colhead{R.A.} & \colhead{Decl.} & \colhead{Livetime}	\\
\colhead{} & \colhead{} & \colhead{} & \twocolhead{(J2000.0)} & \colhead{ks} & \colhead{} & \colhead{} &  \colhead{} & \twocolhead{(J2000.0)} & \colhead{ks}
}
\startdata
\multirow{5}{*}{Circinus} & 60002039002 & 2013 Jan 25 & 213.3888 & -65.3207 & 53.7 & \textit{XMM-Newton} & 0701981001 & 2013 Feb 03 & 213.1538 & -65.3638 & 21.1\\
 & 30002038002 & 2013 Feb 02 & 213.2684 & -65.3867 & 18.3 & \textit{XMM-Newton} & 0792382701 & 2016 Aug 23 & 213.1643 & -65.4209 & 12.3\\
 & 30002038004 & 2013 Feb 03 & 213.2347 & -65.3826 & 40.3 & \ldots & \ldots & \ldots & \ldots & \ldots & \ldots\\
 & 30002038006 & 2013 Feb 05 & 213.2318 & -65.3847 & 35.9 & \ldots & \ldots & \ldots & \ldots & \ldots & \ldots\\
 & 90201034002 & 2016 Aug 23 & 213.1020 & -65.3954 & 39.0 & \ldots & \ldots & \ldots & \ldots & \ldots & \ldots\\
\hline\multirow{3}{*}{Holmberg II} & 30001031002 & 2013 Sep 09 & 124.8517 & 70.6840 & 25.1 & \textit{XMM-Newton} & 0200470101 & 2004 Apr 15 & 124.9008 & 70.6784 & 28.5\\
 & 30001031003 & 2013 Sep 09 & 124.9776 & 70.6930 & 67.0 & \textit{XMM-Newton} & 0724810101 & 2013 Sep 09 & 124.8877 & 70.7336 & 4.0\\
 & 30001031005 & 2013 Sep 17 & 124.9655 & 70.7048 & 93.1 & \textit{XMM-Newton} & 0724810301 & 2013 Sep 17 & 124.8792 & 70.7334 & 5.5\\
\hline\multirow{10}{*}{Holmberg IX} & 30002033002 & 2012 Oct 26 & 149.4906 & 69.0590 & 27.1 & \textit{XMM-Newton} & 0693850801 & 2012 Oct 23 & 149.4712 & 69.0920 & 5.7\\
 & 30002033003 & 2012 Oct 26 & 149.5251 & 69.0561 & 76.4 & \textit{XMM-Newton} & 0693850901 & 2012 Oct 25 & 149.4716 & 69.0921 & 4.9\\
 & 30002033005 & 2012 Nov 11 & 149.5509 & 69.0710 & 33.2 & \textit{XMM-Newton} & 0693851001 & 2012 Oct 27 & 149.4682 & 69.0923 & 3.9\\
 & 30002033006 & 2012 Nov 11 & 149.4688 & 69.0646 & 30.0 & \textit{XMM-Newton} & 0693851701 & 2012 Nov 12 & 149.4493 & 69.0916 & 6.2\\
 & 30002033008 & 2012 Nov 14 & 149.5119 & 69.0671 & 10.7 & \textit{XMM-Newton} & 0693851801 & 2012 Nov 14 & 149.4477 & 69.0912 & 6.6\\
 & 30002033010 & 2012 Nov 15 & 149.5060 & 69.0666 & 38.5 & \textit{XMM-Newton} & 0693851101 & 2012 Nov 16 & 149.4466 & 69.0908 & 2.6\\
 & 30002034002 & 2014 May 02 & 149.3835 & 69.0447 & 49.7 & \ldots & \ldots & \ldots & \ldots & \ldots & \ldots\\
 & 30002034004 & 2014 Nov 15 & 149.4594 & 69.0778 & 54.5 & \ldots & \ldots & \ldots & \ldots & \ldots & \ldots\\
 & 30002034006 & 2015 Apr 06 & 149.4038 & 69.0677 & 44.3 & \ldots & \ldots & \ldots & \ldots & \ldots & \ldots\\
 & 30002034008 & 2015 May 16 & 149.3901 & 69.0381 & 50.3 & \ldots & \ldots & \ldots & \ldots & \ldots & \ldots\\
\hline\multirow{4}{*}{IC 342} & 30002032002 & 2012 Aug 10 & 56.5500 & 68.0815 & 17.5 & \textit{XMM-Newton} & 0693850601 & 2012 Aug 11 & 56.4305 & 68.1026 & 24.3\\
 & 30002032003 & 2012 Aug 10 & 56.5393 & 68.1033 & 80.6 & \textit{XMM-Newton} & 0693851301 & 2012 Aug 17 & 56.4315 & 68.1038 & 27.8\\
 & 30002032005 & 2012 Aug 16 & 56.5439 & 68.1020 & 112.8 & \ldots & \ldots & \ldots & \ldots & \ldots & \ldots\\
 & 90201039002 & 2016 Oct 16 & 56.4568 & 68.1026 & 44.4 & \ldots & \ldots & \ldots & \ldots & \ldots & \ldots\\
\hline\multirow{16}{*}{M31} & 50026001002 & 2015 Feb 06 & 10.8508 & 41.3004 & 106.4 & \textit{Chandra} & 17008 & 2015 Oct 06 & 11.0654 & 41.3876 & 49.1\\
 & 50026002001 & 2015 Feb 08 & 11.0826 & 41.3762 & 109.2 & \textit{Chandra} & 17011 & 2015 Oct 08 & 11.3757 & 41.7235 & 49.4\\
 & 50026003002 & 2015 Feb 11 & 11.3306 & 41.5763 & 107.0 & \textit{Chandra} & 17010 & 2015 Oct 19 & 11.2461 & 41.5343 & 49.4\\
 & 50026001004 & 2015 Mar 01 & 10.8535 & 41.2919 & 104.3 & \textit{Chandra} & 17009 & 2015 Oct 26 & 11.0174 & 41.5775 & 49.4\\
 & 50026002003 & 2015 Mar 06 & 11.0821 & 41.3688 & 104.0 & \ldots & \ldots & \ldots & \ldots & \ldots & \ldots\\
 & 50026003003 & 2015 Mar 08 & 11.3277 & 41.5753 & 15.3 & \ldots & \ldots & \ldots & \ldots & \ldots & \ldots\\
 & 50111001002 & 2015 Jun 26 & 10.8878 & 41.3078 & 104.1 & \ldots & \ldots & \ldots & \ldots & \ldots & \ldots\\
 & 50110001002 & 2015 Jul 25 & 10.8817 & 41.3043 & 52.2 & \ldots & \ldots & \ldots & \ldots & \ldots & \ldots\\
 & 50110002002 & 2015 Jul 27 & 11.1122 & 41.3753 & 34.0 & \ldots & \ldots & \ldots & \ldots & \ldots & \ldots\\
 & 50110002004 & 2015 Jul 29 & 11.1126 & 41.3749 & 30.6 & \ldots & \ldots & \ldots & \ldots & \ldots & \ldots\\
 & 50110001004 & 2015 Aug 01 & 10.8822 & 41.2990 & 71.4 & \ldots & \ldots & \ldots & \ldots & \ldots & \ldots\\
 & 50110002006 & 2015 Aug 05 & 11.1047 & 41.3758 & 38.0 & \ldots & \ldots & \ldots & \ldots & \ldots & \ldots\\
 & 50110003002 & 2015 Aug 17 & 11.3425 & 41.5610 & 96.4 & \ldots & \ldots & \ldots & \ldots & \ldots & \ldots\\
 & 50101001002 & 2015 Sep 13 & 10.6413 & 41.2401 & 98.6 & \ldots & \ldots & \ldots & \ldots & \ldots & \ldots\\
 & 50111002002 & 2015 Oct 10 & 11.1285 & 41.3694 & 94.8 & \ldots & \ldots & \ldots & \ldots & \ldots & \ldots\\
 & 50111003002 & 2015 Oct 23 & 11.3704 & 41.5913 & 105.4 & \ldots & \ldots & \ldots & \ldots & \ldots & \ldots\\
\hline\multirow{1}{*}{M81} & 60101049002 & 2015 May 18 & 148.8056 & 69.0481 & 181.5 & \textit{XMM-Newton} & 0111800101 & 2001 Apr 22 & 148.8992 & 69.0380 & 76.8\\
\hline\multirow{17}{*}{M82} & 80002092002 & 2014 Jan 23 & 148.8576 & 69.6987 & 59.7 & \textit{XMM-Newton} & 0206080101 & 2004 Apr 21 & 148.9516 & 69.6523 & 41.4\\
 & 80002092004 & 2014 Jan 25 & 148.8647 & 69.7010 & 77.4 & \textit{XMM-Newton} & 0560590301 & 2009 Apr 29 & 148.9692 & 69.6508 & 11.9\\
 & 80002092006 & 2014 Jan 28 & 148.8511 & 69.7064 & 271.4 & \textit{XMM-Newton} & 0657802301 & 2011 Nov 21 & 148.9263 & 69.7053 & 8.1\\
 & 80002092007 & 2014 Feb 04 & 148.8804 & 69.6876 & 260.0 & \textit{Chandra} & 16580 & 2014 Feb 03 & 148.9116 & 69.6716 & 46.8\\
 & 80002092008 & 2014 Feb 10 & 148.8779 & 69.6835 & 26.5 & \textit{Chandra} & 17578 & 2015 Jan 16 & 148.8175 & 69.7153 & 9.1\\
 & 80002092009 & 2014 Feb 11 & 148.8538 & 69.6884 & 98.2 & \textit{Chandra} & 16023 & 2015 Jan 20 & 148.9160 & 69.6913 & 10.0\\
 & 80002092011 & 2014 Mar 03 & 148.8121 & 69.6800 & 98.2 & \textit{Chandra} & 17678 & 2015 Jun 21 & 149.1251 & 69.6900 & 9.3\\
 & 50002019002 & 2015 Jan 15 & 148.9231 & 69.7165 & 23.6 & \textit{Chandra} & 18062 & 2016 Jan 26 & 149.0273 & 69.7373 & 23.2\\
 & 50002019004 & 2015 Jan 19 & 148.9173 & 69.7115 & 134.0 & \textit{Chandra} & 18063 & 2016 Feb 24 & 148.9199 & 69.7380 & 23.2\\
 & 90101005002 & 2015 Jun 20 & 148.9232 & 69.6533 & 30.9 & \textit{Chandra} & 18064 & 2016 Apr 05 & 148.8330 & 69.7179 & 23.2\\
 & 80202020002 & 2016 Jan 26 & 148.9121 & 69.6979 & 30.2 & \textit{Chandra} & 18068 & 2016 Apr 24 & 149.0902 & 69.6418 & 23.2\\
 & 80202020004 & 2016 Feb 23 & 148.8794 & 69.6918 & 23.7 & \textit{Chandra} & 18069 & 2016 Jun 03 & 149.1305 & 69.6724 & 23.2\\
 & 80202020006 & 2016 Apr 05 & 148.8780 & 69.6804 & 26.5 & \textit{Chandra} & 18067 & 2016 Jul 01 & 149.1182 & 69.7019 & 24.1\\
 & 30101045002 & 2016 Apr 15 & 148.8957 & 69.6743 & 163.3 & \ldots & \ldots & \ldots & \ldots & \ldots & \ldots\\
 & 80202020008 & 2016 Apr 24 & 148.8943 & 69.6729 & 35.7 & \ldots & \ldots & \ldots & \ldots & \ldots & \ldots\\
 & 90202038002 & 2016 Oct 07 & 149.0146 & 69.6776 & 38.7 & \ldots & \ldots & \ldots & \ldots & \ldots & \ldots\\
 & 90202038004 & 2016 Nov 30 & 148.9574 & 69.6900 & 36.3 & \ldots & \ldots & \ldots & \ldots & \ldots & \ldots\\
\hline\multirow{6}{*}{M83} & 50002043002 & 2013 Aug 07 & 204.2200 & -29.8677 & 42.3 & \textit{XMM-Newton} & 0723450101 & 2013 Aug 07 & 204.2760 & -29.8969 & 41.6\\
 & 50002043004 & 2013 Aug 09 & 204.2270 & -29.8744 & 79.7 & \textit{XMM-Newton} & 0723450201 & 2014 Jan 11 & 204.2626 & -29.8407 & 25.1\\
 & 50002043006 & 2013 Aug 21 & 204.2160 & -29.8675 & 42.5 & \textit{Chandra} & 16024 & 2014 Jun 07 & 204.2509 & -29.8767 & 29.6\\
 & 50002043008 & 2014 Jan 19 & 204.3021 & -29.8458 & 81.0 & \ldots & \ldots & \ldots & \ldots & \ldots & \ldots\\
 & 50002043010 & 2014 Jun 04 & 204.2388 & -29.8793 & 70.4 & \ldots & \ldots & \ldots & \ldots & \ldots & \ldots\\
 & 50002043012 & 2014 Jun 07 & 204.2380 & -29.8791 & 109.4 & \ldots & \ldots & \ldots & \ldots & \ldots & \ldots\\
\hline\multirow{5}{*}{NGC 1313} & 30002035002 & 2012 Dec 16 & 49.6268 & -66.5225 & 100.2 & \textit{XMM-Newton} & 0693850501 & 2012 Dec 16 & 49.6311 & -66.4993 & 60.9\\
 & 30002035004 & 2012 Dec 21 & 49.6351 & -66.5268 & 126.4 & \textit{Chandra} & 14676 & 2012 Dec 17 & 49.5101 & -66.5812 & 9.8\\
 & 80001032002 & 2014 Jul 05 & 49.5764 & -66.4456 & 63.0 & \textit{XMM-Newton} & 0693851201 & 2012 Dec 22 & 49.6330 & -66.5027 & 61.3\\
 & 90201050002 & 2017 Mar 29 & 49.5047 & -66.4958 & 64.1 & \textit{Chandra} & 15594 & 2012 Dec 23 & 49.5102 & -66.5812 & 9.8\\
 & \ldots & \ldots & \ldots & \ldots & \ldots & \textit{XMM-Newton} & 0742590301 & 2014 Jul 05 & 49.5016 & -66.4904 & 50.6\\
\hline\multirow{3}{*}{NGC 253} & 50002031002 & 2012 Sep 01 & 11.9200 & -25.2719 & 156.5 & \textit{Chandra} & 13830 & 2012 Sep 02 & 11.8901 & -25.2803 & 19.7\\
 & 50002031004 & 2012 Sep 15 & 11.9143 & -25.2805 & 157.6 & \textit{Chandra} & 13831 & 2012 Sep 18 & 11.8941 & -25.2818 & 19.7\\
 & 50002031006 & 2012 Nov 16 & 11.8900 & -25.3178 & 124.5 & \textit{Chandra} & 13832 & 2012 Nov 16 & 11.8967 & -25.2921 & 19.7\\
\hline\multirow{3}{*}{NGC 4945} & 60002051002 & 2013 Feb 10 & 196.4060 & -49.4605 & 45.0 & \textit{XMM-Newton} & 0204870101 & 2004 Jan 10 & 196.3473 & -49.4435 & 18.3\\
 & 60002051004 & 2013 Jun 15 & 196.3271 & -49.4610 & 54.4 & \textit{Chandra} & 14985 & 2013 Apr 20 & 196.3682 & -49.4664 & 68.7\\
 & 60002051006 & 2013 Jul 05 & 196.3334 & -49.4907 & 34.2 & \textit{Chandra} & 14984 & 2013 Apr 25 & 196.3686 & -49.4668 & 128.8\\
\hline\multirow{4}{*}{NGC 5204} & 30002037002 & 2013 Apr 19 & 202.3756 & 58.4400 & 95.7 & \textit{XMM-Newton} & 0693851401 & 2013 Apr 21 & 202.3691 & 58.3993 & 12.3\\
 & 30002037004 & 2013 Apr 29 & 202.3658 & 58.4413 & 77.3 & \textit{Chandra} & 14675 & 2013 Apr 21 & 202.3968 & 58.4149 & 9.8\\
 & \ldots & \ldots & \ldots & \ldots & \ldots & \textit{XMM-Newton} & 0693850701 & 2013 Apr 29 & 202.3733 & 58.3966 & 6.1\\
 & \ldots & \ldots & \ldots & \ldots & \ldots & \textit{Chandra} & 15603 & 2013 May 01 & 202.4004 & 58.4124 & 9.8\\
\enddata
\tablecomments{Simultaneous/archival \nustar\ and \chandra/\xmmn\ observations for galaxies in the sample (Table \ref{tab:gals}). Livetime represents the exposure time of the cleaned event file (Section \ref{sec:data}).} 

\end{deluxetable*}
}

\section{Data Analysis}	\label{sec:data}

\subsection{NuSTAR}

\nustar\ data were reduced using \textsc{heasoft} v6.19/$\textsc{nustardas}$ v1.6.0 along with CALDB version 20161021. We reprocessed all Level 1 event files using \textsc{nupipeline} to obtain cleaned level 2 event files. This script measured the alignment of the mast connecting the focal plane detectors and optics, applied gain and dead time correction, flagged bad/hot pixels, filtered good time intervals and screened events based on grade and status, and converted raw detector positions into sky coordinates. The script also filtered out observational data during passages through the South Atlantic Anomaly that caused periods of high background, accomplished by setting the parameters $\texttt{SAAMODE}$ to $\textit{strict}$ and $\texttt{TENTACLE}$ to $\textit{yes}$. 
While reducing exposure times by $\sim10$\%, these parameters decreased the uncertainty associated with our background calculations. 
We also inspected light curves to ensure no flares were present. We only used observing mode 01 event data for both focal plane modules A and B throughout our analysis. The resulting total exposure times for each observation after applying all these corrections/filters are listed in Table \ref{tab:obsid-info}. 

We computed the background for each telescope (FPMA/B) in each observation for each galaxy using the publicly available tool \textsc{nuskybgd} \citep{wik09-14}. The \nustar\ background is comprised of stray light (from the cosmic X-ray background or bright sources outside the FOV), reflected solar X-rays, the focused cosmic X-ray background, and the instrumental background. Due to the spectral and spatial variation of the background across even individual detectors, accurate modeling is required to produce background images at source locations. For each observation, we created four source-free\footnote{Created by masking out visually identifiable sources in an image} box regions for each of the detectors ($0-3$) of each telescope (A/B) for fitting a background model (see \citealt{wik09-14} for an example). 
Stray light from bright sources within approximately $1\degr-5$\degr\ of the optical axis can cause significant contamination in addition to the aperture background component. Stray light was only an issue for M83 (see \citealt{yukita06-16}). We were able to overcome this issue by excluding data from telescope B in the Jan 2014 observation and excluding telescope A data for the remaining observations.

\subsection{Chandra and XMM-Newton}

Reduction of \chandra\ ACIS observations was performed using the \textsc{chandra interactive analysis of observations (ciao)} tools package version 4.7.2 \citep{fruscione07-06} and the \textsc{chandra calibration database (caldb)} version 4.8 \citep{graessle07-06}. \chandra\ data were reduced using the \textsc{chandra\_repro} script. Events files were filtered using the standard (ASCA) grades ($0, 2-4, 6$), status bits ($0$), good time intervals, and CCD chips (I0-I3 for ACIS-I and S3 for ACIS-S). 
We then created exposure maps and exposure-corrected images using \textsc{fluximage} with a binsize of 1 in the $4-8$ keV energy band. Source lists were created with \textsc{wavdetect} using the $\sqrt{2}$ series from 1 to 8 for the $\textit{scales}$ parameter and corresponding exposure maps to reduce false positives. Default settings were used for all other parameters.

\xmmn\ data were reduced using \textsc{sas} v.16.0.0. Level 1 event data were processed using the \textsc{epchain} and \textsc{emchain} scripts. High-background intervals were filtered using the \textsc{pn-filter} and \textsc{mos-filter} scripts. We created $4-10$ keV images using single- and double-pixel events (PATTERN $0-4$) for the PN detector and single- to quadruple-pixel events (PATTERN $0-12$) from the MOS detector. 
Source lists were created using \textsc{edetect\_chain} with 16 spline nodes and likelihood threshold of 6 to include faint sources.

\subsection{NuSTAR Point Source Detection via PSF Fitting}		\label{sec:psd}

Point source detection in \nustar\ images can be complicated in regions with a high spatial density of comparably bright point sources such as those present in many galaxies. The moderate-quality 18\arcsec\ PSF core FWHM can lead to source confusion and/or PSF contamination by sources within 58\arcsec\ (corresponding to the \nustar\ PSF half-power diameter). 
Therefore we used simultaneous or archival \chandra\ and \xmmn\ observations to create point source lists to localize and determine source characteristics in the \nustar\ observations. For each galaxy we merged the \nustar\ imaging data from telescopes A and B to increase the signal-to-noise ratio. Exposure times for \chandra\ and/or \xmmn\ observations were sufficient to reach below the sensitivity limits of the combined \nustar\ observations for each galaxy. The methodology outlined here follows that in \citet{wik12-14}.

\subsubsection{PSF and Response File Generation}

Due to \nustars\ 58\arcsec\ PSF half-power diameter and 18\arcsec\ PSF core FWHM, source confusion is an issue in crowded fields. To accurately determine a source's count rate, we modelled the PSF of each source to deconvolve the contribution from nearby sources. The PSF shape changes more dramatically once sources are $>3$\arcmin\ off-axis, such that pointing variations over the course of an observation can induce errors in the PSF shape. We created composite PSFs for each source using PSF model images from the CALDB that were weighted by the time spent at each off-axis angle. The same procedure was applied to determine the vignetting function, which represents the effective area and is dependent on both off-axis angle and energy. The average vignetting of an image with a given energy band is found by weighting the vignetting function over that energy range by a power law spectrum typical of XRBs with $\Gamma=1.7$, to ensure the calculated vignetting function is appropriately weighted for the sources we are studying. The weighting was used to prevent the higher-energy vignetting from influencing our results, due to \nustar\ having a strongly energy-dependent vignetting function \citep[e.g.][]{harrison06-13, madsen09-15}. ARFs were created by multiplying the on-axis ARF from the \textsc{CALDB} by the weighted vignetting function. RMFs were created using the appropriate response file from the \textsc{CALDB}. ARFs and RMFs were created for each source and were used to obtain corrected count rates. The overall result was count rates that were the same as those expected for an on-axis source. 

\subsubsection{PSF Fitting and Astrometric Alignment}	\label{sec:psffit}

Using the previously described techniques for generating data products, we determined the astrometric shifts for every observation and count rates via PSF fitting. We used the \chandra/\xmmn\ source positions as the reference coordinate system and used the brightest few sources to determine the ($x/y$) image shifts of the \nustar\ data. This was completed for every ObsID in the galaxy sample using the \full keV images. These shifts were then applied to the images and the PSF fitting routine was executed again to determine count rates without fitting for image shifts. To reach the lowest sensitivity limits for each galaxy, we merged data from both \nustar\ telescopes A and B and combined all observations.

\begin{figure*}[!ht]
\includegraphics[width=1.0\textwidth]{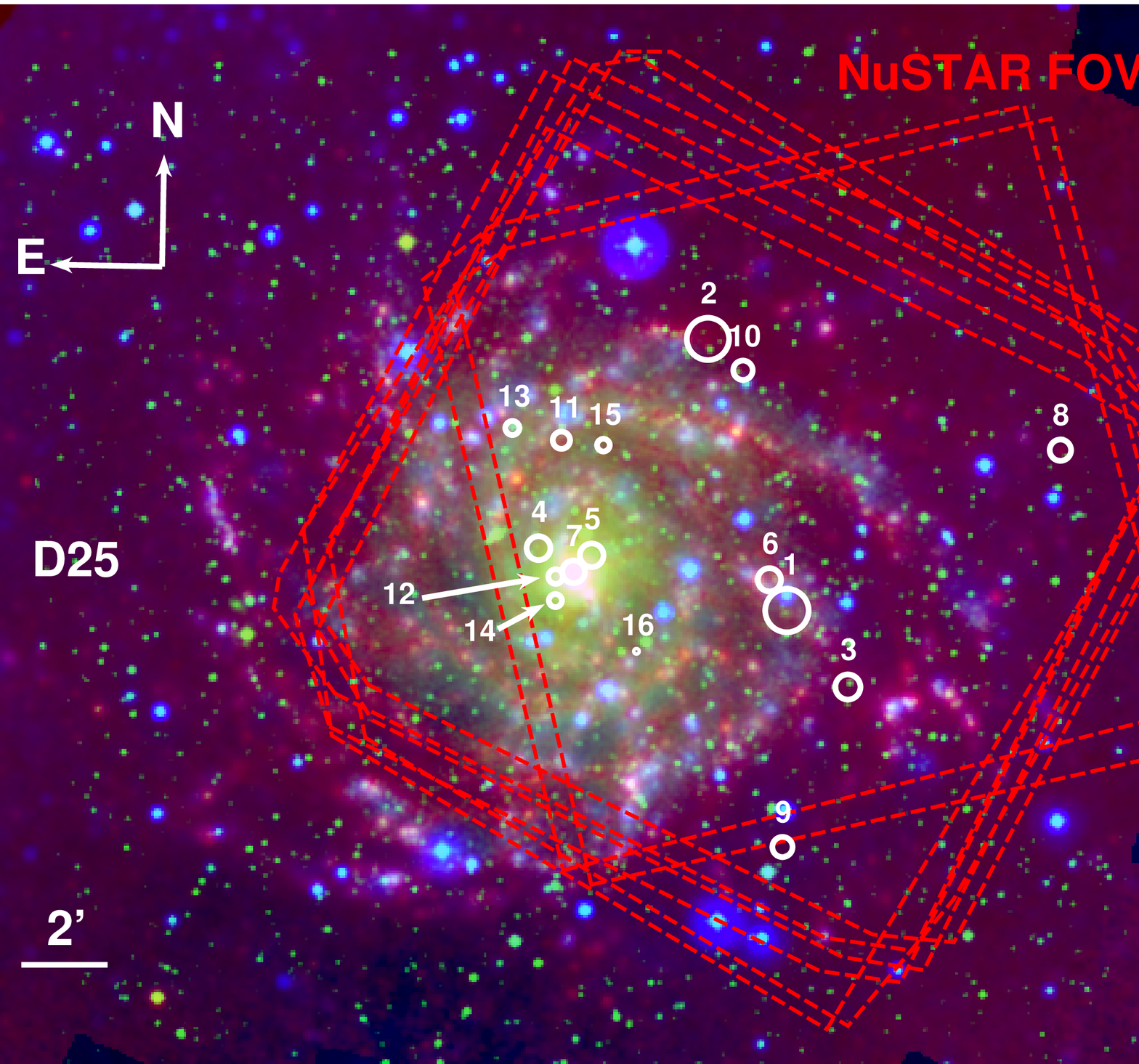}
\caption{Example images for the \nustar\ observations of IC 342. Left: Three-color image of IC 342 from \galex\ NUV (blue), H$\alpha$ (green), and \spitzer\ 24 $\micron$ (red). The \nustar\ observations of IC 342 are outlined in red (Table \ref{tab:obsid-info}, telescopes A and B), whereas the D$_{25}$ ellipse (white) is larger than the FOV (Table \ref{tab:gals}) and cuts through the top right corner. Numbers represent the detected \nustar\ X-ray point sources from Table \ref{tab:rates}, where circle sizes are proportional to \full keV count rate. Right: False-color \nustar\ image smoothed by a Gaussian of 7 pixels. Magenta numbers are identical to the left panel.} \label{fig:ic342-img}
\end{figure*}

The \chandra/\xmmn\ source positions for the brightest sources were used as inputs for the fitting procedure. The composite PSFs and response files were created for each source in a rectangular region that included all sources with overlapping PSFs in the region. A background image was created at each source location using the background model computed for each focal plane module in an observation. A model image was created by combining the PSF and background images, which was then fit to the actual data using the Cash statistic \citep{cash03-79}. The Cash statistic was minimized using the Amoeba algorithm \citep{press-02}, which is ideal for models without derivatives. Count-rate errors were estimated by completing 1000 Monte Carlo simulations of the best-fitting model. During the process if a better fit was found then the original model parameters were reset and fitting was repeated. The 90\% uncertainty range was calculated from the inner 900 sorted values for each simulated parameter. We determined count rates in the soft ($S$, $4-6$ keV), medium ($M$, $6-12$ keV), hard, ($H$, \hard keV), and full ($F$, \full keV) \nustar\ energy bands because they provided the most robust separation between sources (see Section \ref{sec:diag}). The same source positions were used when fitting each energy band. 
We omitted all sources with count rates $<10^{-4}$ counts s$^{-1}$ in each energy band and required that a source was detected (count rate above the 90\% confidence threshold) in at least one energy band. In Figure \ref{fig:ic342-img} we show an example of the \nustar\ data and detected \nustar\ point sources for IC 342.

\subsubsection{Simultaneous PSF Fitting}	\label{sec:simul}

Our goals of identifying the accretion states and compact object types of \nustar\ point sources rely on using hardness-intensity and color-color diagrams. However, the hardness ratio uncertainties can be prohibitively large due to error propagation from count rates. To improve our methodology we used the technique developed by D. Wik et al. 2018 (in prep.). Briefly, the soft, medium, and hard energy band images were fit simultaneously -- with the PSF models described in Section \ref{sec:psffit} -- using hardness ratios and the full (\full keV) count rate as free parameters instead of the $S$, $M$, and $H$ rates themselves. The hardness ratios $HR1=(M-S)/(M+S)$ and $HR2=(H-M)/(H+M)$, as well as the full energy band $F=S+M+H$ were free parameters in the fit instead of fitting individual energy bands to determine count rates. In order to use $F$, $HR1$, and $HR2$ as free parameters we defined the variables $C=(1-HR1)/(1+HR1)$ and $D=(1+HR2)/(1-HR2)$ to convert to count rates in each energy band:
\begin{align}
S & = \frac{FC}{1+C+D},\ M = \frac{F}{1+C+D},\ H = \frac{FD}{1+C+D} 	\label{eq:crate}
\end{align}
By changing the free parameters, we were able to directly calculate uncertainty ranges on the hardness ratios from the data itself, which avoids error propagation (and any assumptions behind that method) from introducing new systematic uncertainties. This method allowed uncertainties for fainter sources to be calculated more accurately, which allowed better limits to be derived when a source was not detected in one of the energy bands. We used the same source positions that were used for fitting each energy band in Section \ref{sec:psffit} such that there was no variation in source positions between methods. 
In Table \ref{tab:rates} we list the count rates in each energy band ($S$, $M$, $H$, and $F$) and their 90\% upper and lower confidence intervals that were derived from individual PSF fitting of each energy band (Section \ref{sec:psffit}). The $S$, $M$, and $H$ count rates were not derived from simultaneous PSF fitting using equation \ref{eq:crate} because their propagated uncertainties are poorly constrained compared to individual PSF fitting of each energy band. Sources were grouped by galaxy and numbered in order of decreasing \full keV count rate. We also show the hardness ratios $HR1$ and $HR2$ and their uncertainties from the simultaneous PSF fitting of the soft, medium, and hard energy bands summarized in this Section. The \full keV luminosity was estimated by converting the \full keV count rate (derived from simultaneous PSF fitting summarized in this Section) assuming a spectral model typical of XRBs, with $\Gamma=1.7$ and $N_{\rm{H}}$ values from Table \ref{tab:gals}. We estimated the influence of using $\Gamma=1$ for sources classified as pulsars (see Section \ref{sec:diag}) and found a $<10\%$ difference in count rates, corresponding to $<0.05$ shift in color-space for pulsars. This does not change source classifications and is smaller than the uncertainties on the hardness ratios. In Table \ref{tab:rates} we presented the count rates from individual PSF fitting in each energy band ($S$, $M$, $H$, and $F$) but we used $HR1$, $HR2$, and $F$ from simultaneous PSF fitting for our X-ray source diagnostics (Section \ref{sec:diag}) due to the improved constraints on uncertainties.

\section{Results and Discussion}	\label{sec:resdis}

In this section we classify BH and NS using \nustar\ hardness-intensity and color-color diagrams. With this information we study the characteristics of compact object types/accretion states and trends with sSFR. We investigate the correlation of XRB luminosity with SFR and stellar mass. Lastly, we study the XLF of the \nustar\ sample and determine the ratio of BH to NS using BH and NS-only XLFs.

\subsection{NuSTAR XRB Diagnostic Diagrams}		\label{sec:diag}

It has long been understood that there are significant observable changes in the X-ray spectra of accreting BH and NS systems, which give an indication of changes in the underlying accretion state (e.g.\ the extent of the accretion disk that dominates in the softer X-rays versus non-thermal/coronal components that are more X-ray hard; see review by \citealt{done12-07}). These changes may, in large part, be directly linked to accretion physics phenomena and have advanced our understanding of the overall geometry of accreting compact objects. These states form the basis of the diagnostic diagrams we use in our work with \nustar, so we begin with a short review of current understanding of such spectral state changes.

\uhuru\ observations of Cygnus X-1 by \citet{tananbaum10-72} revealed a state change where the $2-6$ keV X-ray intensity decreased by a factor of 4 and the $10-20$ keV X-ray intensity doubled. Along with the simultaneous brightening of the radio counterpart, this result indicated that spectral changes signified important changes in accretion physics of BH XRBs. Following 14 years of extensive monitoring with \rxtet, there are now thousands of high signal-to-noise spectra and fairly well-understood phenomena for outbursts and state transitions for BH/NS as a population \citep[e.g][]{maccarone01-03, mcclintock04-06, done12-07, church03-14, tetarenko02-16}. BH in the hard state produce hard thermal Comptonized spectra that can be described by a power law with a photon index $\Gamma\sim1.7$ with a cutoff at $\sim100$ keV. The BH soft state has a spectrum that is dominated by a disk blackbody component that peaks at $\sim1$ keV and a weak power law tail extending to 500 keV with photon index $\Gamma\sim2$. The BH intermediate state is a transitional stage between the hard and soft states \citep[e.g.][]{mcclintock04-06}, exhibiting a soft spectrum as the thermal disk component appears with increased mass accretion rate. In addition, the hard power-law component steepens to $\Gamma\sim2-2.5$. Almost all Galactic BH XRBs were found to follow the same hysteresis pattern in a hardness-intensity diagram \citep[][see below for a more detailed discussion]{maccarone01-03, done12-07}. The only sources that do not fit this pattern are Cygnus X-1 and X-3 \citep{smith04-02}, where Cygnus X-1 happens to be the only bright BH HMXB in the Galaxy (Cygnus X-3 is a BH candidate HMXB). Given that our sample is comprised of late-type galaxies, many with ongoing star-formation, the majority of sources we detect will be bright HMXBs. Therefore we must exercise caution when interpreting BH accretion states in our sample. The drastic spectral changes that occur in BH allow them to be uniquely identified by their accretion state using hardness-intensity and color-color diagnostics \citep[e.g.][]{remillard09-06, done12-07}.

NS occupy a much narrower band in hardness-intensity and color-color diagrams when compared with BH. Due to the small dynamic range of NS colors and the uncertainties associated with extragalactic sources, we are unable to robustly separate NS accretion states. Instead, we group all Z/Atoll sources (non-magnetized NS) together, which are distinct from the harder spectra of young, magnetized accreting pulsars. The spectra of accreting pulsars are usually best described by a hard power law with photon index $\Gamma\sim1$ and a cutoff at $\sim20$ keV \citep[e.g.][]{white07-83}. Z-track sources are named based on the Z pattern they trace out in a hardness-intensity diagram, through the horizontal, normal and flaring branches \citep{hasinger11-89, schulz11-89}. Atoll sources are less luminous ($<10^{38}$ \es) and display island and banana (lower and upper) states. The spectra of Z/Atoll NS vary, with a non-thermal Comptonized component dominating their emission. The high-energy cutoff of the Comptonized emission is $\approx6$ keV for sources $>10^{37}$ \es. There is also a thermal disk component that peaks between $1-2$ keV, which for Atoll sources is weak in the island state and can be very strong in the banana state, but for bright Z sources it can rise to 50\% of the total luminosity \citep{church03-14}. This rich phenomenology enables us to classify X-ray point sources as BH or NS based on luminosities and colors.

A. Zezas (private communication) has completed detailed spectral fitting of $\sim2500$ Galactic \rxtet\ PCA observations of 6 extensively-studied BH and 9 pulsars, where the compact object and orbital properties are extremely well constrained. The spectral library for BH \citep{sobolewska04-09} and pulsars \citep[e.g.][]{reig03-11} were used to characterize each source class/state. The best-fitting results were converted from the \rxtet\ to the \nustar\ energy bands. We selected the soft ($S$, $4-6$ keV), medium ($M$, $6-12$ keV), hard, ($H$, \hard keV), and full ($F$, \full keV) \nustar\ energy bands because they provided the most robust separation between sources. 
Following work from \citet{wik12-14} and \citet{yukita06-16}, we created \nustar\ diagnostic diagrams to determine global properties of the point source population, specifically the distribution of compact object types and accretion states.

In Figure \ref{fig:empty} we show the hardness-intensity (left) and color-color (right) diagrams. Due to the overlap between different accretion states and source types in both diagnostic diagrams, there remain some ambiguities in these diagnostics for some sources, even when the statistics are excellent. For the \full keV count rates, sources near the detection limit naturally have larger uncertainties compared to the brightest sources. 

In Figures \ref{fig:ic342}-\ref{fig:m82} we show the hardness-intensity and color-color plots for each galaxy in our sample. We grouped galaxies in Figures \ref{fig:holmiiix5204} and \ref{fig:cirngc4945} with similar sSFR. Uncertainties shown represent the 90\% confidence interval and delineated regions on the color-color diagram are approximations to isolate different accretion states and source types. Numbers label point sources by decreasing \full keV count rate. 

Due to the many observations with varying cadence that have been co-added for each galaxy, the count rates and hardness ratios represent averages, potentially hiding any variability. \citet{wik12-14} was able to study the multi-epoch properties of the brightest 8 sources in NGC 253 and found only one underwent a state transition, while two sources varied slightly in flux. \citet{yukita06-16} found no statistically significant variability among M83 sources over 3 epochs. However, while most extragalactic sources studied may be persistent, galaxies such as M82 with longer exposure times and high cadence require more detailed investigation. In Figure \ref{fig:allgals} we show all point sources from all galaxies in the sample. The left panel shows that most point sources overlap with the BH XRB intermediate state, which is degenerate with the Z/Atoll source loci. However, the color-color plot allows for clearer separations between source types and shows that some of these sources are likely NS accretors.

\begin{subfigures}

\begin{figure*}
\begin{tabular}{cc}
\includegraphics[width=1.0\columnwidth]{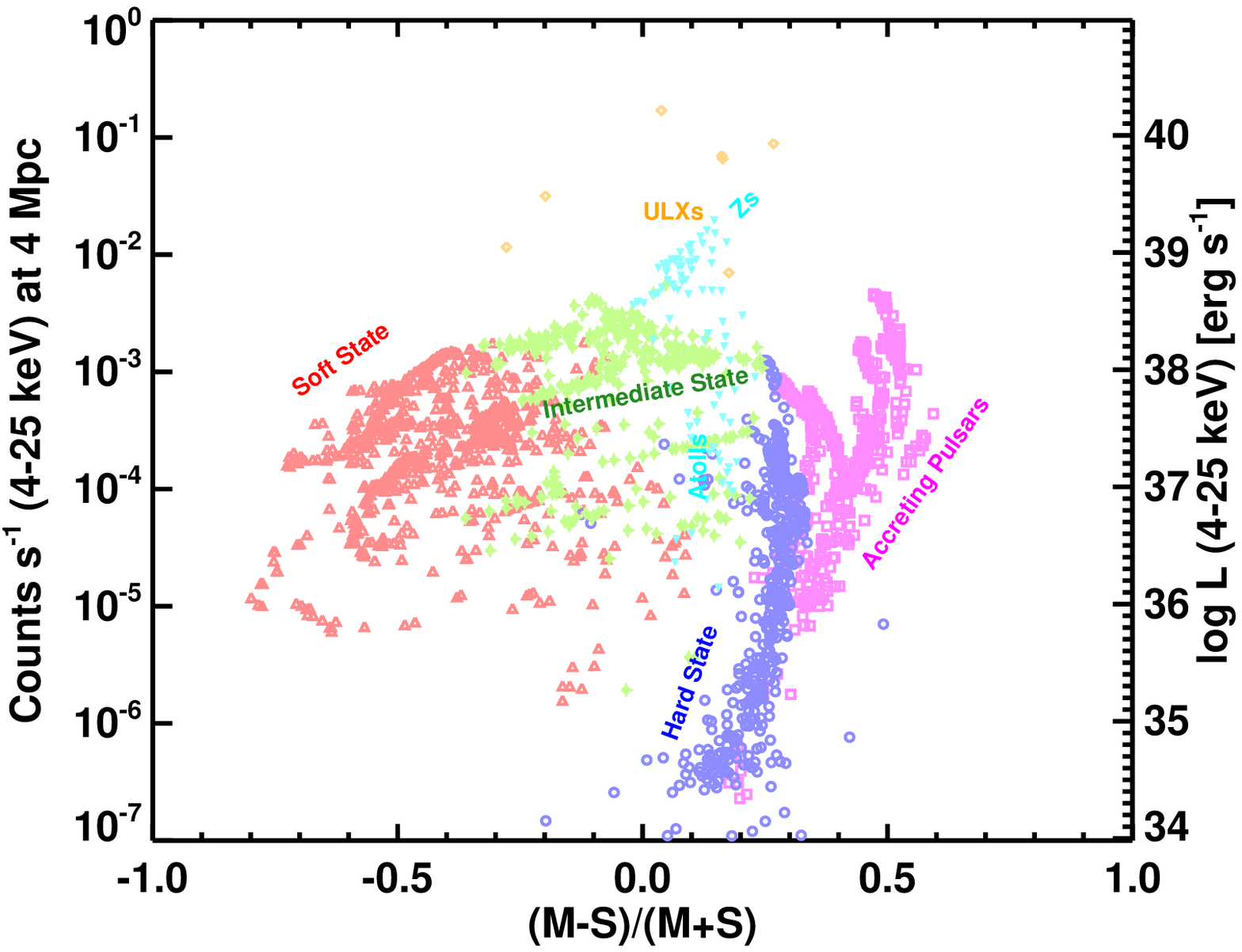}
\includegraphics[width=1.0\columnwidth]{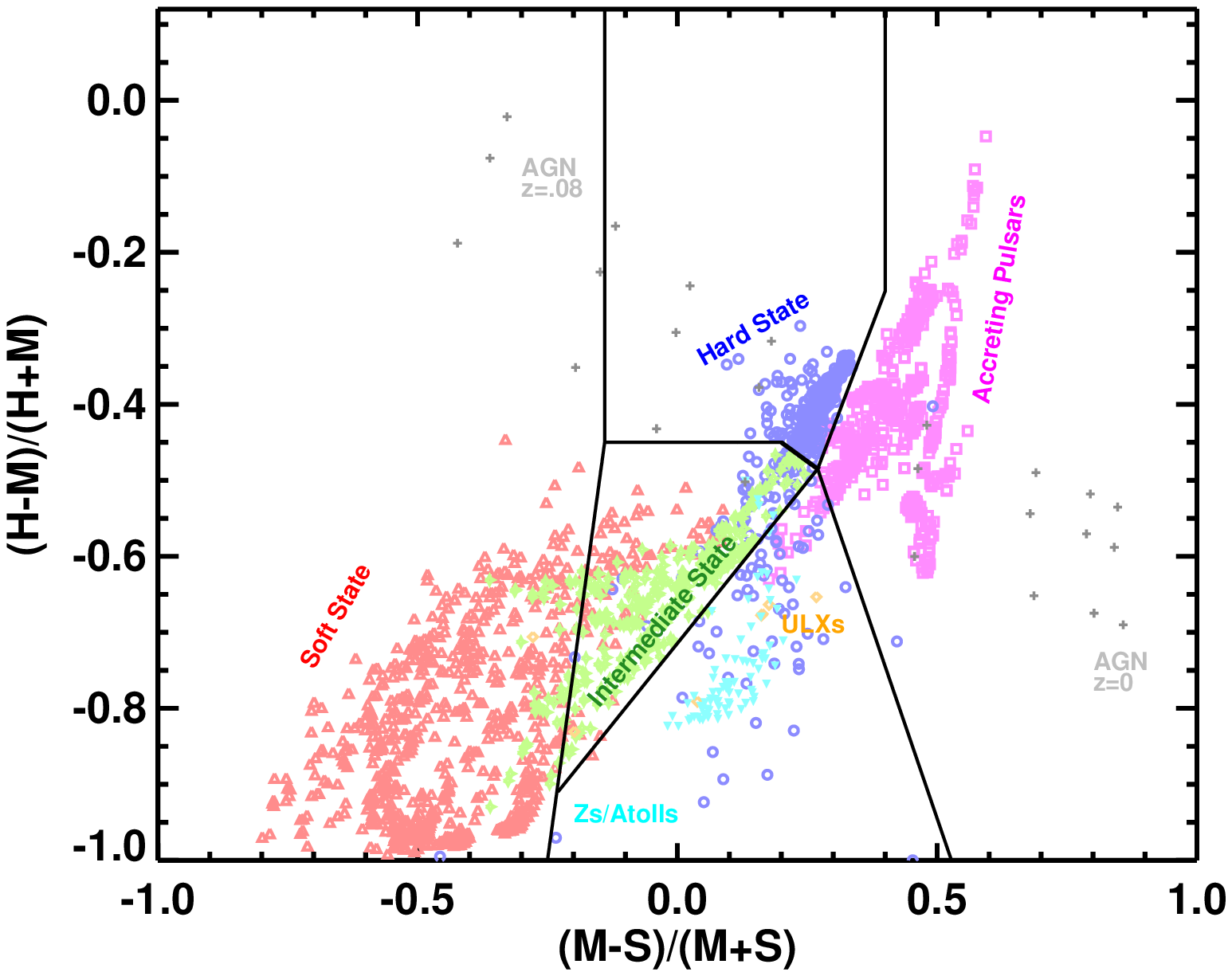}
\end{tabular}
\caption{\nustar\ hardness-intensity (left) and color-color (right) diagrams for Galactic XRBs. The data points indicate different accretion states and compact object types: accreting pulsars (magenta squares), hard state BH XRBs (blue circles), intermediate state BH XRBs (green stars), and soft state BH XRBs (red triangles). These data points were based on spectral fits to thousands of \rxtet\ PCA observations of Galactic XRBs in well-defined accretion states and with known compact object types. Z/Atoll NS are shown as inverted cyan triangles and are based on spectral fits to \rxtet\ and \bepposax\ observations of Galactic LMXBs \citep{church03-14}. ULXs (orange diamonds) were derived from spectral fits from various studies: \citet[][NGC 1313 X-1 and X-2]{bachetti12-13}, \citet[][Circinus ULX5]{walton12-13}, \citet[][Holmberg IX X-1]{walton09-14}, \citet[][IC 342 X-1 and X-2]{rana02-15}. Soft (S), medium (M), and hard (H) correspond to the $4-6$ keV, $6-12$ keV, and \hard keV energy bands. Delineated regions on the color-color diagram are approximations to isolate different accretion states. We also included implied colors of AGN (gray filled plusses) from the \nustar\ extragalactic survey (see Section \ref{sec:agncontam}). The count rate axis was converted to a luminosity axis assuming $\Gamma=1.7$ and $N_{\rm{H}}=10^{20}$ cm$^{-2}$ and normalizing to a distance of 4 Mpc.}	\label{fig:empty}
\end{figure*}

\begin{figure*}
\begin{tabular}{cc}
\includegraphics[width=1.0\columnwidth]{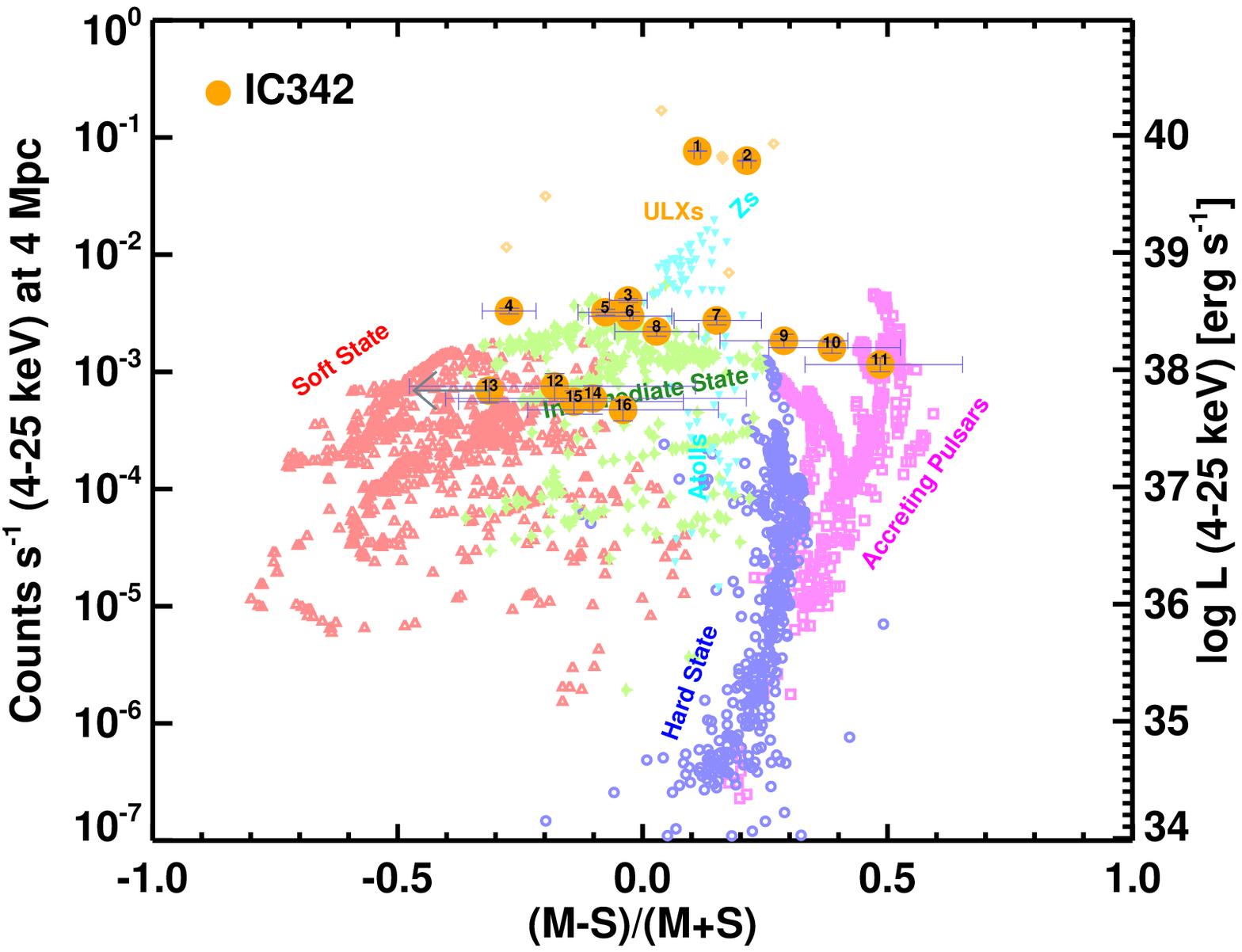}
\includegraphics[width=1.0\columnwidth]{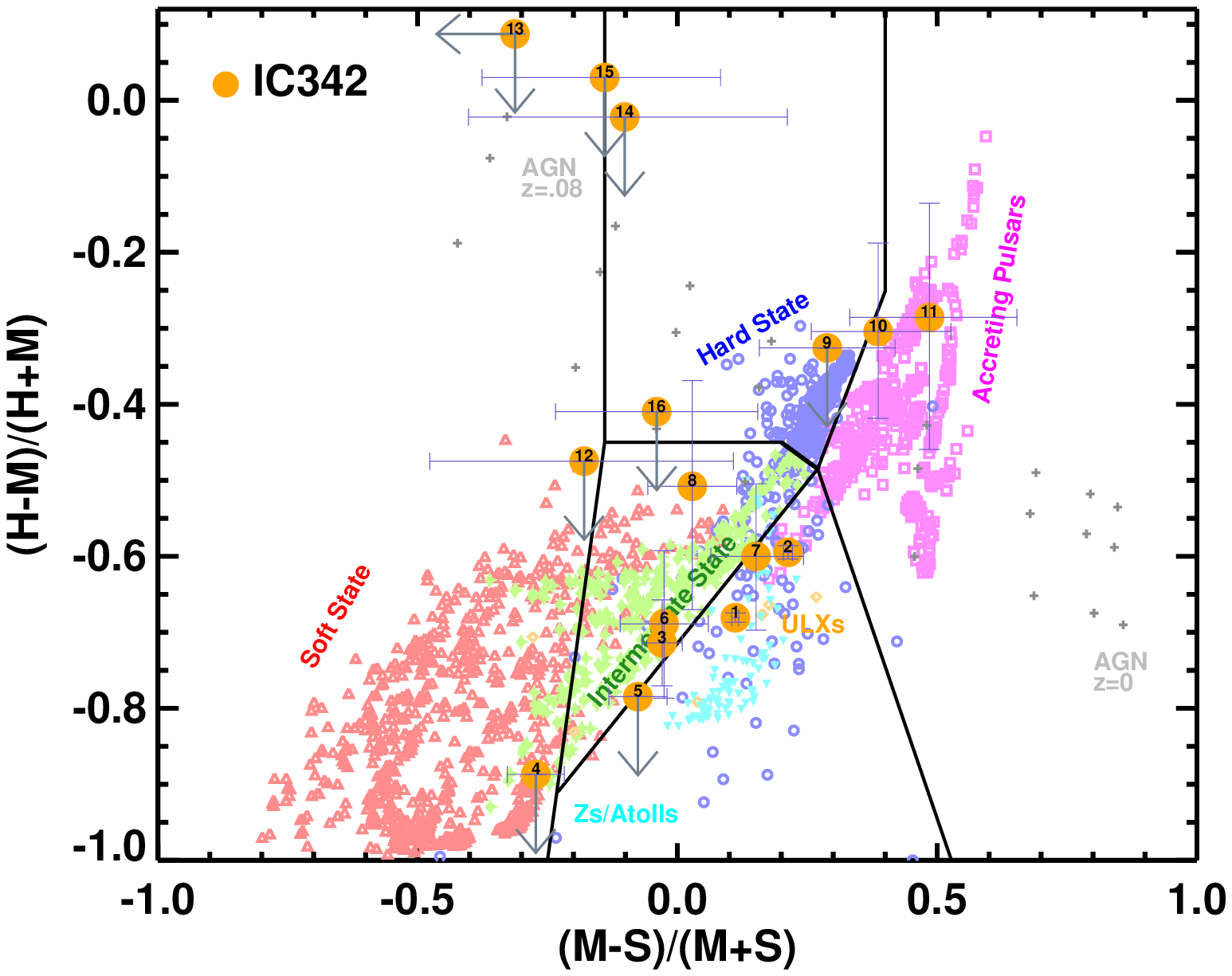}
\end{tabular}
\caption{Hardness-intensity (left) and color-color (right) diagrams for IC 342. Uncertainties shown represent the 90\% confidence interval. Numbers label point sources by decreasing count rate (see Table \ref{tab:rates}).}	\label{fig:ic342}
\end{figure*}

\begin{figure*}
\begin{tabular}{cc}
\includegraphics[width=1.0\columnwidth]{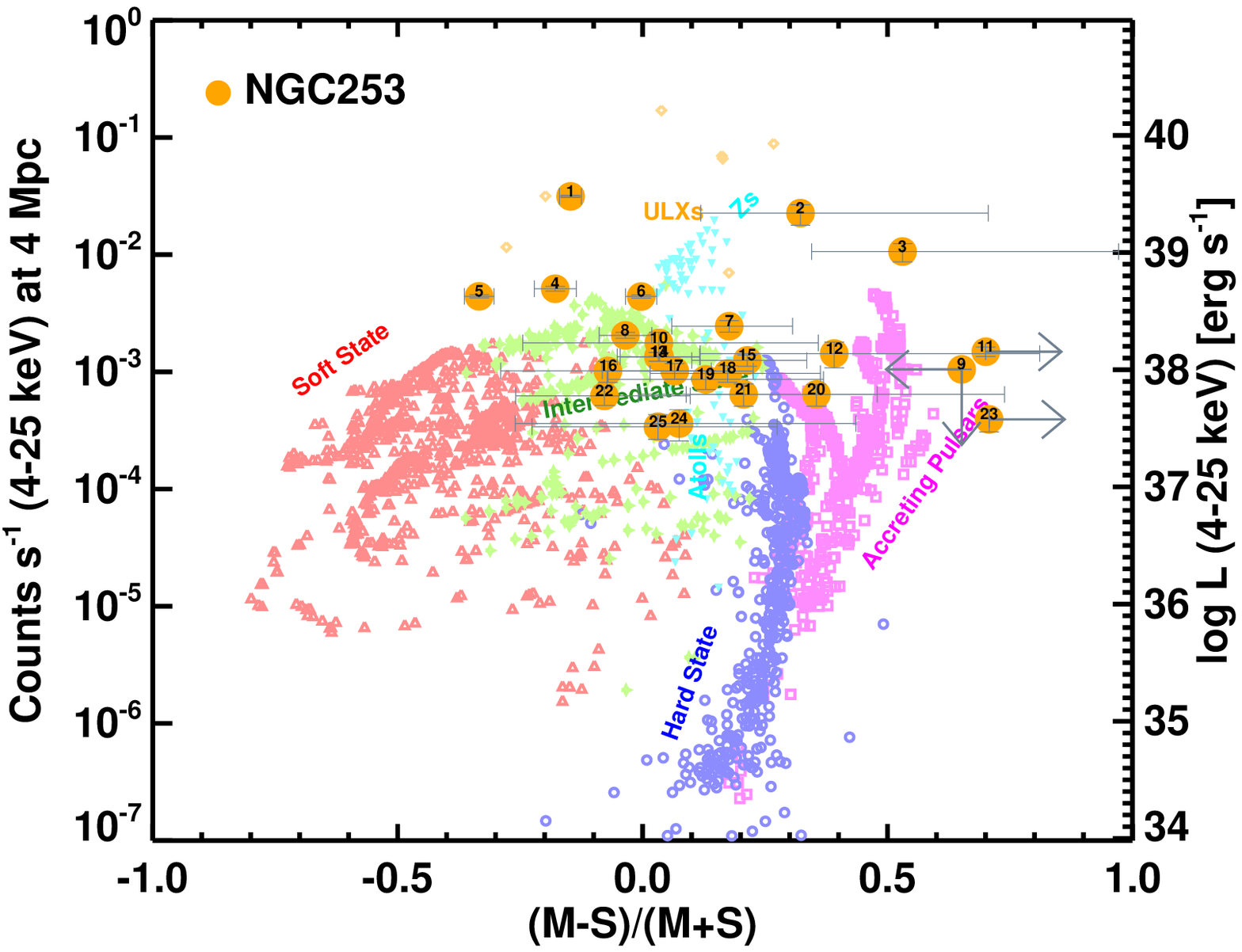}
\includegraphics[width=1.0\columnwidth]{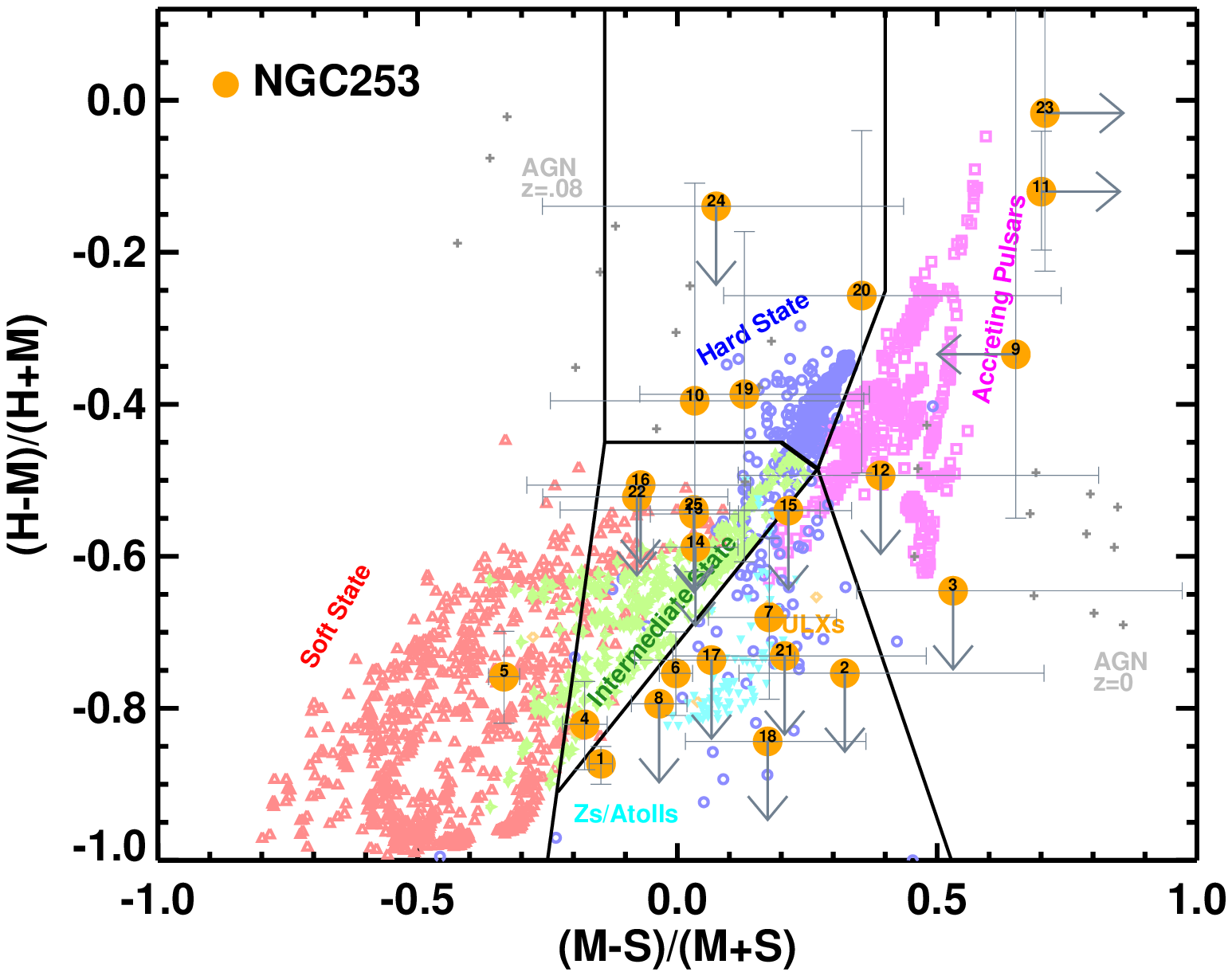}
\end{tabular}
\caption{As in Figure \ref{fig:ic342} for NGC 253.}	\label{fig:ngc253}
\end{figure*}

\begin{figure*}
\begin{tabular}{cc}
\includegraphics[width=1.0\columnwidth]{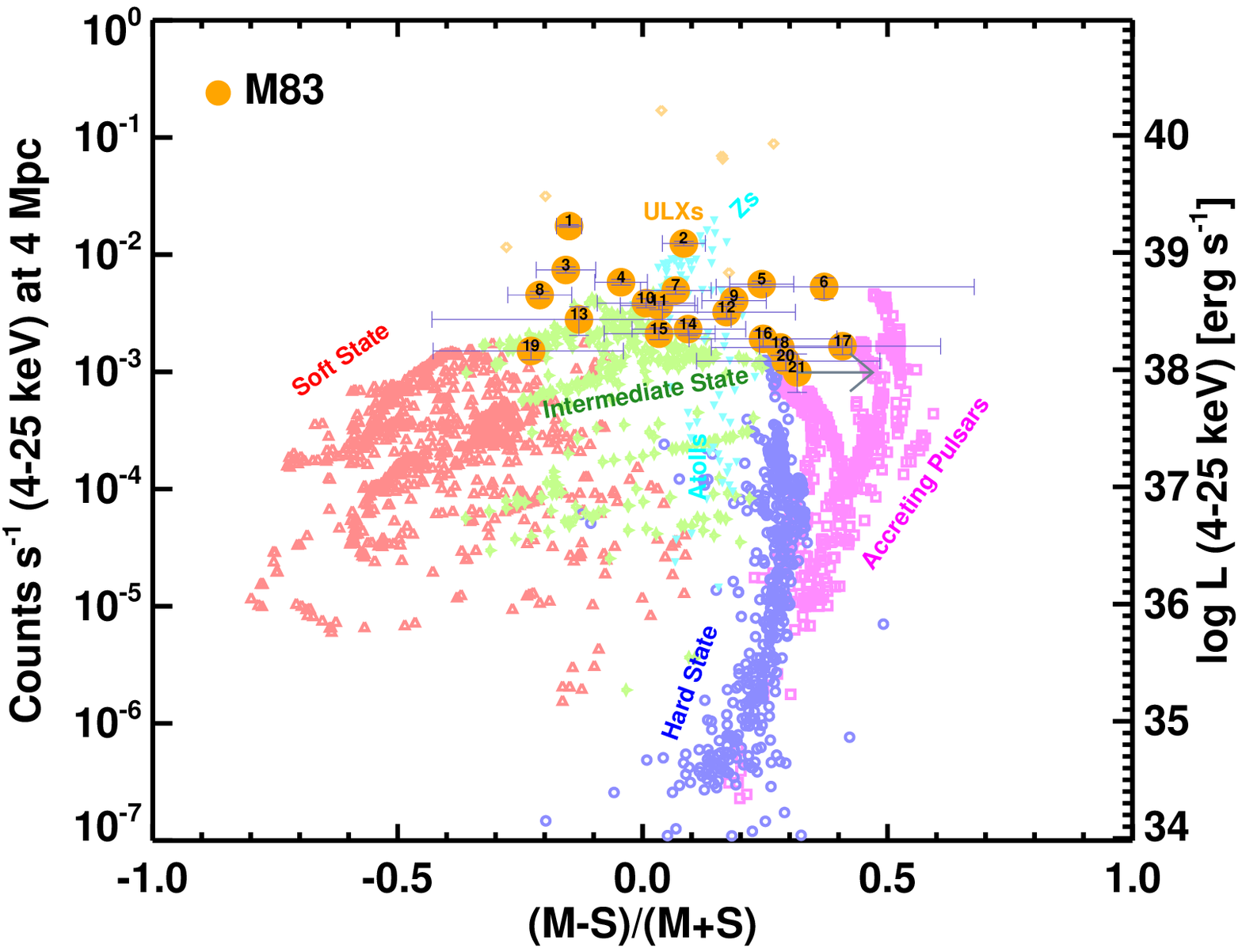}
\includegraphics[width=1.0\columnwidth]{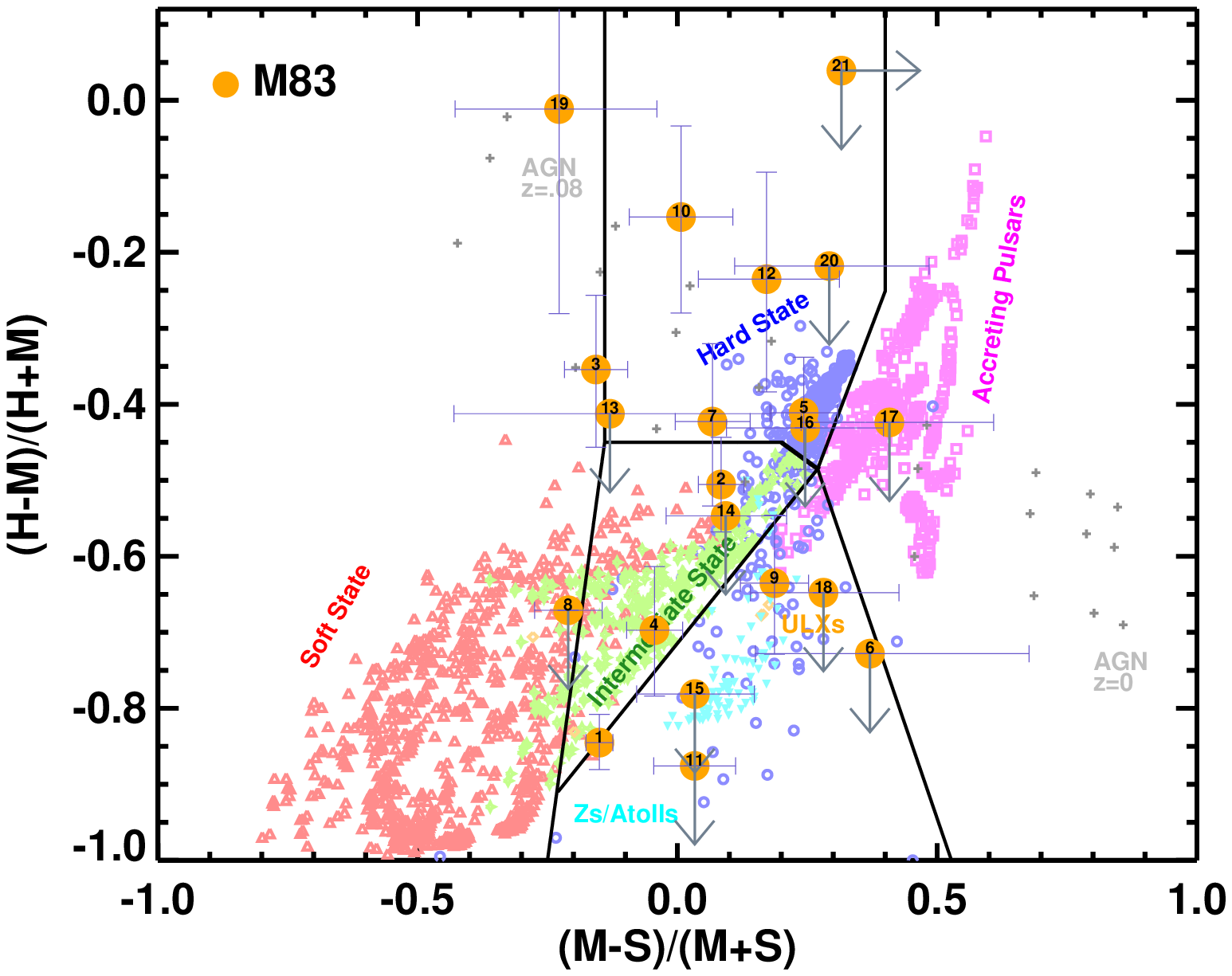}
\end{tabular}
\caption{As in Figure \ref{fig:ic342} for M83.}	\label{fig:m83}
\end{figure*}

\begin{figure*}
\begin{tabular}{cc}
\includegraphics[width=1.0\columnwidth]{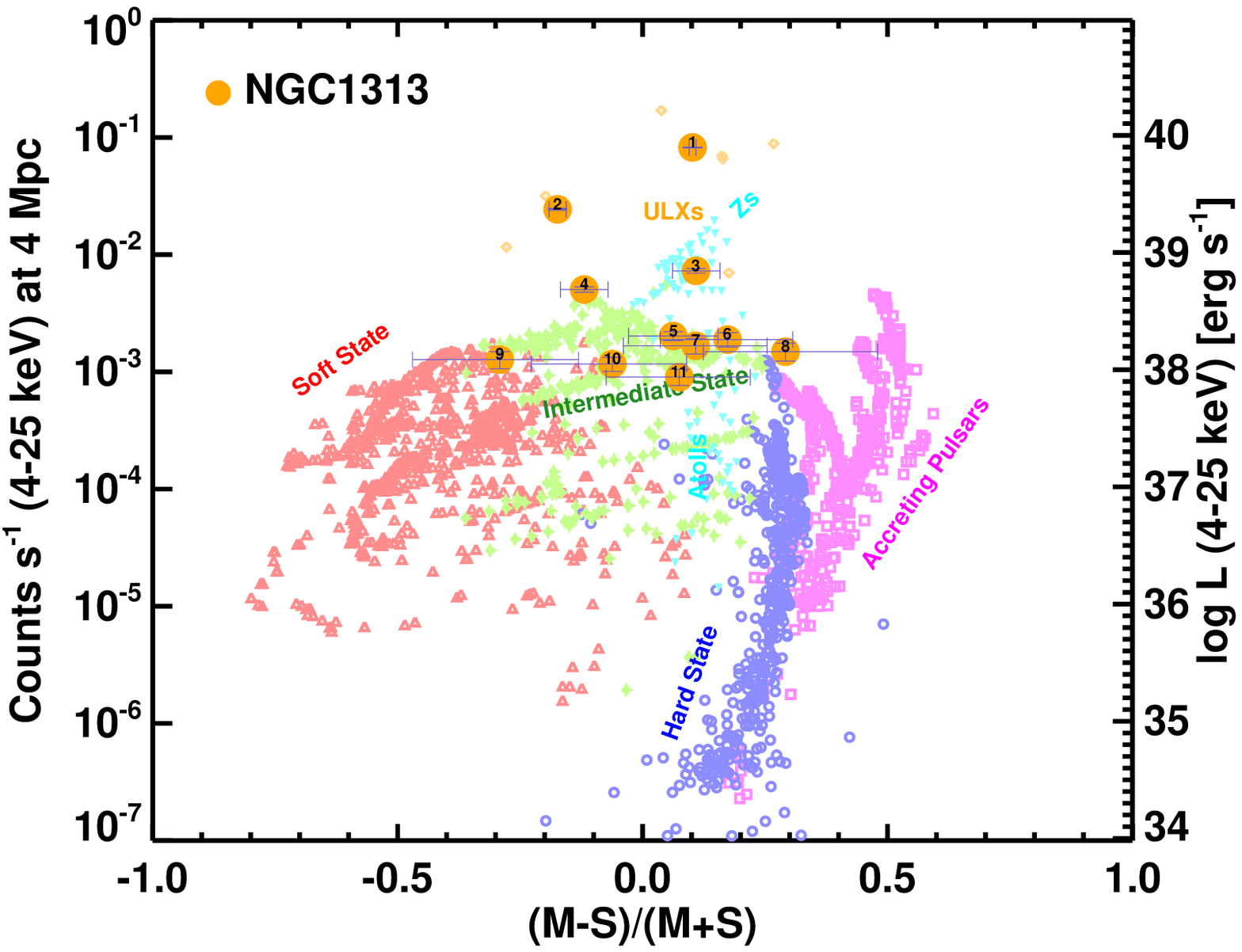}
\includegraphics[width=1.0\columnwidth]{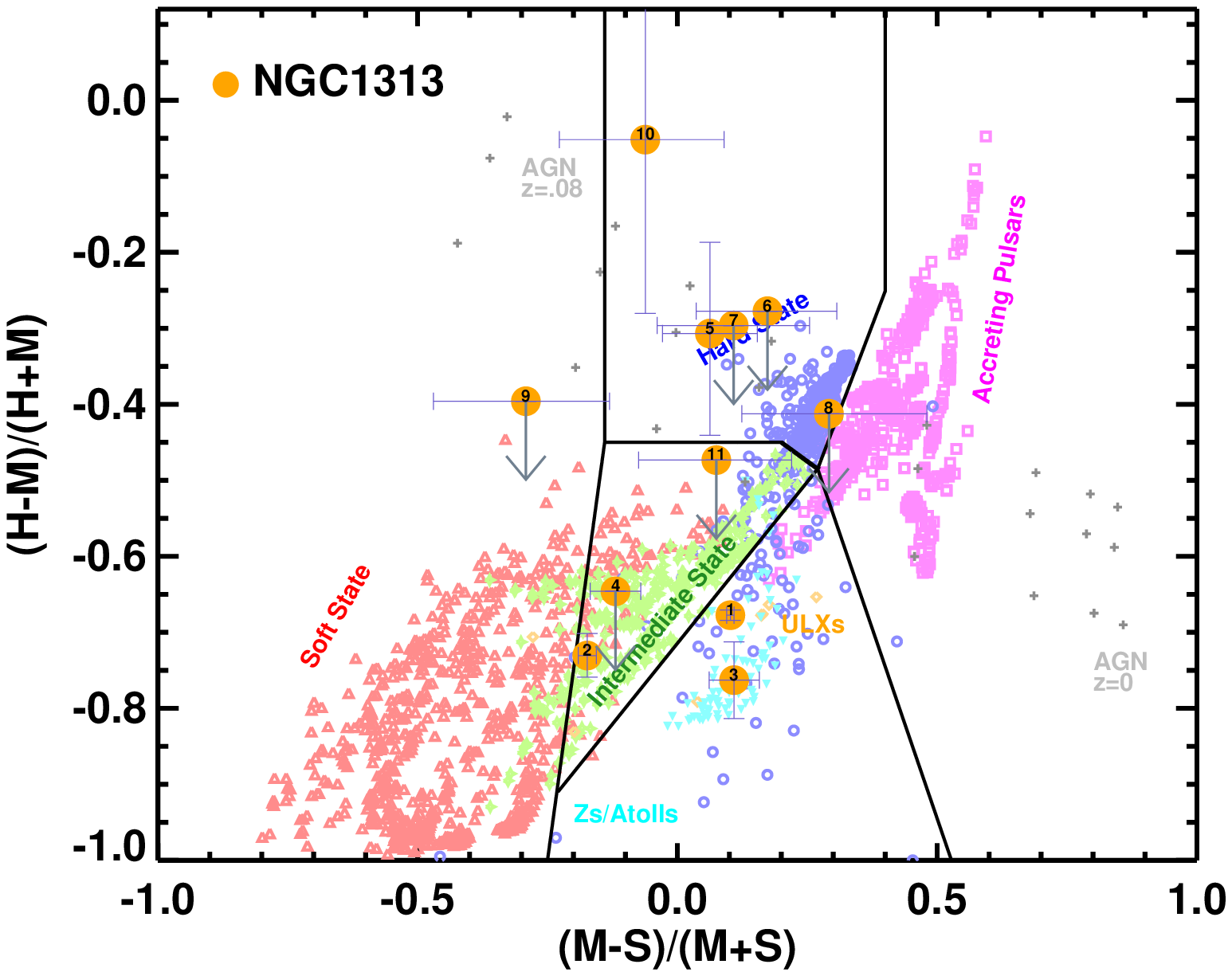}
\end{tabular}
\caption{As in Figure \ref{fig:ic342} for NGC 1313.}	\label{fig:ngc1313}
\end{figure*}

\begin{figure*}
\begin{tabular}{cc}
\includegraphics[width=1.0\columnwidth]{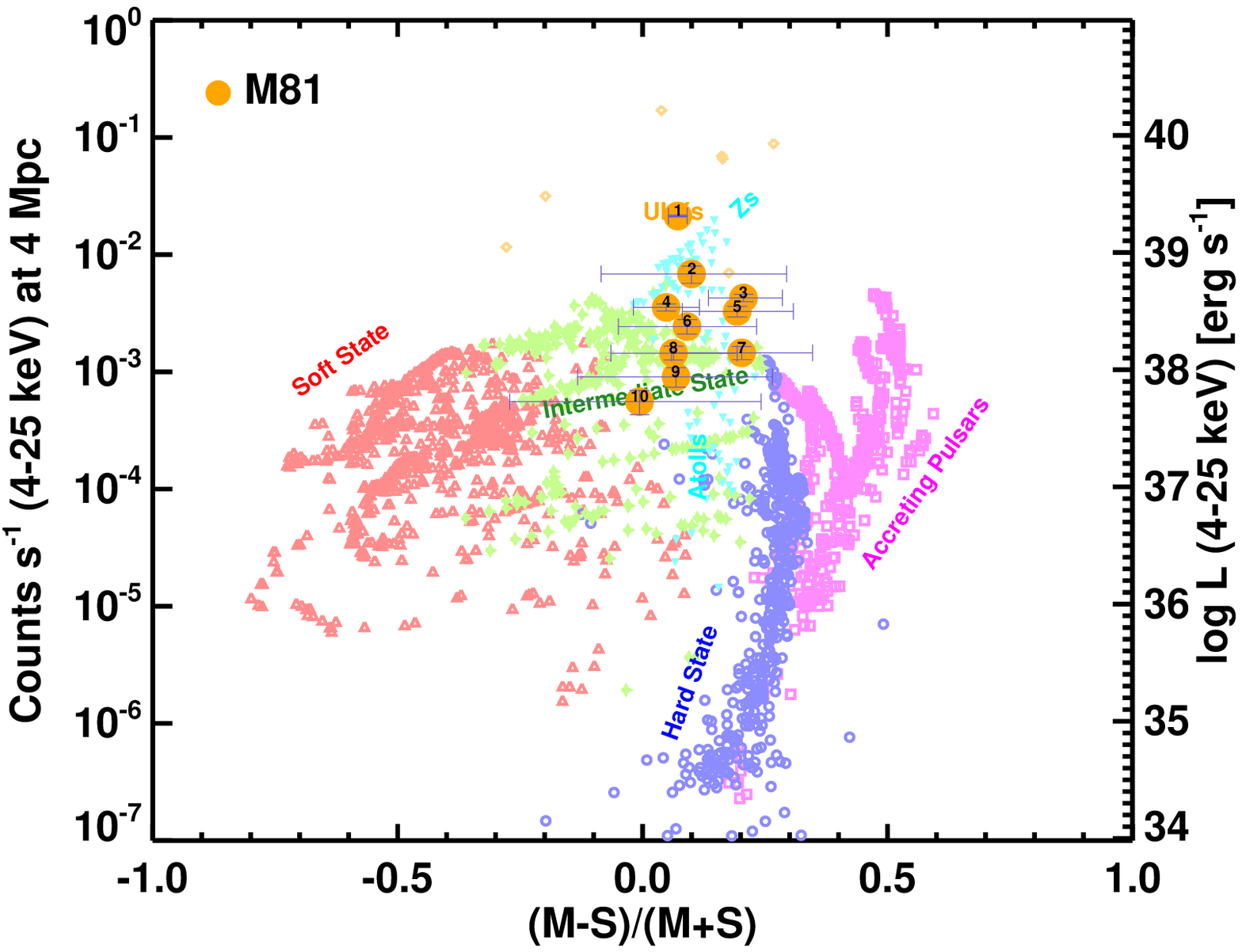}
\includegraphics[width=1.0\columnwidth]{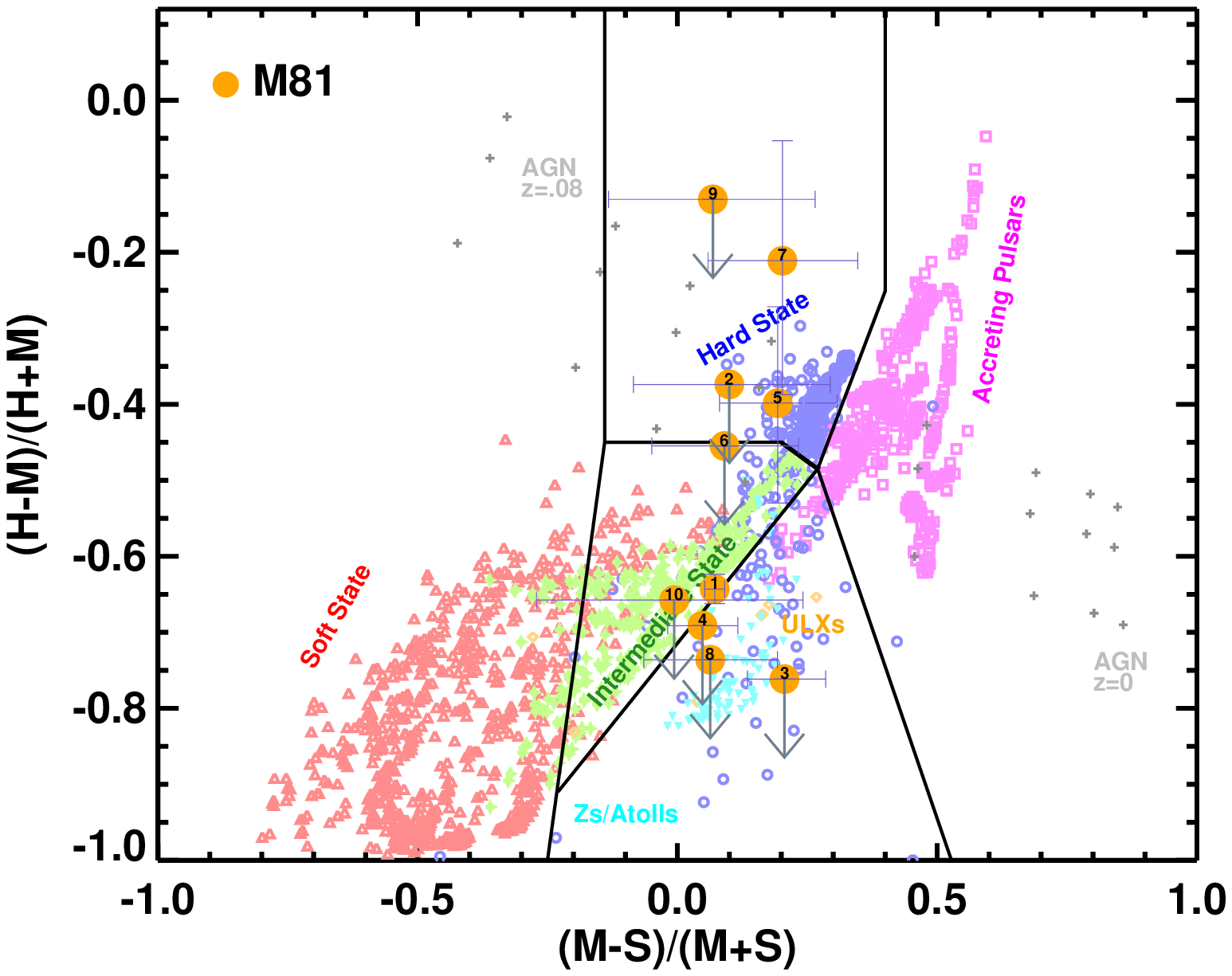}
\end{tabular}
\caption{As in Figure \ref{fig:ic342} for M81.}	\label{fig:m81}
\end{figure*}

\begin{figure*}
\begin{tabular}{cc}
\includegraphics[width=1.0\columnwidth]{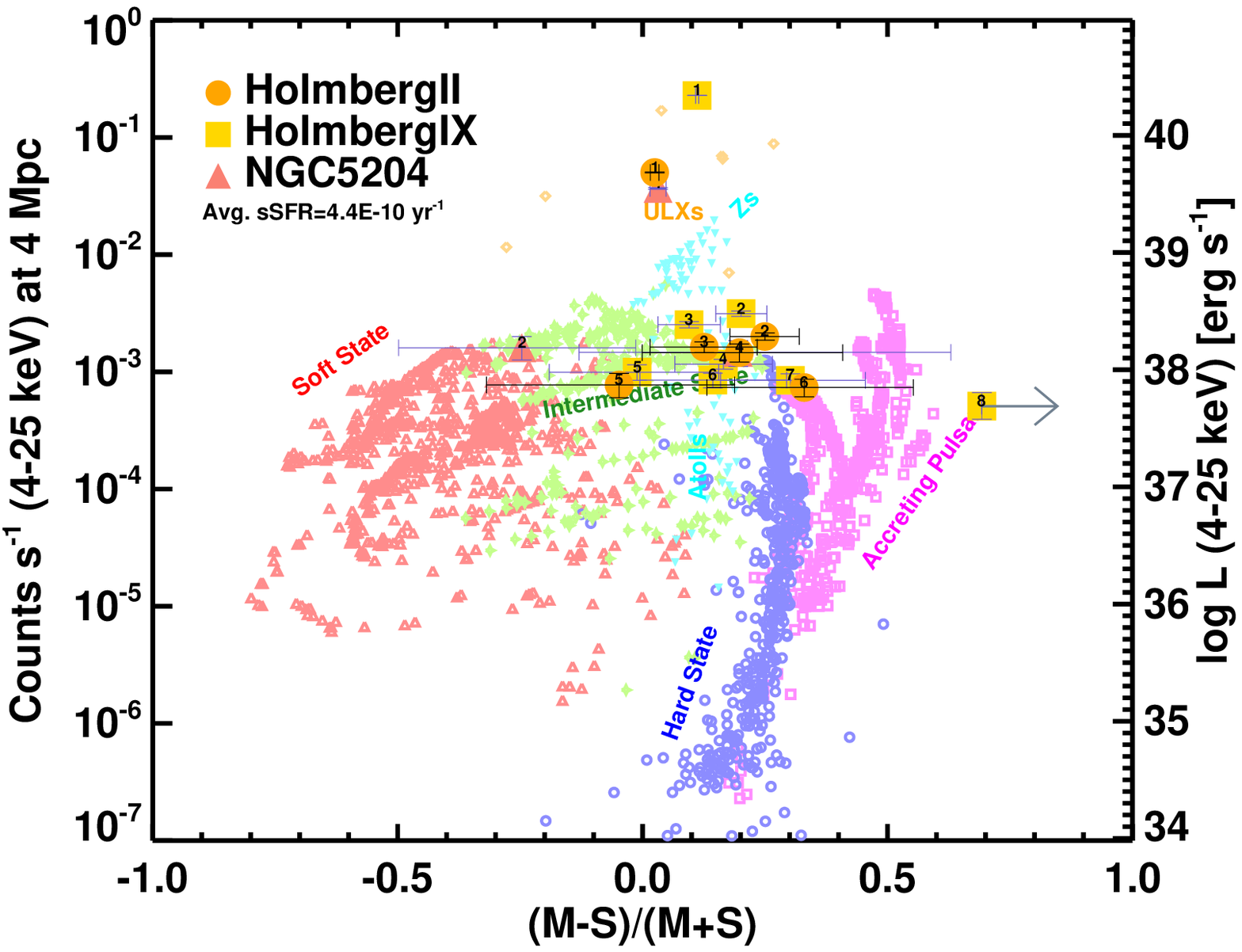}
\includegraphics[width=1.0\columnwidth]{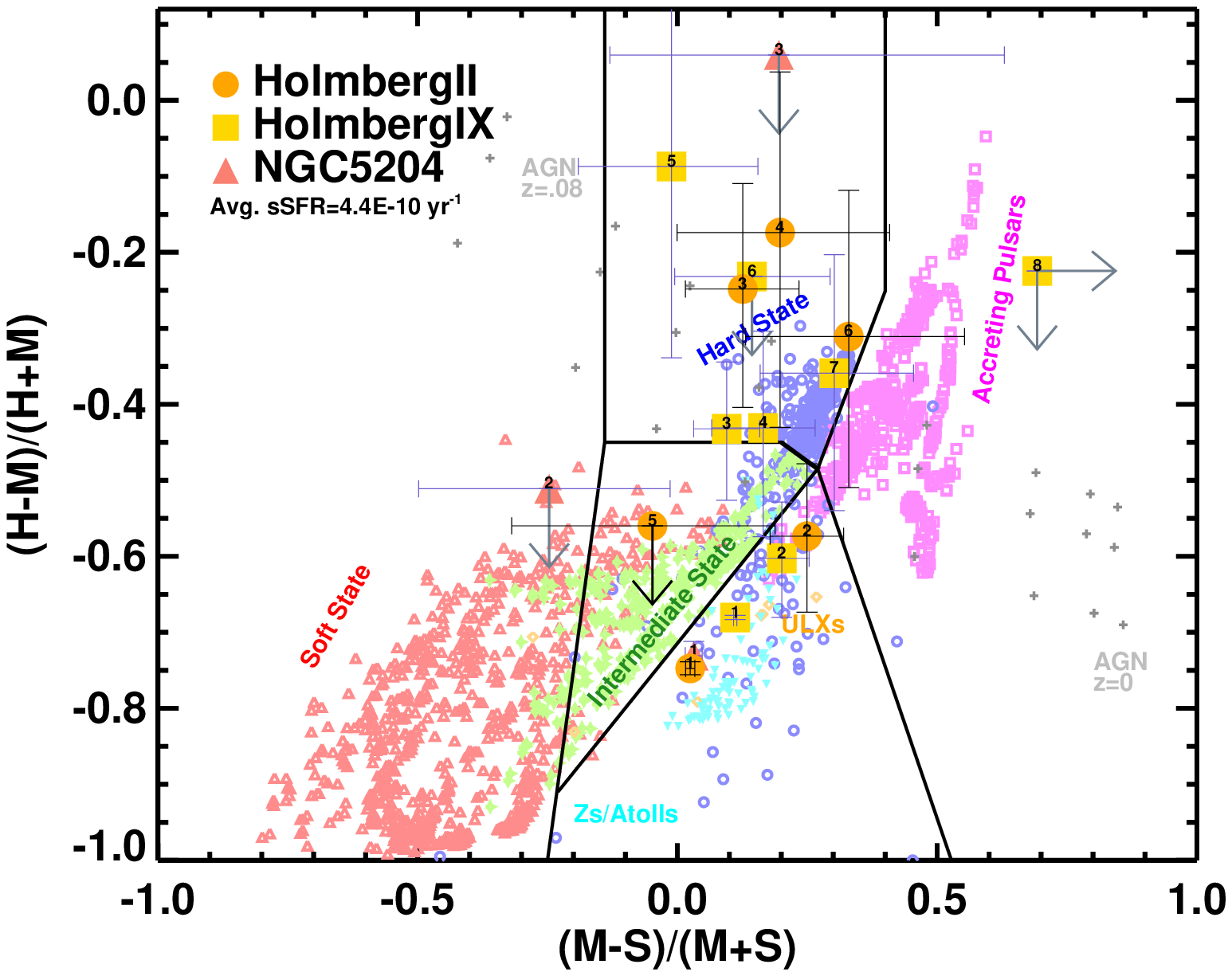}
\end{tabular}
\caption{As in Figure \ref{fig:ic342} for dwarf galaxies Holmberg II, Holmberg IX, and NGC 5204, all with similar sSFR.}	\label{fig:holmiiix5204}
\end{figure*}

\begin{figure*}
\begin{tabular}{cc}
\includegraphics[width=1.0\columnwidth]{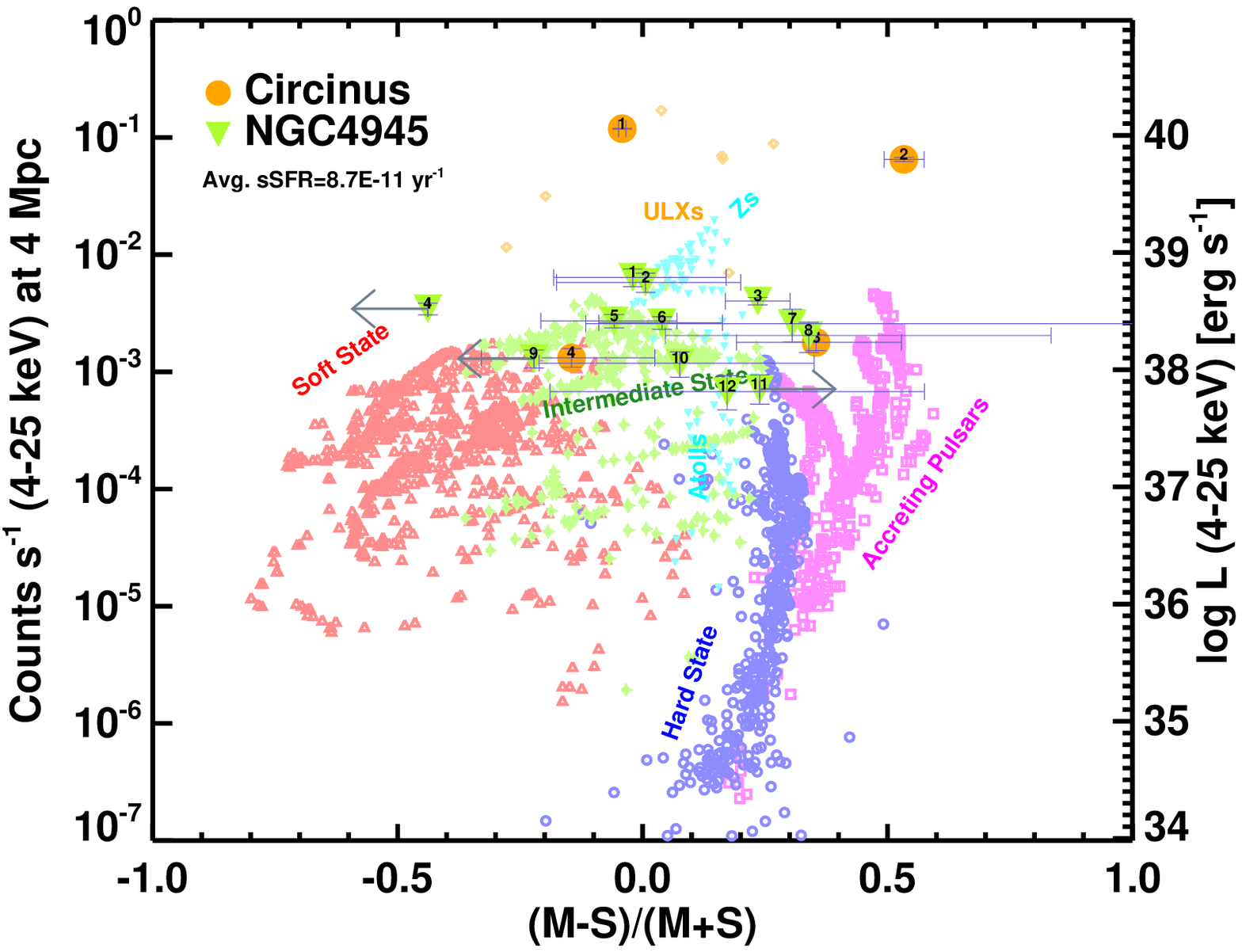}
\includegraphics[width=1.0\columnwidth]{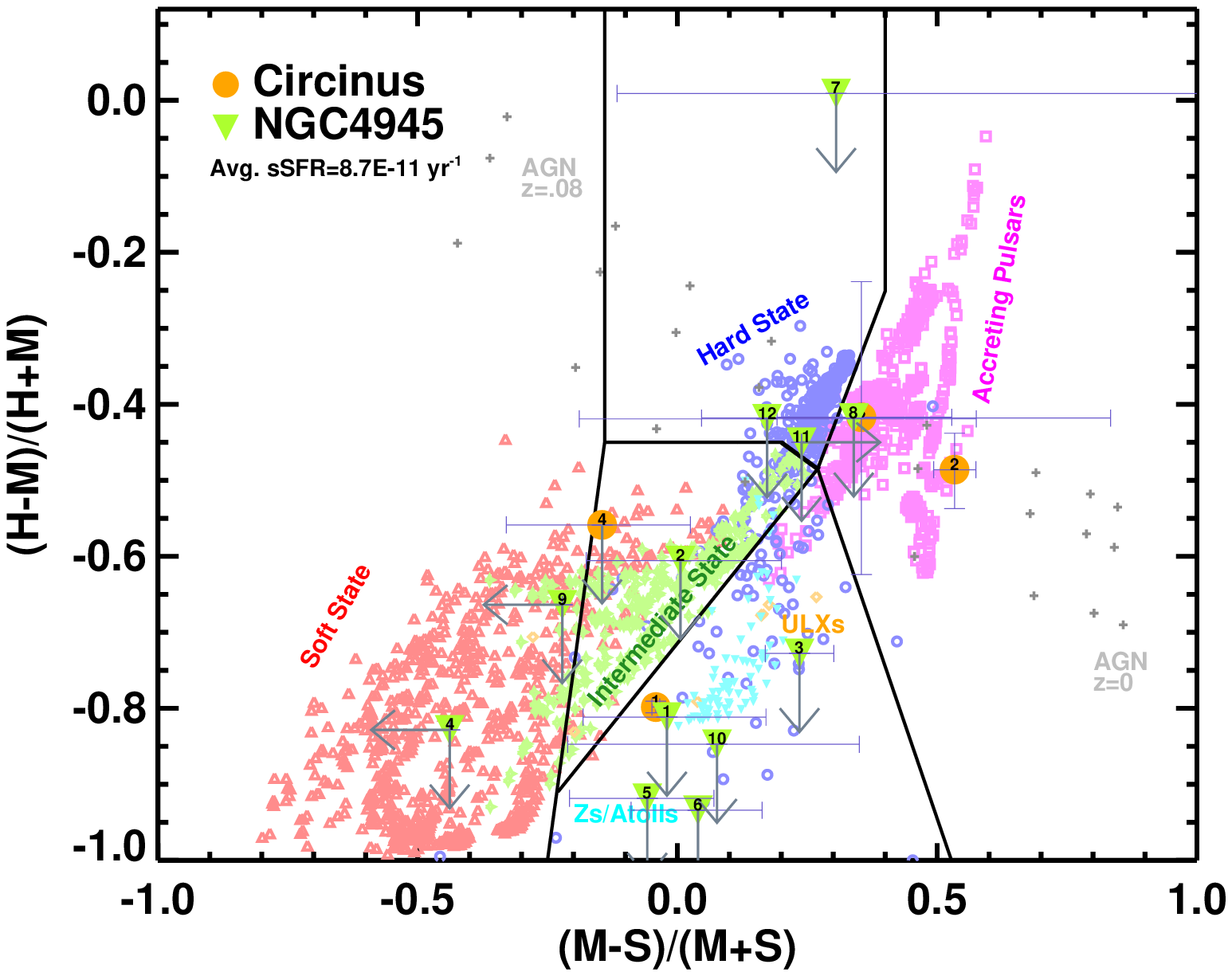}
\end{tabular}
\caption{As in Figure \ref{fig:ic342} for Circinus and NGC 4945.}	\label{fig:cirngc4945}
\end{figure*}

\begin{figure*}
\begin{tabular}{cc}
\includegraphics[width=1.0\columnwidth]{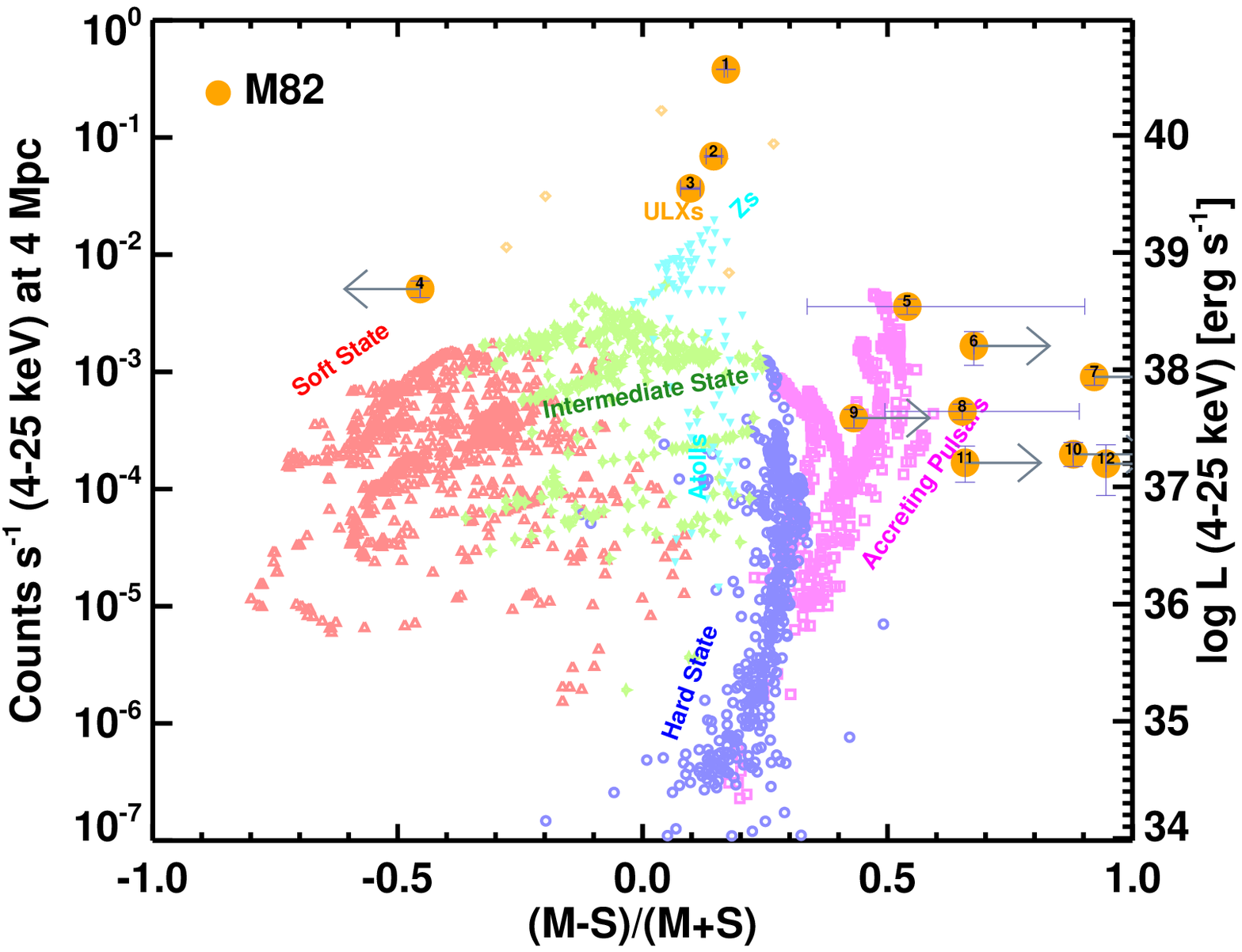}
\includegraphics[width=1.0\columnwidth]{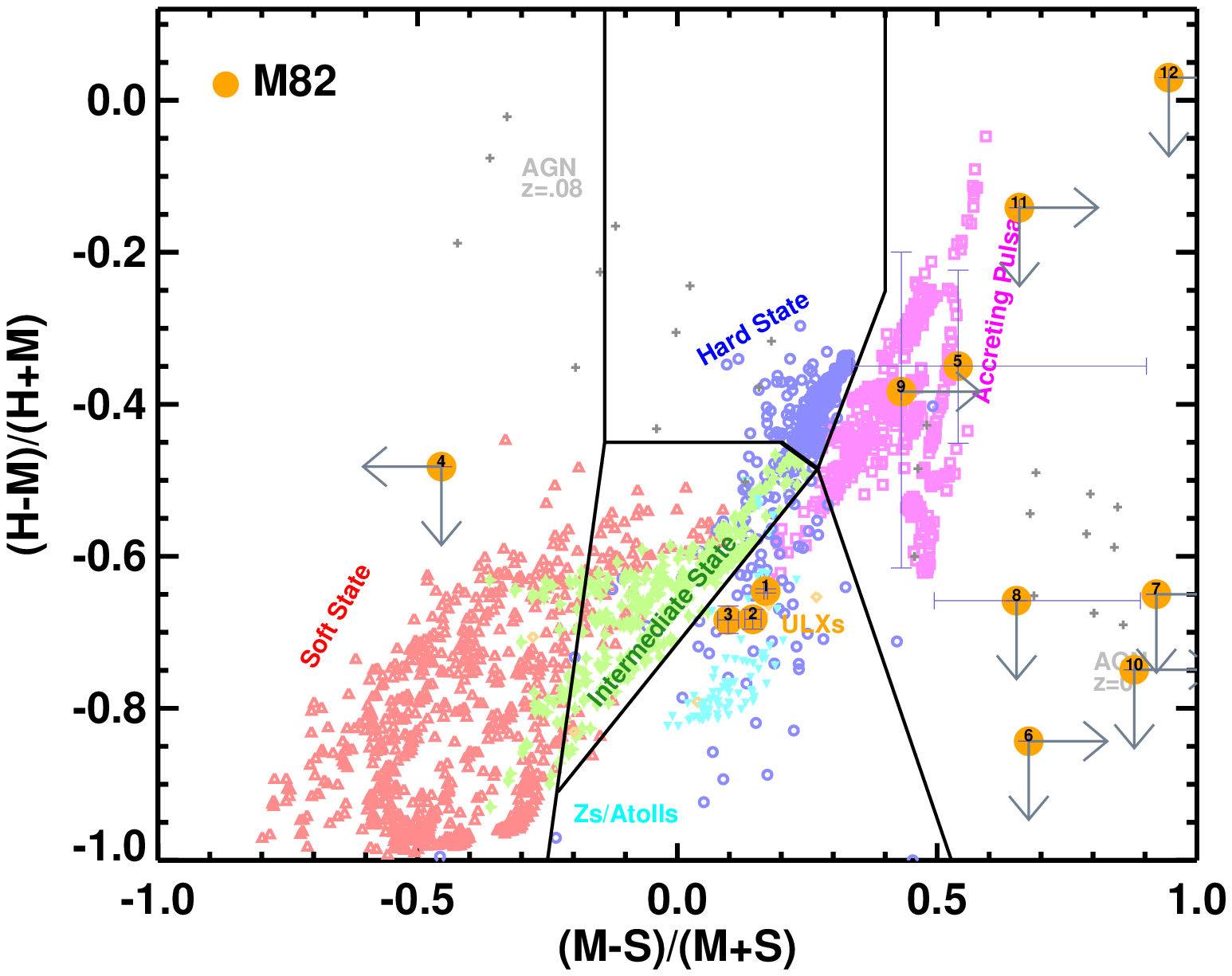}
\end{tabular}
\caption{As in Figure \ref{fig:ic342} for M82.}	\label{fig:m82}
\end{figure*}

\begin{figure*}
\begin{tabular}{cc}
\includegraphics[width=1.0\columnwidth]{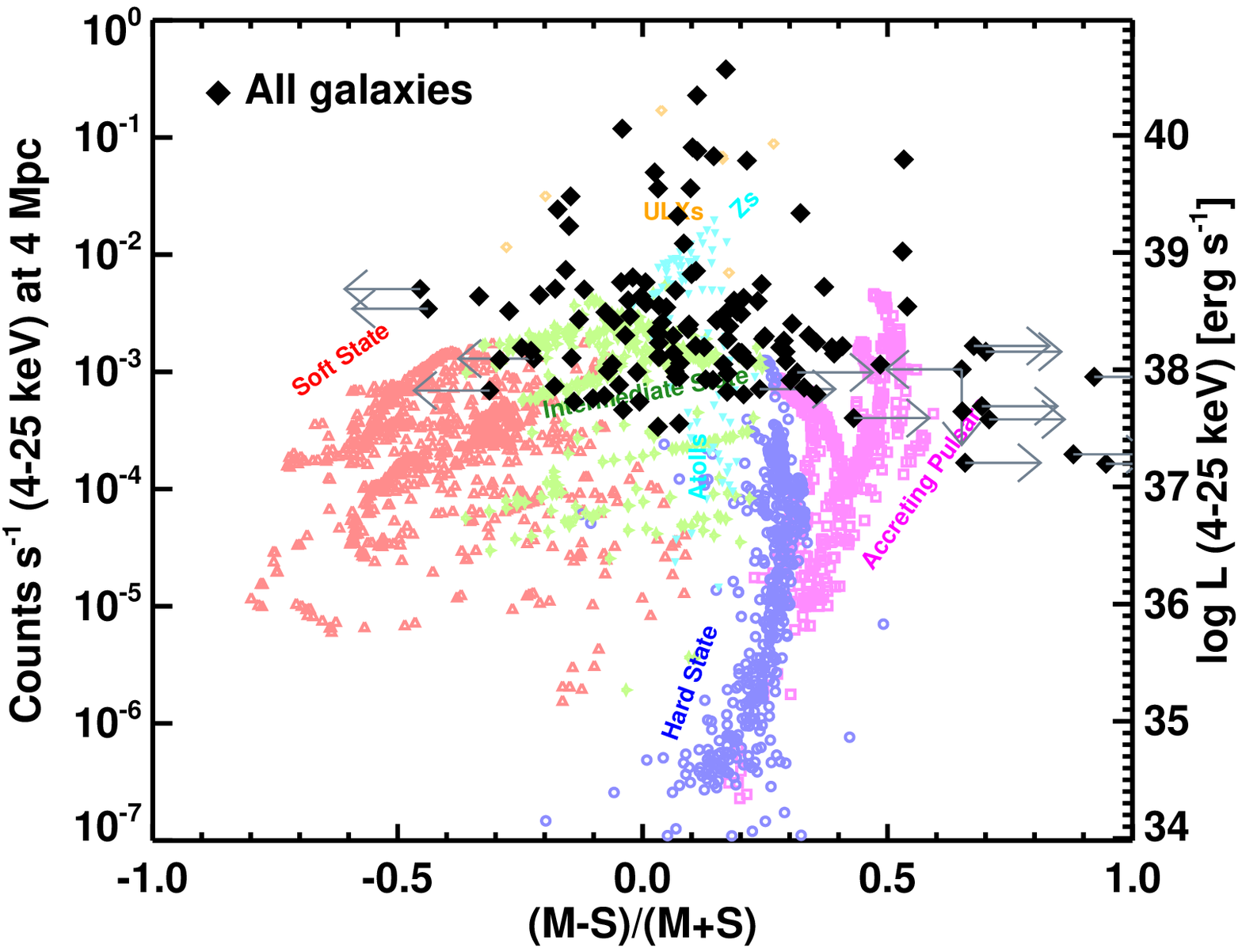}
\includegraphics[width=1.0\columnwidth]{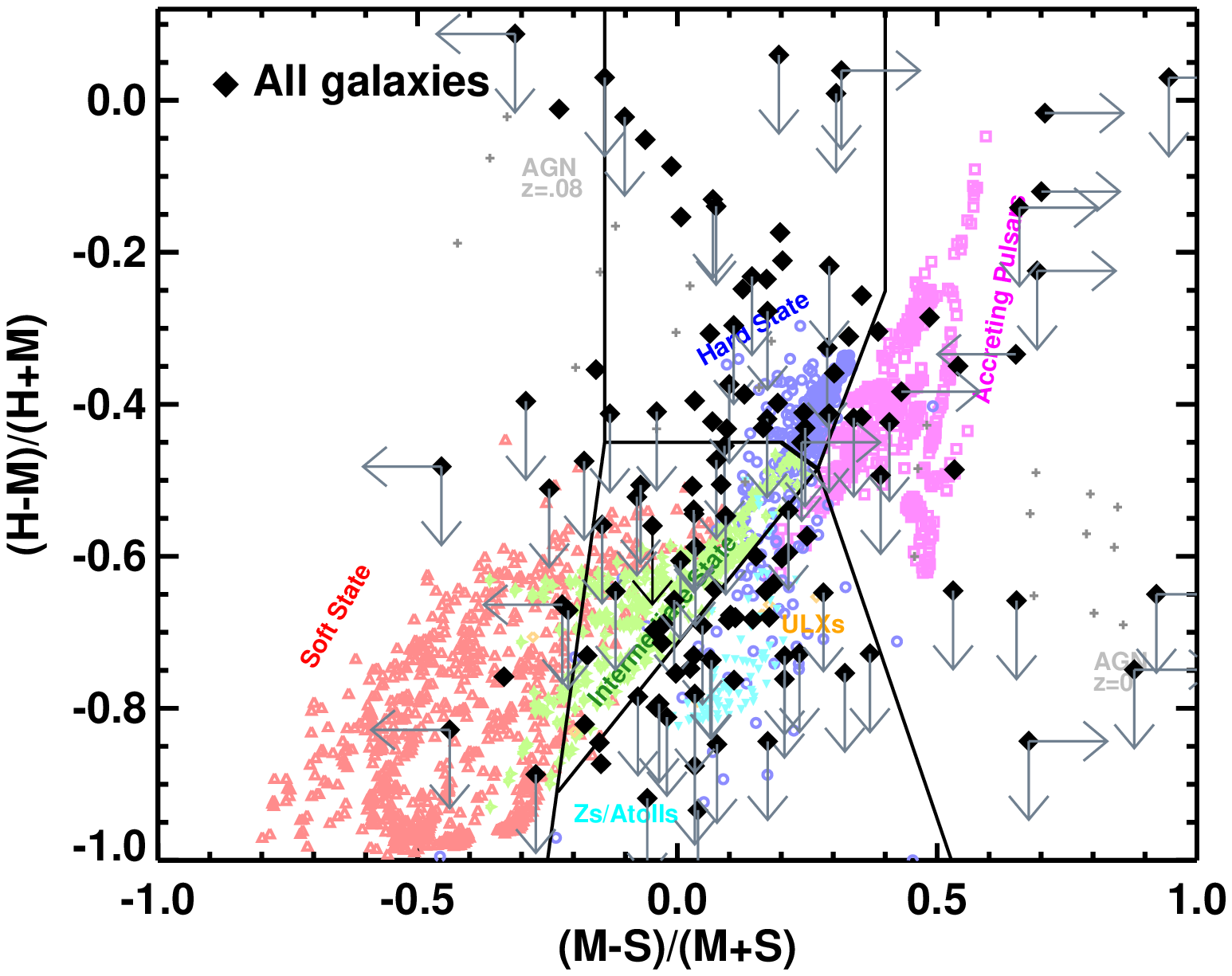}
\end{tabular}
\caption{As in Figure \ref{fig:ic342} for all sources from all galaxies. The left panel shows that most point sources overlap with the BH XRB intermediate state, which is also degenerate with hard state BH and Z/Atoll NS. The characteristics of the soft state are such that our sensitivity limits prevent us from detecting many sources in this region (see Section \ref{sec:ssfrconn}). The right panel shows more distinct separation between accretion states and compact object types, specifically the sources constrained in the Z/Atoll NS region. In particular, there are fewer sources located in the intermediate state due to the constraints from \hard keV. The color-color diagnostic was used for source classification because it was more robust at breaking degeneracies between source types. Sources in the right panel with large hard colors above the hard state are possibly background AGN that can be identified with optical follow-up.}	\label{fig:allgals}
\end{figure*}

\label{fig:diagdiag}
\end{subfigures}

Combining X-ray luminosity with $HR1$ and $HR2$ allows us to constrain X-ray source characteristics via identification of the accretor and accretion state. Color-color plots show more robust separation, particularly the Z/Atoll NS sources that are localized below the BH points with softer $HR1$ colors. This is due to the spectra of non-magnetized NS (and also ULXs), which turn over more quickly than BH \citep[e.g.][]{maccarone06-16}. There are $\sim10$ \nustar\ sources that overlap the isolated Z/Atoll sources (not identified as ULXs based on \lx) and are undetected in $HR2$; however, the upper limits constrain them to the NS region. The scatter about the intermediate state is evident, especially for sources near and above the hard state, which are possibly background galaxies. Given the loci and uncertainties in the diagnostic diagrams, we were able to classify \numclass\ compact objects and \numstate\ accretion states from our sample of \numsrc\ sources. These are candidate classifications that require further detailed study (e.g.\ multiwavelength spectra) to confirm their nature. When source classification between the diagnostic diagrams disagreed, we utilized the color-color result due to its improved source separation. The identifications for each source are shown in Table \ref{tab:rates}, while the total number of sources in each category is shown in Table \ref{tab:types}.

\begin{table}
\tiny
\caption{Compact Object Classification and Accretion States\label{tab:types}}
\begin{tabular}{@{}  c c  c  c  c  c  c  c  @{}}
\hline\hline
	&	Total		&	Hard State	& Intermediate State	&	Soft State		&	ULX	&	Z/Atoll	&	Accreting Pulsar	\\	\hline
BH	&	47	&	16	&	8	&	6	&	15	&	\ldots	&	\ldots
\\ NS	&	43	&	\ldots	&	\ldots	&	\ldots	&	\ldots	&	21	&	21
 \\ \hline
\end{tabular}
\tablecomments{\numclass\ of the \numsrc\ sources in our sample were separated into the BH/NS classification. From these 90 classifications, 2 BH and 1 NS could not be classified to a particular accretion state. We assumed that all ULXs were BH (see Section \ref{sec:corr}).}
\end{table}

\subsubsection{AGN Contamination}	\label{sec:agncontam}

\begin{table*}[!ht]
\caption{Background AGN Contamination\label{tab:agnc}}
\begin{tabular}{@{} c c c c c c c c c c c c @{}}
\hline\hline
Galaxy &	HolmbergII		&	IC342	& M82	&	NGC253	& M81	& NGC4945	& HolmbergIX & Circinus	& NGC1313	& M83	& NGC5204		\\
Background AGN	&	1	&	2.8	&	1.6	&	1.5	& 2	&	1	& 1.3 &1.2 &1.9 &2.3 & 0.7	\\	\hline
\end{tabular}
\tablecomments{The number of background AGN expected in each observed field based on the \nustar\ AGN number counts from \citet{harrison11-16}. These estimates closely match the number of unclassified sources that overlap the AGN region in the color-color diagrams for each galaxy.}
\end{table*}

To estimate AGN colors we used the results from the \nustar\ extragalactic survey of \citet{del-moro11-17}, who calculated the average broadband ($3-24$ keV observed frame) X-ray spectral properties of 182 AGN. They found an average power law photon index of $\Gamma=1.65$, flatter than the typical $\Gamma\approx1.8$ for AGN \citep[e.g.][]{tozzi05-06}. The spectral slope of  broad-line and X-ray unabsorbed AGNs was consistent with typical values ($\Gamma=1.79$), but narrow-line and heavily absorbed sources had values as low as $\Gamma=1.38$. 
We used the best-fitting spectral parameters for composite spectra that were grouped into all AGN, narrow-line AGN, and broad-line AGN, varying the redshift from $z=0.00-0.08$. We show the implied colors of AGN in the right panels of Figure \ref{fig:diagdiag}. AGN are found in a variety of locations, but most are concentrated in regions with large $HR1$ or $HR2$ colors. In particular, M82 has 4 sources located in the bottom right corner of its color-color diagram, which may be absorbed NS sources as opposed to AGN, due to the large values of extinction in M82.

We also used the \nustar\ AGN number counts from \citet{harrison11-16} to determine the number of background galaxies expected for each galaxy in the sample, based on the \nustar\ FOV and the sensitivity limits for detected sources. In Table \ref{tab:agnc} we show the expected number of background AGN for each galaxy. These estimates closely match the number of unclassified sources that overlap the AGN region in the color-color diagrams for each galaxy. Optical follow-up is required for all \nustar-detected sources to determine which are background AGN.

\subsubsection{Connections with sSFR}	\label{sec:ssfrconn}

\begin{figure}[!ht]
\begin{tabular}{cc}
\includegraphics[width=1.0\columnwidth]{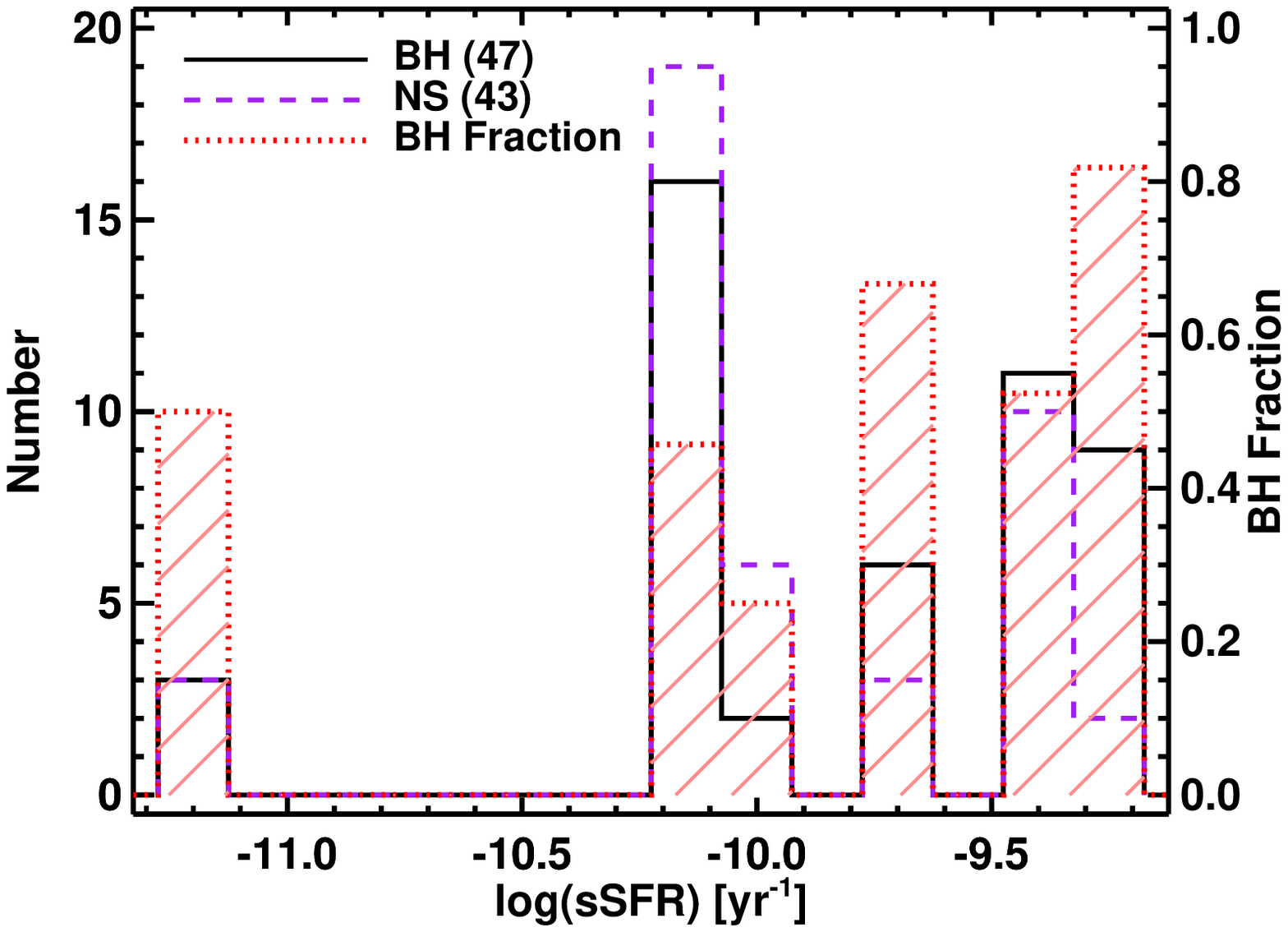}
\end{tabular}
\caption{Histograms showing the distribution of BH and NS for given sSFR based on the \nustar\ galaxy sample. The red dotted line (red line-filled histogram) shows the BH fraction. As shown in Figure \ref{fig:sfrmass}, the sample is not uniform across sSFR and therefore features of the histogram are biased based on the sSFR distribution. There is no statistically significant relation between BH fraction and sSFR (see Section \ref{sec:ssfrconn}).}	\label{fig:idshist}
\end{figure}

In Figure \ref{fig:idshist} we plot the distribution of identified candidate BH and NS as a function of the sSFR of each galaxy. We also plot the BH fraction, defined as $N_{\rm{BH}}$ / ($N_{\rm{BH}}$+$N_{\rm{NS}}$). As shown in Figure \ref{fig:sfrmass}, the galaxy sample is not uniform across sSFR and therefore features of the histogram may be biased based on the sSFR distribution. At large sSFR a higher fraction of BH is evident, which likely represent BH-HMXBs that have formed from recent star formation episodes. Conversely, NS begin to dominate galaxies towards lower sSFR, as BH-HMXBs become less numerous and older (LMXB) populations are more prevalent. M31, which is not shown here, is dominated by NS at $\log(\rm{sSFR/yr}^{-1}) = -11.5$. A Spearman's Rank test on the fraction of BH versus sSFR gave a $p$-value of 0.072 and coefficient $r_{s}=0.56$. While the coefficient indicates weak monotonicity, the $p$-value is too large to claim a correlation. However, when including M31, which is dominated by NS and extends the sSFR to lower values, we obtain a $p$-value of 0.028 and coefficient $r_{s}=0.63$. However, there a number of caveats we must consider. Firstly, most galaxies in the sample have $4-25$ keV sensitivities of $\sim10^{38}$ \es, which is not a sufficient limit from which to draw conclusions between BH fraction and sSFR. In addition, M31 \nustar\ observations extend down to $10^{36}$ \es, $1-2$ orders of magnitude fainter than the rest of the sample. Even among the remaining galaxies, completeness is not uniform (see Figure \ref{fig:sensgals}). These issues are exacerbated due to \nustars\ PSF, which can mask sources under the broad wings of e.g.\ ULXs, thus biasing sensitivity. Therefore, an expanded sample with improved statistics and completeness for X-ray point sources in each galaxy is necessary to draw conclusions between sSFR and the ratio of BH/NS.

Do we expect to find a different ratio of BH/NS based on galaxy type? The preliminary indication that BH fraction may be larger for high-sSFR and decline towards low-sSFR, while not statistically significant, has interesting implications. Firstly, these results are strictly limited to the XRBs we can detect, and therefore the population of BH and NS that are preferentially faint will be missed. Therefore it is difficult to address the issue of compact object formation rates. This type of study is more applicable to the issue of mass accretion rate and conversion of accretion into luminosity. BH HMXBs generally accrete mass via the wind of the donor star at a rate proportional to $M_{\rm{acc}}^{2}$, where $M_{\rm{acc}}$ is the mass of the accreting compact object \citep{hoyle-39, bondi-44}. For BH LMXBs, we can use the analytical approximations of mass transfer rate from \citet{king06-96}, which gives 3 cases that describe the physical processes driving the mass accretion rate:
\begin{enumerate}[1.]
\item evolution based on the nuclear evolution of the donor star, which is independent of accretor mass $M_{\rm{acc}}$
\item magnetic braking, which scales with $M_{\rm{acc}}^{-2/3}$
\item gravitational radiation, which scales with $M_{\rm{acc}}^{2/3}$
\end{enumerate}
We expect high luminosities in the third case only for very short period sources, and in the second case for transient outbursts or a donor of mass $\approx1$ \msun. Most BH are transient sources \citep[e.g.][]{wiktorowicz09-14, corral-santana03-16, belloni-16}, and there is only one strong candidate persistent BH LMXB in the Milky Way, 4U 1957+115 \citep[e.g.][]{ricci07-95, nowak12-08, gomez08-15}. Some sources could be long duration outburst transients, such as GRS 1915+105, which may be difficult to distinguish from persistent sources without extremely long light curves. Thus, the most luminous (persistent) LMXBs are more likely to be NS. We would then expect to detect a lower BH fraction in galaxies with low-sSFR dominated by LMXBs. However, this effect would not necessarily extend down the luminosity function, where numerous transient BH LMXBs reside.

\begin{figure*}[!ht]
\begin{tabular}{cc}
\includegraphics[width=1.0\columnwidth]{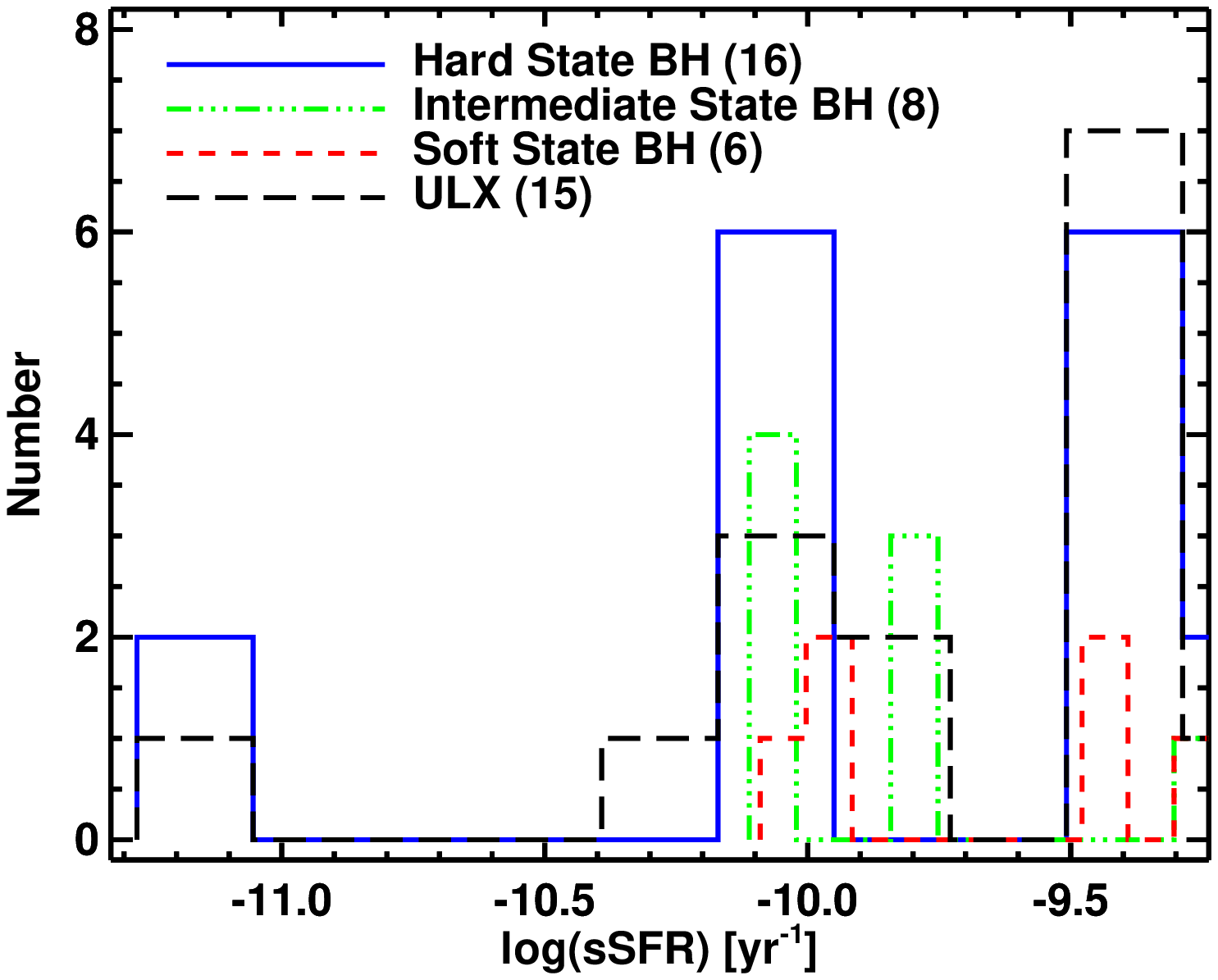}
\includegraphics[width=1.0\columnwidth]{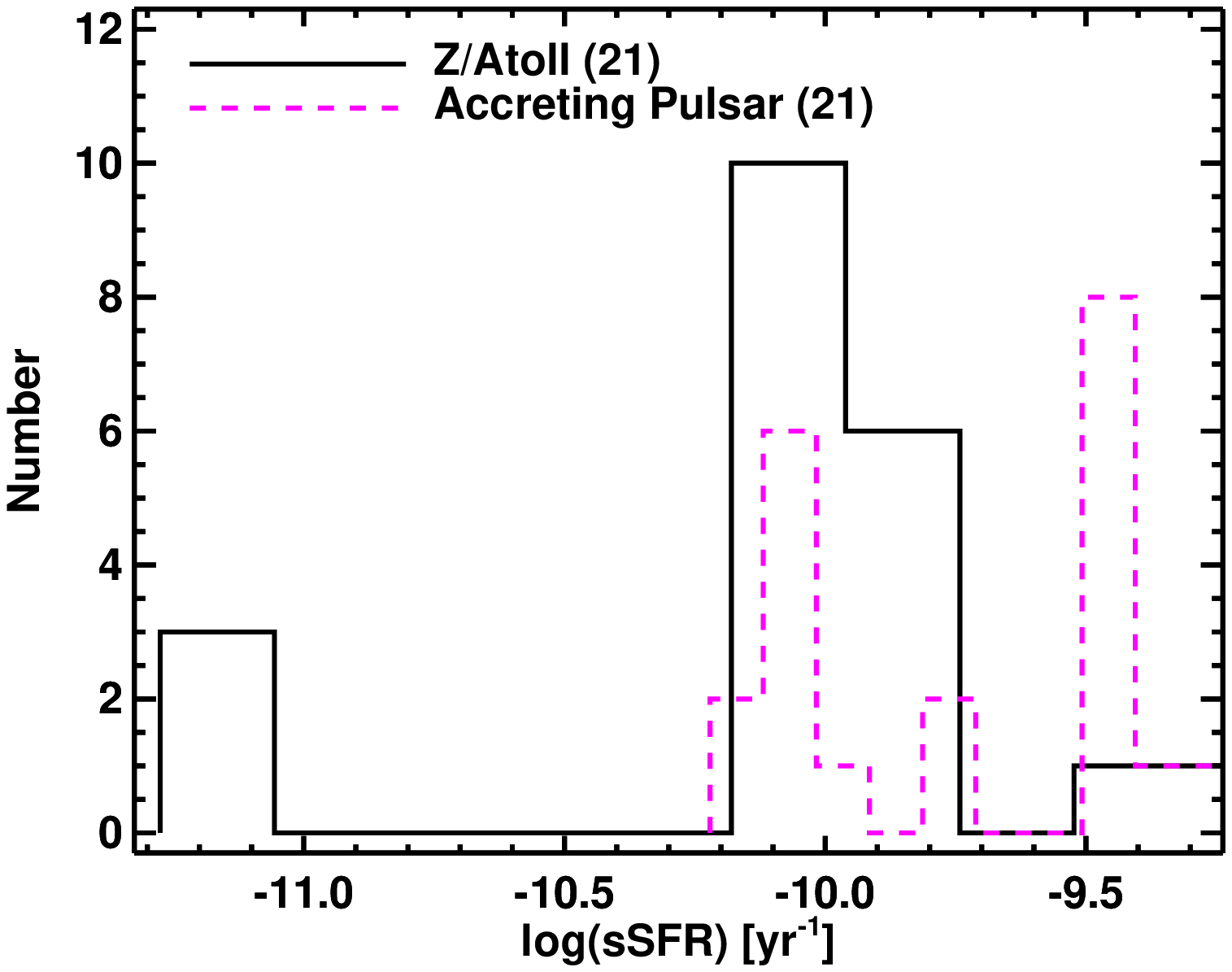}
\end{tabular}
\caption{Histograms showing the distribution of various BH accretion states (left) and accreting NS sources (right) for given sSFR based on the \nustar\ galaxy sample. Bin sizes vary based on the number of sources at each sSFR. Two BH and one NS source could not be separated by accretion state. As shown in Figure \ref{fig:sfrmass}, the sample is not uniform across sSFR and therefore features of the histogram are biased based on the sSFR distribution. ULXs are prevalent at high-sSFR as expected due to elevated star formation, and accreting pulsars are prevalent in the starburst galaxy M82 at $\log(\rm{sSFR/yr}^{-1}) = -9.4$.}	\label{fig:stateshist}
\end{figure*}

In Figure \ref{fig:stateshist} we show the distribution of BH accretion states (left) and accreting NS sources (right) as a function of the sSFR of each galaxy. Two BH and one NS source could not be separated by accretion state. All sources with \lx\ (\full keV) $\gtrsim1.3\times10^{39}$ \es, the Eddington limit for a 10 \msun\ BH, are classified as ULXs. Eight of fifteen ULXs are found at high-sSFR as expected due to elevated star formation. Hard state BH compose a large proportion of sources at all sSFR, indicating a common XRB accretion state independent of sSFR. With optical follow-up it may become apparent that a fraction of hard state sources are background AGN, which would explain their prevalence at all values of sSFR. Intermediate state sources are found at intermediate sSFR. The accreting pulsar population is prevalent in the starburst galaxy M82 at $\log(\rm{sSFR/yr}^{-1}) = -9.4$, whereas the fraction of Z/Atoll NS increases towards lower sSFR as expected. M31, which is not shown here, is dominated by Z/Atoll sources.

Hard state BH appear as a large fraction of all identified accretion states at all sSFR in the left panel of Figure \ref{fig:stateshist}. However, we do not expect there to be a relation between BH accretion states and sSFR. Instead, the accretion state depends on parameters such as orbital separation and disk instabilities. \citet{tetarenko02-16} recently produced an updated catalog of Galactic BH XRBs that provides a convenient comparison for our extragalactic sample. The study found that $38^{+6.0}_{-5.6}$\% of 132 transient outbursts detected in the Galaxy in the last 19 years do not complete the hysteresis pattern, skipping the transition from the hard to soft state. Either long periods were spent in the hard state or the source would reach the intermediate state then transition back to the hard state. They found that this hard state-only behaviour was not limited to recurrent transients but also seen in long-term transient and persistent accreting BH \citep[][and references therein]{tetarenko02-16}. The duration of Galactic BH outbursts vary depending on whether they were successful (i.e.\ state transition occurred) or hard state-only, having mean values of $\approx247$ and $\approx391$ days, respectively \citep{tetarenko02-16}. These hard state outburst sources generally have peak luminosities $\lesssim0.11$ $L_{\rm{Edd}}$, near the transition region to soft state luminosities. The hard state-only phenomenon is thought to occur in sources with low-level mass accretion rates. These factors may also explain the lack of identified soft state sources. While we need increased sensitivity to detect soft state sources, \nustar\ observations of M31 reaching $\sim10^{36}$ \es\ have shown that most sources are NS compared with only $2-3$ potential BH hard state candidates. Of these BH candidates, none were soft state sources.\footnote{\citet{maccarone06-16} demonstrated the likely NS nature of 5 M31 GCs, contrary to the previous BH candidate classification for 2 of these sources. They argued that BH candidates that are persistent and luminous are more likely NS \citep[e.g. 50 M31 BH candidates in][]{barnard08-14}.} For the Galactic BH hard state sources in \nustar\ hardness-intensity diagrams (blue circles), most are found at \lx\ $ < 3\times10^{37}$ \es, although they do extend just above $10^{38}$ \es. However, the \nustar\ sample sources classified as hard state BH have \lx\ $ \gtrsim10^{38}$ \es, similar to the persistent Galactic BH-HMXB Cygnus X-1\footnote{Not part of the \rxtet\ sample Galactic BH sources used to create \nustar\ diagnostic diagrams}, which spends most of its time in the hard state \citep{grinberg06-13}.

The accreting pulsars used to create the diagnostic diagrams in Figure \ref{fig:diagdiag} are Galactic NS with strong magnetic fields, accreting from high-mass companions. 
We detected accreting pulsar candidates in M82 (8), NGC 253 (5), Circinus (2), IC 342 (2), and one in each of M83, NGC 1313, NGC 4945, and Holmberg IX. These types of sources are produced by recent bursts of star formation, thus it is no surprise that M82 and NGC 253, galaxies with the largest SFR in our sample, contain a large number of pulsars. 
X-ray pulsars have also been found in the Large \citep[e.g.][]{antoniou06-16} and Small \citep[e.g.][]{antoniou06-10, haberl02-16} Magellanic Clouds. Both the SMC \citep{harris03-04} and LMC \citep{harris11-09} have SFR of $\approx0.3$ \sfr\ and sSFR similar to that of M82. To date, various studies have identified 64 pulsars in the SMC \citep{haberl02-16, vasilopoulos10-17} and 19 in the LMC \citep{clark07-15, antoniou06-16, haberl02-17, vasilopoulos03-18}, with sensitivity limits $\sim10^{33}$ \es\ in the $0.5-10$ keV energy band. All but two of the \numap\ pulsars identified in this work have \lx\ (\full keV) $\leq 3\times10^{38}$ \es, with a peak at the Eddington limit for NS. Conversely, the populations in the Magellanic Clouds have \lx\ ($0.2-12$ keV) $\leq10^{38}$ \es, with the majority of sources at least an order of magnitude fainter \citep{yang03-17}. This hampers comparison to the Magellanic Clouds because the detection limit of our survey mainly probes accreting pulsars undergoing very luminous type II outbursts, which are very rare \citep{reig03-11}. 

Due to the dependence of the pulsar rate (number formed per SFR) on the time since the star formation episode, and varying sensitivity limits, it is difficult to directly compare results. Pulsars in the SMC were found in regions having a peak in star formation history at $\approx42$ Myr and duration of 33 Myr \citep{antoniou06-10}. However, using spatially resolved maps of star formation history in the LMC, \citet{antoniou06-16} found the region around pulsars peaked earlier at $\approx13$ Myr with duration 32 Myr. 
There is indication that the dominant stellar population in M82's nucleus is $\sim10$ Myr old, and reaches 100 Myr at 500 pc \citep{rodriguez-merino01-11}. The star-forming complex in the central region of NGC 253 is thought to be $<8$ Myr old and have been formed in the last $\sim30$ Myr \citep{engelbracht10-98, davidge02-16}. These estimates are broadly in agreement with time-scales from the Magellanic Clouds.

Many galaxies in our sample have rich muti-wavelength data sets that allow these sources to be investigated using UV/optical/IR catalogs, which, in combination with the \nustar-\chandra/\xmmn\ data we analyzed, can help confirm the nature of these sources (e.g.\ to determine if they are located in globular clusters). These cross-correlations will help to confirm the accuracy of \nustar\ diagnostics. We do not perform this analysis here as it is beyond the scope of this paper (see \citealt{lazzarini06-18} for detailed optical counterpart identification).

\subsection{Correlation of X-ray Luminosity with SFR and Stellar Mass}	\label{sec:corr}

The connection between X-ray luminosity and the SFR and stellar mass of a galaxy has been well studied. \chandra, in combination with multiwavelength data, has constrained the correlation between HMXBs and SFR as well as LMXBs and stellar mass \citep[e.g.][]{ranalli02-03, grimm03-03, gilfanov01-04, colbert02-04, gilfanov03-04, persic02-07, lehmer11-10, mineo01-12, lehmer07-16}. Despite variations in other galaxy properties such as stellar age, metallicity, dynamics, etc., these global relations are remarkably consistent. The well-known relation between galaxy X-ray luminosity and SFR can be parametrized as follows:
\begin{align}
\log L_{\text{X}}	&=\log A+B\log \text{SFR}	\label{eq:lxsfr}
\end{align}
where \lx\ is in units of \es\ and SFR is in units of \sfr. The X-ray luminosity of galaxies at energies $\gtrsim2$ keV is dominated by XRBs. While non-linear scalings exist for the individual \lx\ (LMXB) and \lx\ (HMXB) relations based on low stellar mass and low-SFR regimes, respectively (e.g.\ \citealt{gilfanov07-04}), we also adopt the combined form to constrain the total XRB emission from a galaxy:
\begin{gather}	\label{eq:lxsfrm}
L_{\mathrm{X}} (\text{XRB}) = L_{\mathrm{X}} (\mathrm{LMXB}) + L_{\mathrm{X}} (\mathrm{HMXB}) = \alpha M_{\star} + \beta \mathrm{SFR} 	\\	\label{eq:lxssfr}
L_{\mathrm{X}} (\text{XRB})/\mathrm{SFR} = \alpha(\mathrm{SFR}/M_{\star})^{-1} + \beta  	
\end{gather}
Previous surveys have investigated X-ray emission in the $0.5-2$, $2-10$, and $0.5-8$ keV energy bands. X-ray emission from normal galaxies in the $2-10$ keV band is dominated by XRBs and therefore is a cleaner correlation. We utilized the broad \full keV and the hard \hard keV energy bands to study the \lx\ relation with SFR and stellar mass. The broad band is an ideal comparison to previous $2-10$ keV studies because of the similar flux in each band for XRB spectra with $\Gamma\sim1.7$ and average values of extinction. The hard band provides a clean sample of XRB-only emission and insight into how hard X-ray luminosity from XRBs varies with SFR and stellar mass.

The \nustar\ sample includes 12 galaxies, with three of these being dwarf galaxies (Holmberg II, Holmberg IX, NGC 5204). A fourth, NGC 1313, is intermediate between dwarf and $L^{\star}$ galaxies\footnote{An $L^{\star}$ galaxy has a luminosity similar to the Milky Way}. The sample is slightly biased towards intermediate sSFR (Figure \ref{fig:sfrmass}) as a result of the relative lack of nearby starbursts and massive elliptical galaxies. The absence of elliptical galaxies from our sample means that the correlations will be dominated by the star-forming component that produces bright HMXBs. The exceptions to this rule are M31 and M81, where the low SFR and thus lack of bright HMXBs results in LMXBs (globular cluster sources in the case of M31) dominating the integrated emission.

\begin{figure*}[!ht]
\begin{tabular}{cc}
\includegraphics[width=1.0\columnwidth]{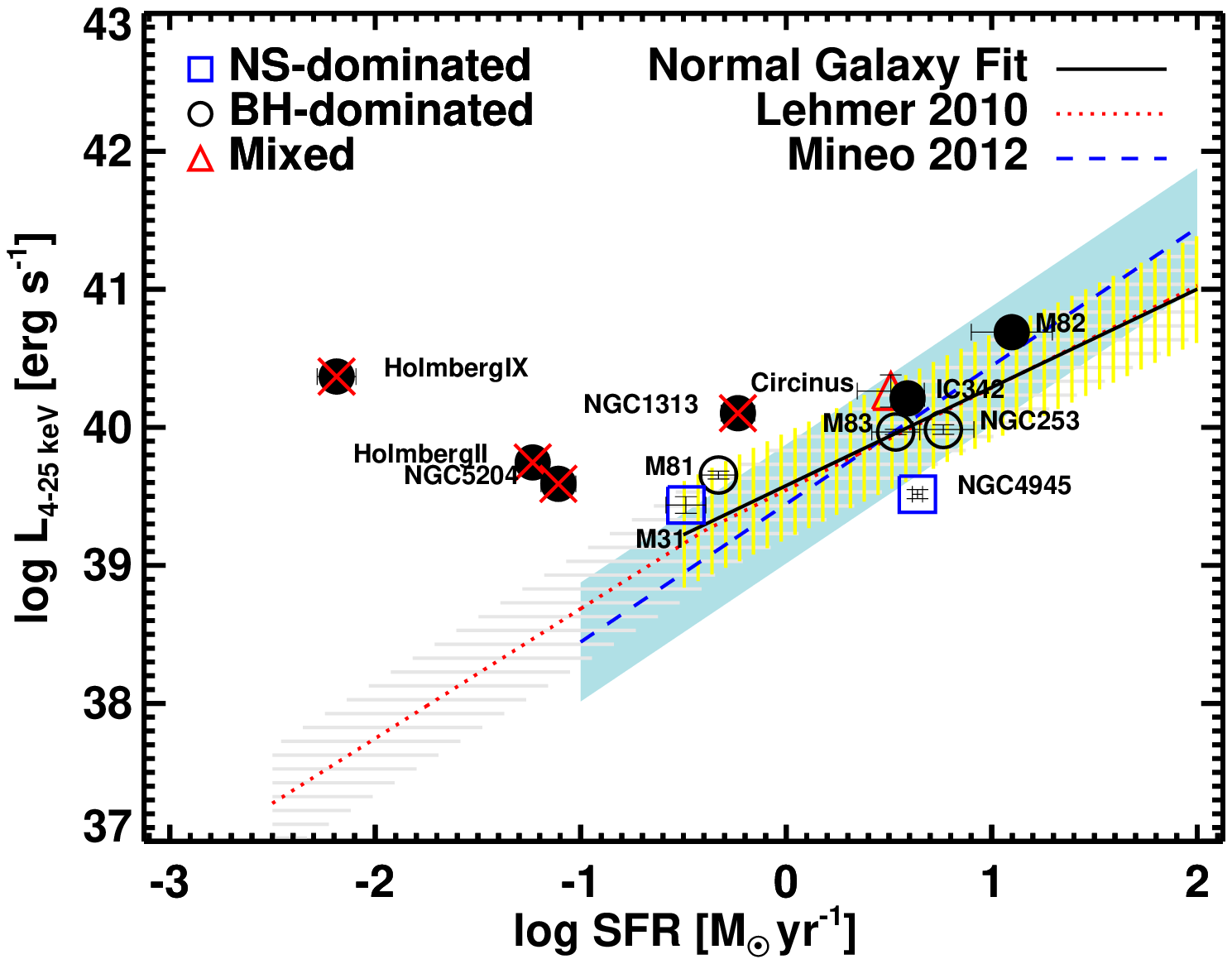}
\includegraphics[width=1.0\columnwidth]{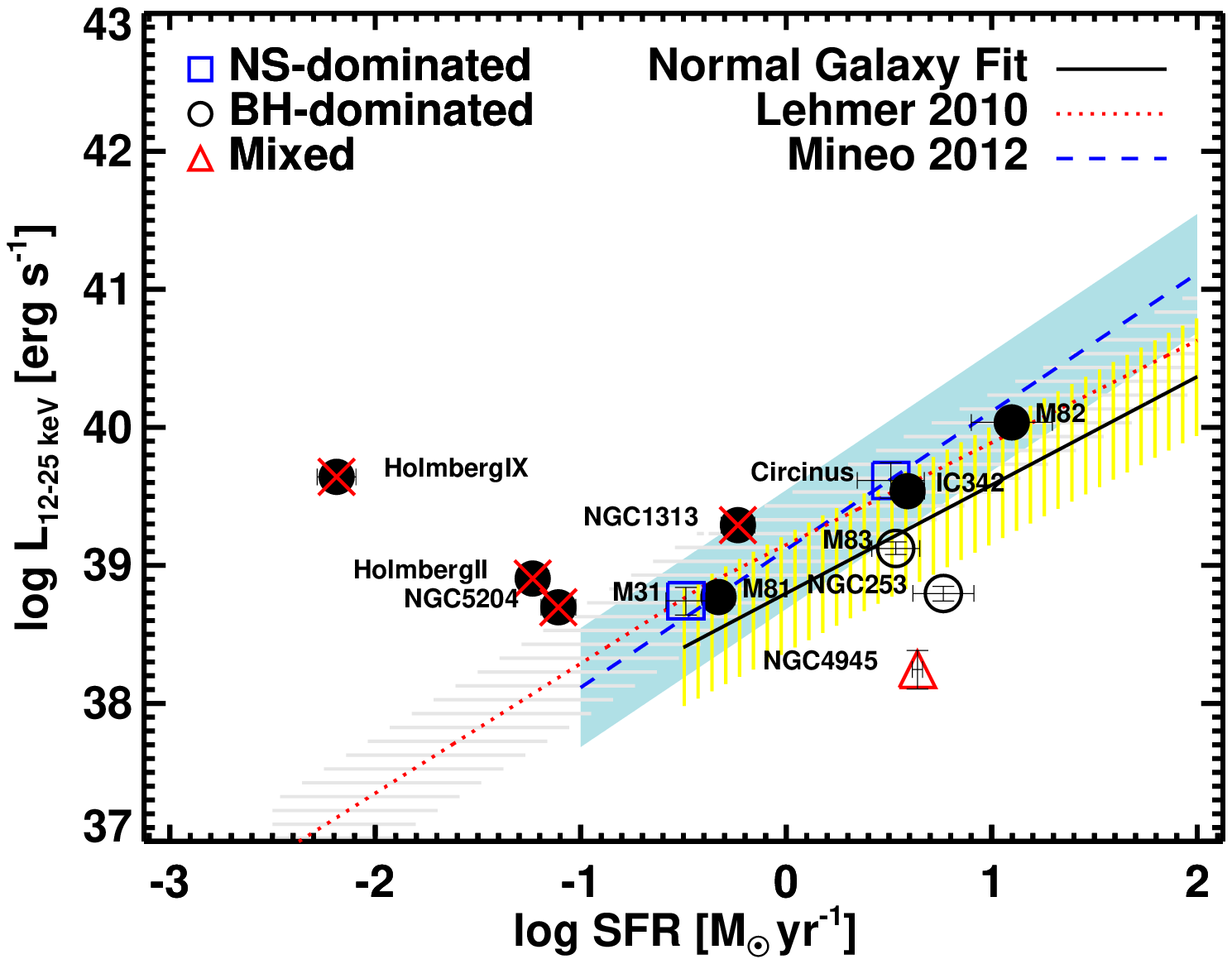}
\end{tabular}
\caption{The integrated \full keV (left) and \hard keV (right) point source emission (based on sources in Table \ref{tab:rates}) as a function of the SFR of that galaxy is shown for the \nustar\ sample. The same spectral model for conversion from count rate to \lx\ was assumed as in Figures \ref{fig:empty}-\ref{fig:allgals}. Galaxies were classified as NS (blue squares), BH (black circles) and ULX-dominated (filled) if $>$70\% of their \full keV (left) or \hard keV (right) point source emission came from one of these groups (see Table \ref{tab:percent}). Galaxies that were not BH or NS-dominated were classified as mixed (red triangles). The dashed blue line shows the \lx-SFR relation for 29 nearby star-forming galaxies from \citet{mineo01-12}, with the dispersion shaded light blue. The dotted red line represents the \lx-SFR relation for 66 normal galaxies from \citet{lehmer11-10}, with horizontal gray lines showing the dispersion. We converted these relations to the \full and \hard keV bandpasses using the same spectral model as in Figures \ref{fig:empty}-\ref{fig:allgals}. The solid black line shows the best fit for the 8 normal (Milky Way-type) galaxies in the sample (see Section \ref{sec:corr} and Table \ref{tab:params}, red crosses indicate dwarfs excluded from the fit), with yellow vertical lines showing the dispersion. The \full keV fit matches the result from \citet{lehmer11-10} whereas the \hard keV fit is offset based on the assumed spectral model.}	\label{fig:lx-sfr}
\end{figure*}

While we expect \lx\ to scale with both stellar mass and SFR, the sample galaxies are mostly intermediate sSFR (not LMXB-dominated), so we also investigated the \lx-SFR correlation \citep[e.g.][]{lehmer11-10,mineo01-12}. In Figure \ref{fig:lx-sfr} we show the integrated \full keV (left) and \hard keV (right) point source emission (based on sources in Table \ref{tab:rates}) as a function of the SFR for each galaxy in the \nustar\ sample. 

We grouped galaxies in the sample using results from the \nustar\ diagnostic diagrams (Figures \ref{fig:empty}-\ref{fig:allgals}). Specifically, for all the sources in a galaxy that were classified as BH or NS, we determined what percentage of the \full and \hard keV luminosity came from each population. Galaxies with $\gtrsim70$\% of their \lx\ from BH (NS) were classified as BH-dominated (NS-dominated) and shown as black circles (blue squares). Galaxies that did not meet either of these criteria were classified as mixed and are shown as red triangles. Galaxies that had $\gtrsim70$\% of their \lx\ from ULXs (defined as sources with \lx\ (\full keV) $\gtrsim 1.3\times10^{39}$ \es, the Eddington limit for a 10 \msun\ BH) were classified as ULX-dominated (filled symbols). 

In Table \ref{tab:percent} we show the proportion of \full and \hard keV luminosity from each of these populations. While most ULXs have historically been presumed to be BH, recent work using \nustar/\xmmn\ has shown that some sources exhibit pulsations and are in fact NS \citep{bachetti10-14, fuerst11-16, israel02-17, israel03-17}. Only the confirmed ULX pulsar M82 X-2 is in the \nustar\ galaxy sample we analyzed\footnote{We were not able to separate emission from ULXs M82 X-1 and M82 X-2, which are 5\arcsec\ apart}, but we cannot rule out the possibility that other ULX sources may be proven to be NS as opposed to BH. A binary population synthesis study by \citet{fragos03-15} found that only 13\% of galaxies that have a similar star formation history to M82 are likely to have ULXs with a NS accretor. Therefore we assumed that galaxies dominated by ULXs have BH accretors.

\begin{table*}

\caption{Galaxy Classifications Based on Source Type\label{tab:percent}}
\begin{tabular}{c c c c c c c c c c c c c c c c}
\hline\hline
Galaxy	&	NS per cent	&	BH per cent	& ULX per cent		&	BH+NS	&	NS per cent	&	BH per cent	& ULX per cent		&	BH+NS	\\	\cmidrule(lr){2-5} \cmidrule(lr){6-9}
	&	\multicolumn{4}{c}{(\full keV)}	&	\multicolumn{4}{c}{(\hard keV)}	\\	\hline
Circinus &    36 (2) &    64 (2) &    64 (1) & 4 &    75 (2) &    25 (1) &    25 (1) & 3
\\ IC342 &     3 (3) &    97 (6) &    89 (2) & 9 &     4 (3) &    96 (4) &    94 (2) & 7
\\ NGC4945 &    80 (6) &    20 (2) &     0 (0) & 8 &     0 (0) &     0 (0) &     0 (0) & 0
\\ HolmbergII &     0 (0) &   100 (3) &    94 (1) & 3 &     0 (0) &   100 (2) &    91 (1) & 2
\\ M81 &    26 (3) &    74 (3) &    60 (1) & 6 &     0 (0) &   100 (3) &    79 (1) & 3
\\ HolmbergIX &     2 (2) &    98 (5) &    96 (1) & 7 &     1 (1) &    99 (1) &    99 (1) & 2
\\ NGC5204 &     0 (0) &   100 (2) &    96 (1) & 2 &     0 (0) &   100 (1) &   100 (1) & 1
\\ NGC1313 &     7 (2) &    93 (6) &    85 (2) & 8 &     5 (1) &    95 (3) &    94 (2) & 4
\\ M83 &    25 (7) &    75 (8) &    23 (1) & 15 &    26 (2) &    74 (6) &    10 (1) & 8
\\ NGC253 &    28 (10) &    72 (6) &    59 (2) & 16 &    26 (3) &    74 (3) &    27 (1) & 6
\\ M82 &     2 (8) &    98 (4) &    97 (3) & 12 &     2 (2) &    98 (3) &    98 (3) & 5
\\ \hline Total & 43 & 47 & 15 & 90 & 14 & 27 & 15 & 41
\\	\hline
\end{tabular}
\tablecomments{Percentage of \full and \hard keV luminosity from BH, NS, and ULXs based on the total luminosity for all classified sources. The number of sources in a category is shown in parentheses, and the total number of sources in each category is shown in the last row. The total number of BH and NS in each galaxy is shown in the BH+NS column. These values were used to categorize galaxies in Figures \ref{fig:lx-sfr} and \ref{fig:lxsfr-ssfr}, where galaxies were classified as NS, BH, and ULX-dominated if $>$70\% of their point source emission came from one of these groups.}
\end{table*}

We fit the log-linear model from equation \ref{eq:lxsfr} to the data in Figure \ref{fig:lx-sfr}. We performed fitting using the generalized linear model ($glm$) in $\bf{R}$ \citep{R-2008}. The solid black line shows the best fit for the 8 normal\footnote{Galaxies with stellar masses $M_{\star}>10^{10}$ \msun} (Milky Way-type) galaxies in the sample. We only used the 8 normal ($L^{\star}$) galaxies in the sample, excluding Holmberg II, Holmberg IX, NGC 5204, and NGC 1313. Fitting the relation using different morphological types biases the underlying physical assumption of late-type galaxies having X-ray point source emission dominated by HMXBs. The best-fitting parameters are shown in Table \ref{tab:paramlxsfr}. For comparison we show the \lx-SFR relation for 29 nearby star-forming galaxies from \citet{mineo01-12} (dashed blue line), converted from $0.5-8.0$ keV. We also show the relation for 66 normal galaxies from \citet{lehmer11-10} (dotted red line), converted from $2-10$ keV. We used the same spectral model as in Figures \ref{fig:empty}-\ref{fig:allgals} to convert to the \full and \hard keV ranges. The fit from \citet{lehmer11-10} is separated into distinct curves above and below an SFR of 0.4 \sfr. Our \full keV best fit matches the result from \citet{lehmer11-10}, whereas the \hard keV fit, while still consistent, is offset based on the assumed spectral model used to convert the \citet{lehmer11-10} relation to \hard keV. Contrary to these non-linear \lx-SFR fits, \citet{mineo01-12} found a linear correlation, which agrees within uncertainties due to the large scatter in the relations. They argued the dispersion was not a result of measurement uncertainties or the CXB/LMXB sources but instead of physical origin.

\begin{table}
\caption{Best-fitting \lx-SFR Parameters\label{tab:paramlxsfr}}
\begin{tabular}{@{}  c c c  @{}}
\hline\hline
Energy Band		&	 A	&	B	\\	
(keV)	&		&		\\	\hline
\full	&	$\fitaf\pm\fiterraf$ 	&	$\fitbf\pm\fiterrbf$
\\ \hard 	&	$\fitah\pm\fiterrah$	&	$\fitbh\pm\fiterrbh$
\\ \hline
\end{tabular}
\tablecomments{Best-fitting parameters and $1\sigma$ uncertainties for $\log L_{\text{X}}=\log A+B\log \text{SFR}$ (equation \ref{eq:lxsfr}) for the 8 normal (Milky Way-type) galaxies in the \nustar\ sample from Figure \ref{fig:lx-sfr}.}
\end{table}

\begin{figure*}[!ht]
\begin{tabular}{cc}
\includegraphics[width=1.0\columnwidth]{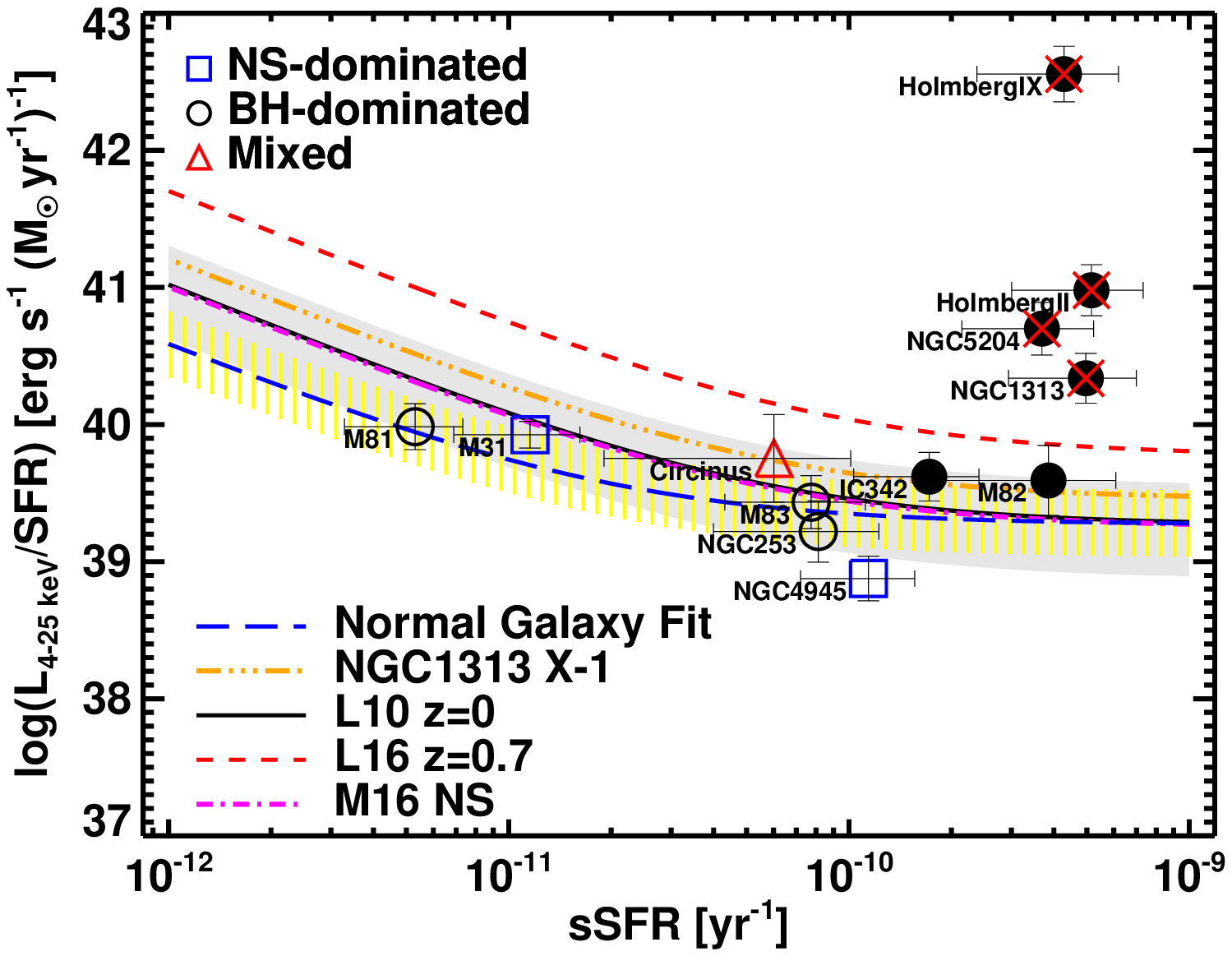}
\includegraphics[width=1.0\columnwidth]{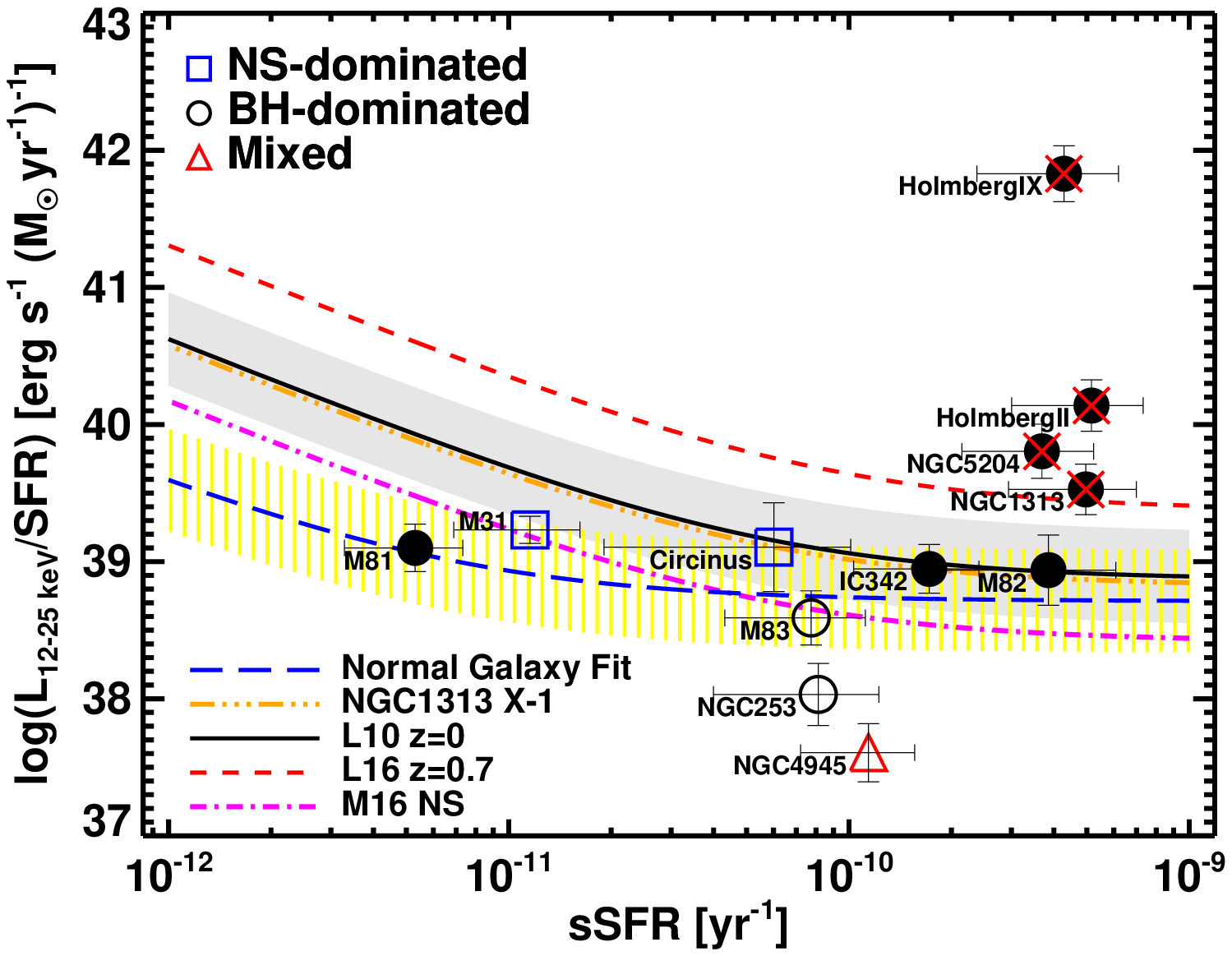}
\end{tabular}
\caption{The integrated \full keV (left) and \hard keV (right) point source emission (based on sources in Table \ref{tab:rates}) normalized by the SFR as a function of the sSFR of that galaxy is shown for the \nustar\ sample. The same spectral model for conversion from count rate to \lx\ was assumed as in Figures \ref{fig:empty}-\ref{fig:allgals}. Galaxies were classified as NS (blue squares), BH (black circles) and ULX-dominated (filled) if $>$70\% of their \full keV (left) or \hard keV (right) point source emission was from one of these groups (see Table \ref{tab:percent}). Galaxies that were not BH or NS-dominated were classified as mixed (red triangles). The solid black line represents the relation from the local ($z=0$) galaxy sample of \citet{lehmer11-10}, with the dispersion shaded light gray. The dashed red line shows the relation from the \chandra\ Deep Field South stacked galaxy sample of \citet{lehmer07-16} with median redshift of 0.7. We converted the original $2-10$ keV luminosities to our bandpasses using the spectral model assumed by \citet{lehmer07-16}. We also converted the \citet{lehmer11-10} local galaxy result based on various spectral models: using the NGC 1313 X-1 ULX spectrum from \citet{bachetti12-13} (dash-dotted orange line) and the \textsc{COMPTT} model for the NS in Bo 185 from \citet{maccarone06-16} (dash-dotted magenta line). The dashed blue line shows the best fit for the 8 normal (Milky Way-type) galaxies in the sample (see Section \ref{sec:corr} and Table \ref{tab:params}, red crosses indicate dwarfs excluded from the fit), with yellow vertical lines showing the dispersion. We have adjusted the stellar mass of M31 and M81 based on the FOV and AGN-dominated emission region, respectively.}	\label{fig:lxsfr-ssfr}
\end{figure*}

In order to constrain the total XRB emission from each galaxy we fit the data using equation \ref{eq:lxssfr}. In Figure \ref{fig:lxsfr-ssfr} we show the integrated \full keV (left) and \hard keV (right) point source emission (based on sources in Table \ref{tab:rates}) normalized by the SFR as a function of the sSFR for each galaxy in the \nustar\ sample. Classifications are the same as in Figure \ref{fig:lx-sfr}. 
To properly analyze XRB emission and its relation with sSFR required us to make corrections to the stellar masses of certain galaxies. We adjusted the stellar masses of M31 and M81 based on the FOV and AGN-dominated emission region, respectively. For M31, we used the updated stellar mass maps from \citet{williams09-17}, which were derived from fits to data from the Panchromatic Hubble Andromeda Treasury Survey, scaled to the FOV of the \nustar\ observations. 
We did not include the M81 AGN in our source list/analysis and as such excluded the stellar mass in a circular region of radius 1.5\arcmin\ centered on the AGN, where no other sources could be detected. Using Table 3 of \citet{tenjes07-98} we calculated the mass within this region to be $2.4\times10^{10}$ \msun\ (25\% of the total) and excluded that value from the total stellar mass of M81. Both NGC 4945 and Circinus have Seyfert nuclei that were also excluded from our source list. However, their low-luminosity did not prevent us from detecting sources in their bulges within $0.5$\arcmin\ of their nuclei, therefore we do not exclude any stellar mass from the total as it is negligible.

As in Figure \ref{fig:lx-sfr}, we used the 8 normal ($L^{\star}$) galaxies in the sample to determine the best fit to equation \ref{eq:lxssfr}. The dashed blue line in Figure \ref{fig:lxsfr-ssfr} shows the ($glm$) best fit, while the parameters are presented in Table \ref{tab:params}. 
The \full keV \lx/SFR for the NS model does not differ appreciably from the \citet{lehmer11-10} local galaxy relation. The ULX model of NGC 1313 X-1 does show elevated \lx/SFR relative to \citet{lehmer11-10} as expected. For the \hard keV panel the NS model has relatively lower \lx/SFR than the NGC 1313 X-1 ULX model, indicating that NS spectra turn over faster than ULXs. The increased scatter in the \hard keV panel compared to the \full keV is evident as in Figure \ref{fig:lx-sfr}, particularly for NGC 4945 and NGC 253 in the \hard keV band. Both panels indicate that \lx\ per unit SFR is larger for lower sSFR. 

Previous studies \citep[e.g.][]{lehmer11-10} found that extinction in the star-forming regions of starburst galaxies could account for the decreased \lx/SFR in the $2-10$ keV band. With \nustar, the \full keV and especially \hard keV energy bands are not subject to the same degree of extinction. Even with extreme extinction of $N_{\rm{H}}=10^{24}$ cm$^{-2}$, the \full and \hard keV energy bands are attenuated\footnote{For photoelectric absorption; scattering further reduces the flux by $\sim50\%$ in each energy band.} by factors of 2 and 1.2, respectively, from standard Galactic values of 10$^{20}$ cm$^{-2}$, assuming a power-law with $\Gamma=1.7$. Thus, we find that \lx/SFR is indeed lower at low sSFR compared to previous studies. A larger sample of galaxies with uniform sSFR is required in order to determine whether these correlations hold for a larger range in sSFR. 

The galaxies in the \nustar\ sample are spirals/dwarfs with recent star formation and are not LMXB-dominated elliptical galaxies. As such, their point source emission should be dominated by HMXBs. The four dwarf galaxies that are BH and ULX-dominated all have elevated \lx/SFR for a given sSFR compared to the normal $L^{\star}$ galaxies. This effect may be a result of the star formation history of a galaxy that leads to a peak in \lx/SFR, similar to the peak in the \be-HMXB distribution $\sim50$ Myr after a star formation episode found in nearby galaxies \citep[e.g.][]{antoniou06-10,williams07-13,antoniou06-16}. However, M82 X-1 has also been detected at a \full keV luminosity\footnote{Converted from $0.3-10$ keV \citep{bachetti10-14}} of $5\times10^{40}$ \es, which would shift it above the best-fitting relation into the dwarf galaxy locus. None of the galaxies in \citet{lehmer11-10} at similar sSFR, which have even higher SFR, have a total \lx\ above $10^{40}$ \es. The transient nature of XRBs, specifically the duration and recurrence times of their outbursts, can introduce complications in studying the relationship of \lx\ with SFR and stellar mass.

\begin{table}
\caption{Best-fitting \lx/SFR-sSFR Parameters\label{tab:params}}
\begin{tabular}{@{}  c c c  @{}}
\hline\hline
Energy Band		&	 $\alpha$	&	$\beta$	\\	
(keV)	&	($10^{28}$ \es\ \msun$^{-1}$)	&	($10^{39}$ \es\ (\sfr)$^{-1}$)	\\	\hline
\full	&	$\fitalphaf\pm\fiterralphaf$ 	&	$\fitbetaf\pm\fiterrbetaf$
\\ \hard 	&	$\fitalphahten\pm\fiterralphahten$	&	$\fitbetahten\pm\fiterrbetahten$
\\ \hline
\end{tabular}
\tablecomments{Best-fitting parameters and $1\sigma$ uncertainties for the relation $L_{\mathrm{X}} (\text{XRB}) = \alpha M_{\star} + \beta \mathrm{SFR}$ (equations \ref{eq:lxsfrm} and \ref{eq:lxssfr}) for the 8 normal (Milky Way-type) galaxies in the \nustar\ sample from Figure \ref{fig:lxsfr-ssfr}.}
\end{table}

\subsection{Hard X-ray Luminosity Functions}	\label{sec:xlf}

XLFs of nearby galaxy point source populations are a powerful tool because they are not subject to the uncertainties associated with Galactic sources (distance, extinction, low number statistics, etc.). \citet{gilfanov03-04} investigated a sample of 11 nearby galaxies with old stellar populations and determined that the total X-ray luminosity and the XLF of LMXBs each scaled with stellar mass. Similarly, \citet{grimm03-03} studied the X-ray populations of nearby spiral/starburst galaxies and found that the total X-ray luminosity and XLF of HMXBs scaled with SFR (see \citealt{fabbiano09-06} for a review of XRB XLFs). The latter discovery is particularly appropriate for our work as our sample is dominated by late-type galaxies with HMXBs. Recent work by \citet{mineo01-12} using \chandra, \spitzer, \galex, and \twomass\ data of 29 nearby star-forming galaxies confirmed and updated the $0.5-8$ keV HMXB-SFR relation. They found an XLF power-law slope of 1.6 in the range \lx $=10^{35-40}$ \es\ with evidence for a break above this limit. They did not find any features near the Eddington limit for a NS or BH. However, a larger study of 343 nearby galaxies (213 late-type) with \chandra\ by \citet{wang09-16} did detect a break dividing NS and BH XRBs in the composite XLF of early and late-type galaxies. For the 213 late-type galaxies the break was located at $(6.3\pm0.3)\times10^{38}$ \es\ with a power-law slope of $1.6\pm0.03$ and $2\pm0.05$ below and above the break, respectively. 
The flat slope of HMXB XLF up to $10^{40}$ \es\ suggests that ULXs are prevalent among high-SFR galaxies. The smooth transition past the Eddington limit for NS may be due to a population of ULX pulsars (see Section \ref{sec:corr}). Recent analysis of \xmmn\ and \nustar\ data favors super-critically accreting NS as the engines of a large fraction of ULXs; although degeneracy between spectral models warrants deeper broadband observations to support this interpretation \citep{koliopanos12-17}. If there is a break in the HMXB XLF of late-type galaxies, separating NS and BH XRBs to create independent XLFs would determine how sources are distributed and help interpret the origin of the break.

Do we expect the XLFs of XRBs to differ at harder energies? XLFs of the Milky Way and nearby galaxies are generally presented in the $0.5-10$ keV energy range. The XLFs in the soft X-ray band ($\sim1-10$ keV) do not accurately represent the total luminosity of spectrally hard and absorbed HMXBs. For LMXBs, the brightest systems emit most of their energy in the standard $\sim1-10$ keV band, whereas faint systems emit a similar amount of energy in the standard and $10-100$ keV energy range \citep{revnivtsev11-08}. Only a few studies have investigated the hard (E $>10$ keV) XLF, where most have focused on AGN \citep[e.g][]{sazonov01-07, paltani07-08, ajello04-12, bottacini08-12, mereminskiy06-16}. 

The first hard XLF was presented by \citet{krivonos11-07} using $17-60$ keV data from the \intg/IBIS all-sky survey, detecting 219 Galactic sources, including 90 LMXBs and 76 HMXBs. \citet{revnivtsev11-08} used this \intg\ catalog \citep{krivonos11-07} to study the XLF of Galactic Center/bulge LMXBs, separating persistent (22) and transient (16) sources. 
The LMXB XLF was probed to a limit of $7\times10^{34}$ \es\ and exhibited a flattening at the faint-end with a differential slope of $0.96\pm0.2$ and $1.13\pm0.13$ for persistent and persistent+transient sources, respectively. 
The authors argued that the drop-off in the hard XLF for $L_{17-60 \rm{\ keV}}\gtrsim10^{37}$ \es\ is a result of the spectral change near this luminosity in the $2-10$ keV band, where sources have very soft spectra and therefore lower luminosities for energies E $>10$ keV.

More recently, a detailed XLF for Galactic sources above 10 keV was compiled by \citet{voss10-10} using $15-55$ keV data taken with \swift-BAT from $2005-2007$. Specifically, they classified 211/228 (93\%) of their sources, including 61 LMXBs and 43 HMXBs, for which they derived XLFs down to $7\times10^{34}$ \es. They found a differential faint slope of 1 for LMXBs, consistent with results from the $2-10$ keV band and \citet{revnivtsev11-08}. 
The HMXB XLF had a differential slope of $1.3^{+0.3}_{-0.2}$, which was flatter but similar to the canonical value of $\approx1.6$ for the Milky Way and other galaxies from the soft X-ray. However, the break at $2\times10^{37}$ \es\ was inconsistent with the single power-law slope from soft X-ray surveys. 
\citet{doroshenko07-14} completed a robust analysis of Galactic LMXB and HMXB XLFs using the \intg\ catalog of \citet{krivonos09-12}. A novel method was used to create XLFs that did not require distance measurements for any of the sources. 
The differential HMXB XLF parameters were $\alpha_{1}=0.3^{+0.8}_{-0.2}$, $\alpha_{2}=2.1^{+3}_{-0.6}$, and $L_{\rm{break}} = 0.55^{+4.6}_{-0.28}\times10^{36}$ \es. 
Both LMXBs and HMXBs had a flatter slope at low luminosities and a lower break luminosity compared to the previous hard X-ray studies.

The only known extragalactic hard XLF was published by \citet{lutovinov08-12}. Using \intg\ data from $2003-2004$ and $2010-2012$ they produced the $20-60$ keV completeness-corrected XLF of HMXBs in the LMC. 
From the 6 HMXBs in their sample, they found a power-law slope of $1.8^{+0.4}_{-0.3}$ and a break at $\sim10^{37}$ \es.
The power-law slope agreed with previous results in the 2-10 keV band from HMXBs in the LMC \citep{shtykovskiy09-05} and predictions from the canonical HMXB XLF \citep{grimm03-03}. 

Following their extragalactic study, \citet{lutovinov05-13} used the 9-year \intg\ All-Sky Survey catalog \citep{krivonos09-12} to study 48 persistent Galactic HMXBs. 
Differential slopes for the HMXB XLF were $\alpha_{1}=1.40\pm0.13 (\rm{stat})\pm0.06(\rm{syst.})$ and $\alpha_{2}>2.2$ with $L_{\rm{break}} = [2.5^{+2.7}_{-1.3}(\rm{stat.})\pm1.0(\rm{syst.})]\times10^{36}$ \es. The statistical significance of the break was at the $2\sigma$ level, but the authors argued that the low-luminosity flattening of the wind-fed NS-HMXB XLF is likely real. 
The break luminosity was different from that of \citet{grimm09-02} and \citet{voss10-10}, primarily due to the absence of BH systems, transients, Roche lobe systems, and varying completeness-correction methods. The lack of bright ($>10^{37}$ \es) HMXB sources in the Galaxy differs from the HMXB XLFs of star-forming galaxies, where the power-law slope continues to higher \lx.

\begin{figure*}[!ht]
\begin{tabular}{cc}
\includegraphics[width=1.0\columnwidth]{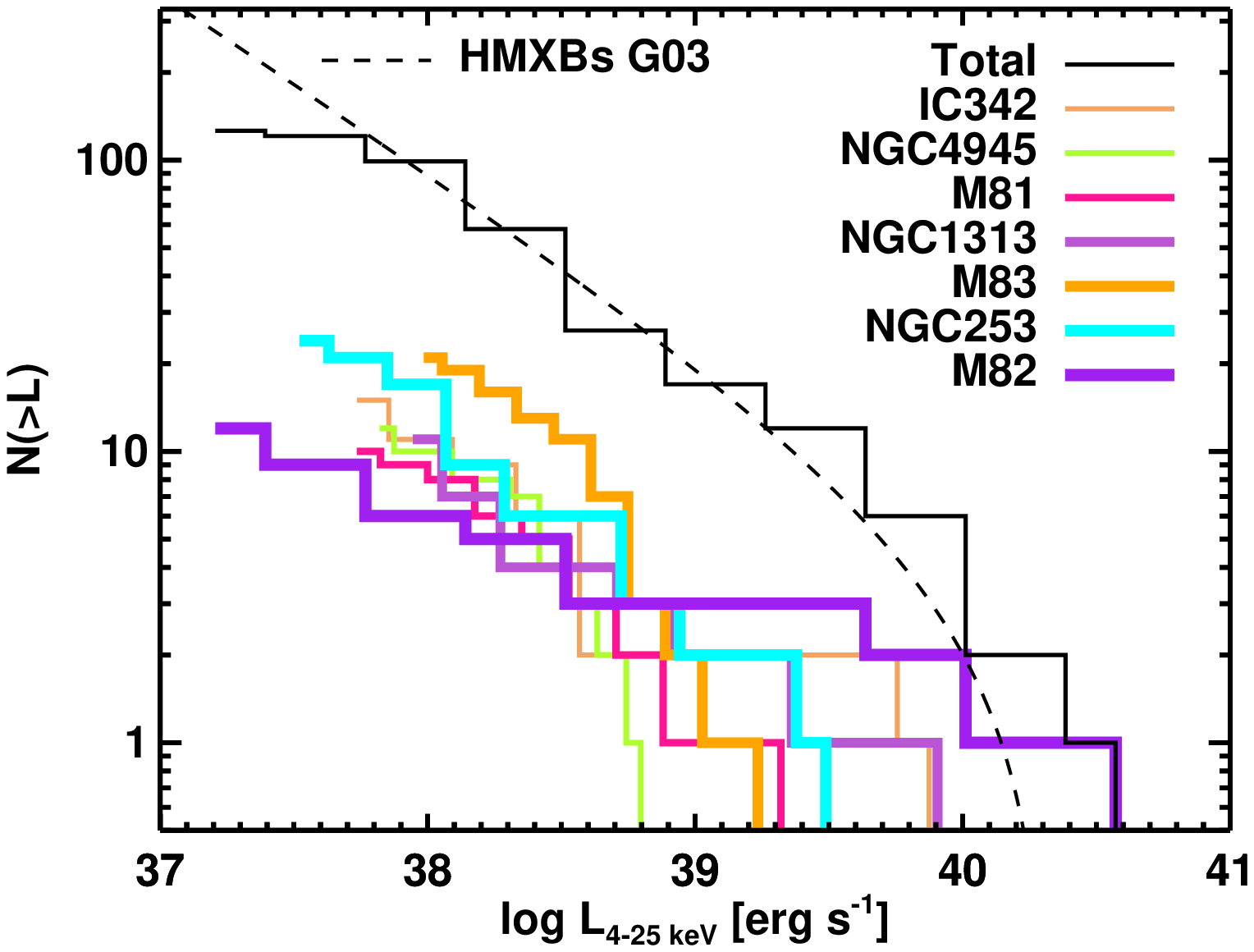}
\includegraphics[width=1.0\columnwidth]{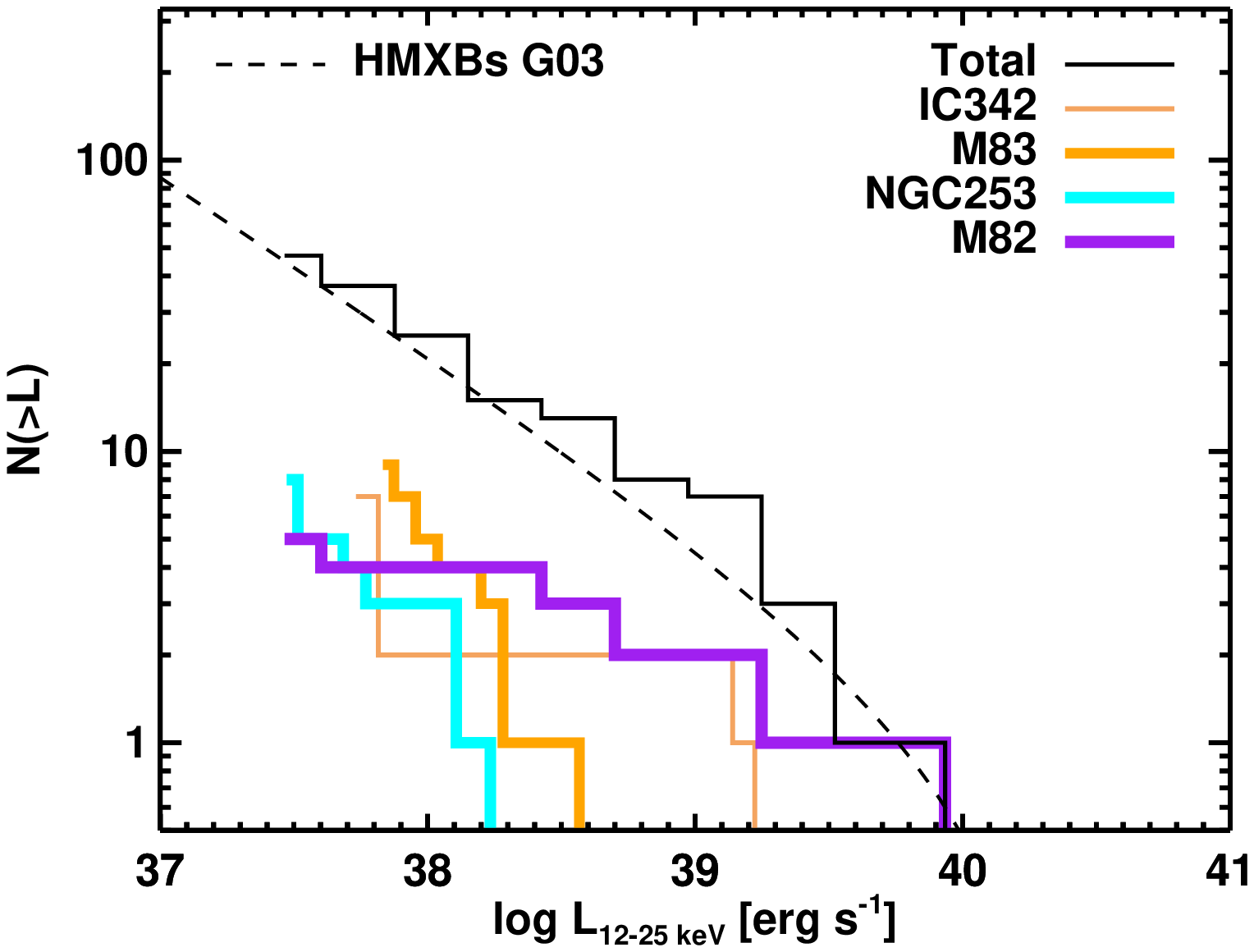}
\end{tabular}
\caption{\nustar\ XLFs in the \full keV (left) and \hard keV (right) energy bands for detected sources. We only included individual XLFs for galaxies that had at least 10 (5) sources for \full (\hard) keV. No completeness correction nor normalization for SFR nor stellar mass has been applied. The total XLFs (solid black line) represent all detected sources (in the given energy band) from all galaxies in the sample, including those whose individual XLFs are not shown here. The dash black line shows the HMXB XLF from \citet{grimm03-03}, normalized to 17 and 4 \sfr\ for \full and \hard keV, respectively. These normalizations were chosen such that the HMXB power law relation was coincident with the total XLF in each panel. The total \full and \hard keV XLFs, which are comprised of only spiral/dwarf galaxies, match well compared to the HMXB XLF from star-forming galaxies.}	\label{fig:galxlfs}
\end{figure*}

To investigate the characteristics of the hard X-ray source population, we plot the \full keV and \hard keV observed XLFs of detected sources in Figure \ref{fig:galxlfs}. The number of sources between panels varies because many sources are undetected in the \hard keV band. 
The total XLF in both panels, which is comprised of only spiral/dwarf galaxies, has a slope that matches the HMXB XLF from the Milky Way and star-forming galaxies. However, the dwarf galaxies, whose sources are included in the total XLFs, were selected based on hosting ULXs, and so represent a biased sample that skews the bright end of the total XLFs. 

\begin{figure*}[!ht]
\begin{tabular}{cc}
\includegraphics[width=1.0\columnwidth]{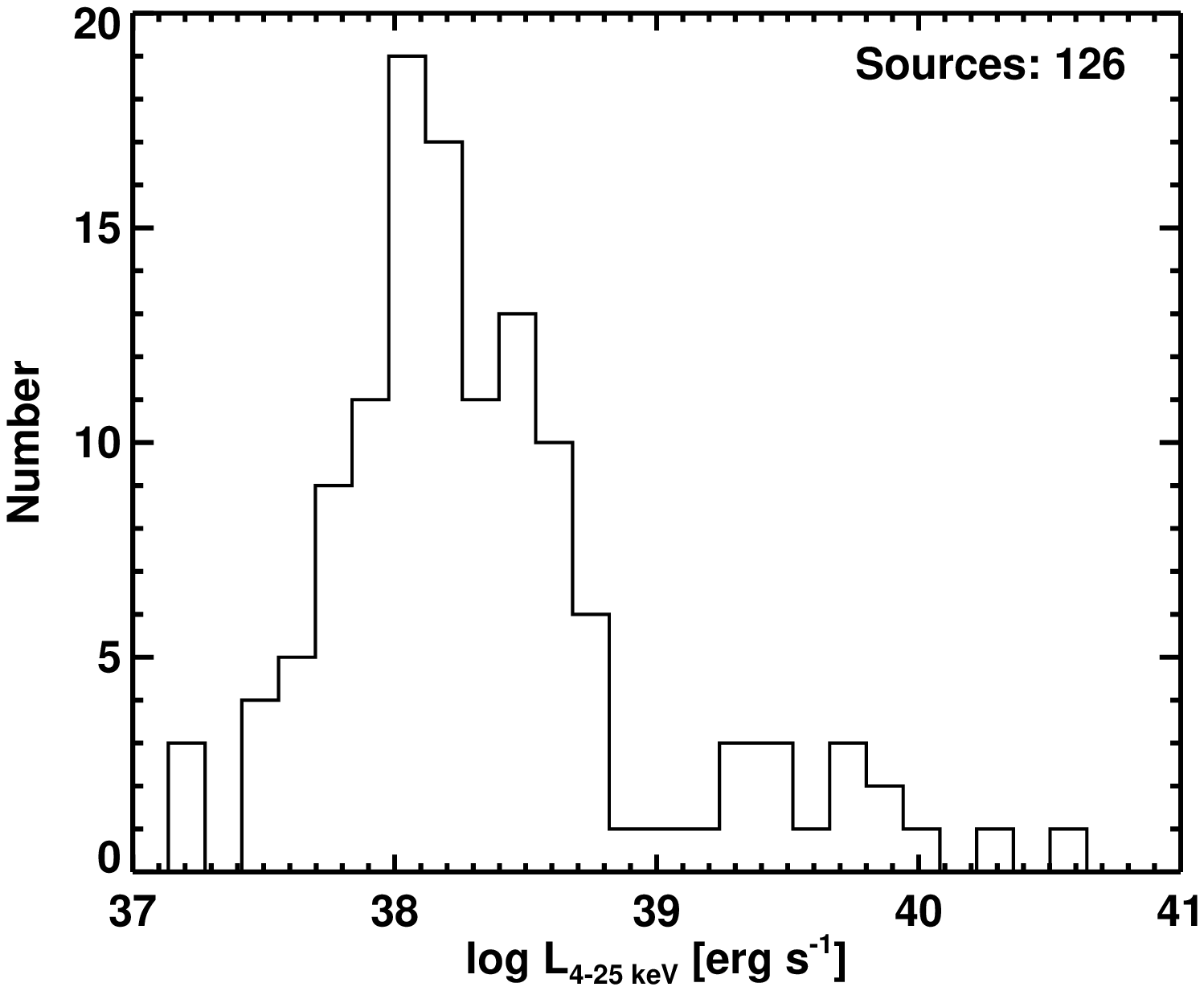}
\includegraphics[width=1.0\columnwidth]{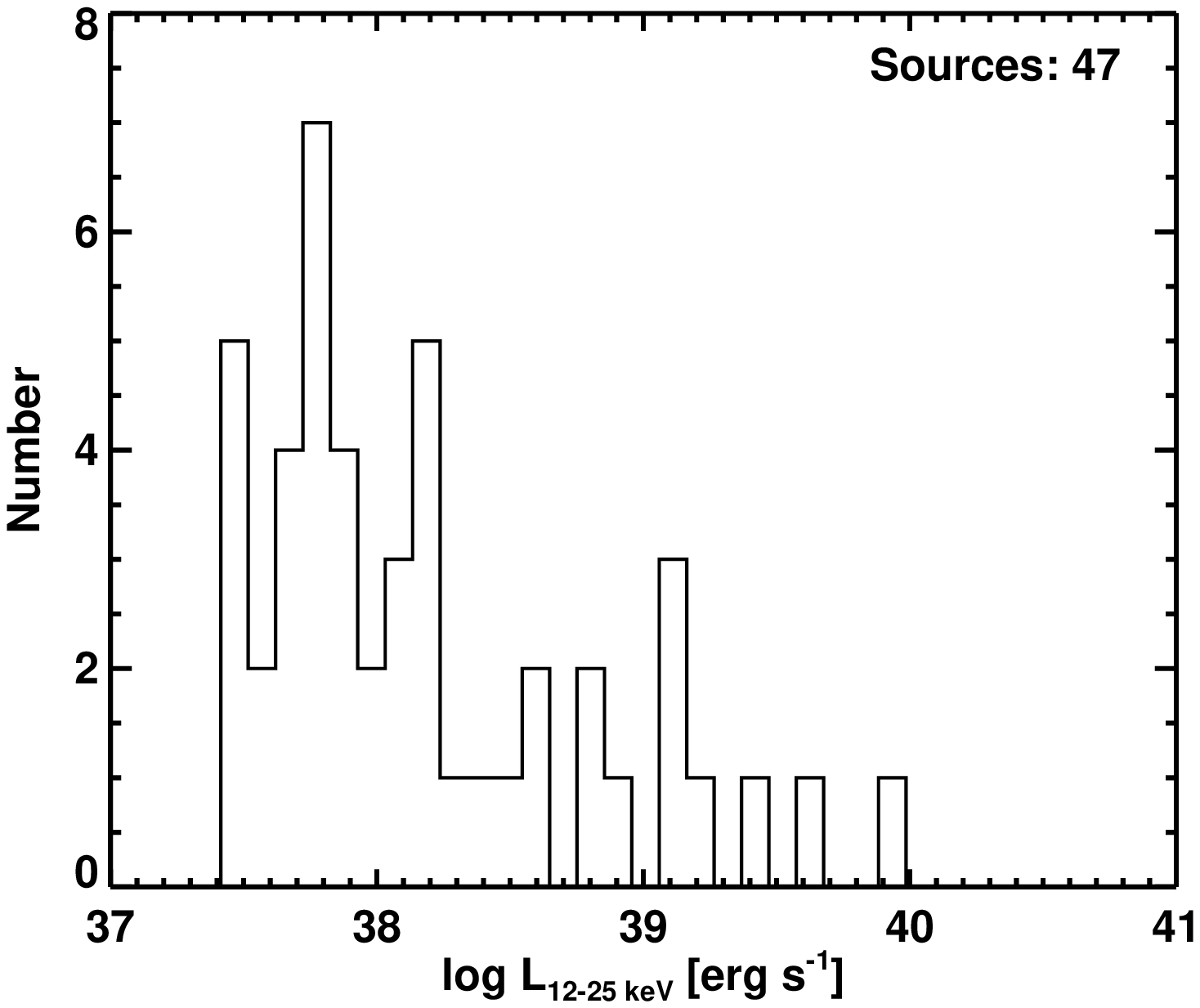}
\end{tabular}
\caption{Histograms showing the luminosity distribution of detected sources in the \full keV (left) and \hard keV (right) energy bands.}	\label{fig:lumhist}
\end{figure*}

Due to low number statistics for individual galaxies, which resulted from a combination of point source sensitivity and source confusion, the total XLF serves as a more robust sample suitable for comparison. 
The \hard keV XLF represents the intrinsic XRB luminosity for each galaxy as no extinction is expected and contribution from other source types is minimal. The shape is consistent with results from the soft $0.5-8$ keV energy band. Even though SFR ranges from $\approx0.01-15$ \sfr\ there is limited scatter in the XLFs, where variations at low-\lx\ are a result of differing sensitivities. 
\citet{mineo07-13} found a monotonic increase in the ULX rate with SFR, and therefore the high-luminosity breaks in our XLFs constrain the ULX rate of the sample galaxies. 
In Figure \ref{fig:lumhist} we show histograms of the luminosity distribution of detected sources in the \full keV (left) and \hard (right) keV energy bands. The \full keV completeness limit of our sample is apparent at $\approx10^{38}$ \es.

\subsubsection{Black Hole and Neutron Star X-ray Luminosity Functions}	\label{sec:bhnsxlfs}

As opposed to previous extragalactic XLF studies with \chandra\ and \xmmn, we were able to separate the population of XRBs into BH and NS. Identifying compact object types using, e.g., dynamical measurements or quasi-periodic oscillations, is unfeasible for galaxy populations. This simple yet powerful methodology enables us to explore hard X-ray characteristics in relation to BH/NS accretion regimes. 

\begin{figure}[!t]
\includegraphics[width=1.0\columnwidth]{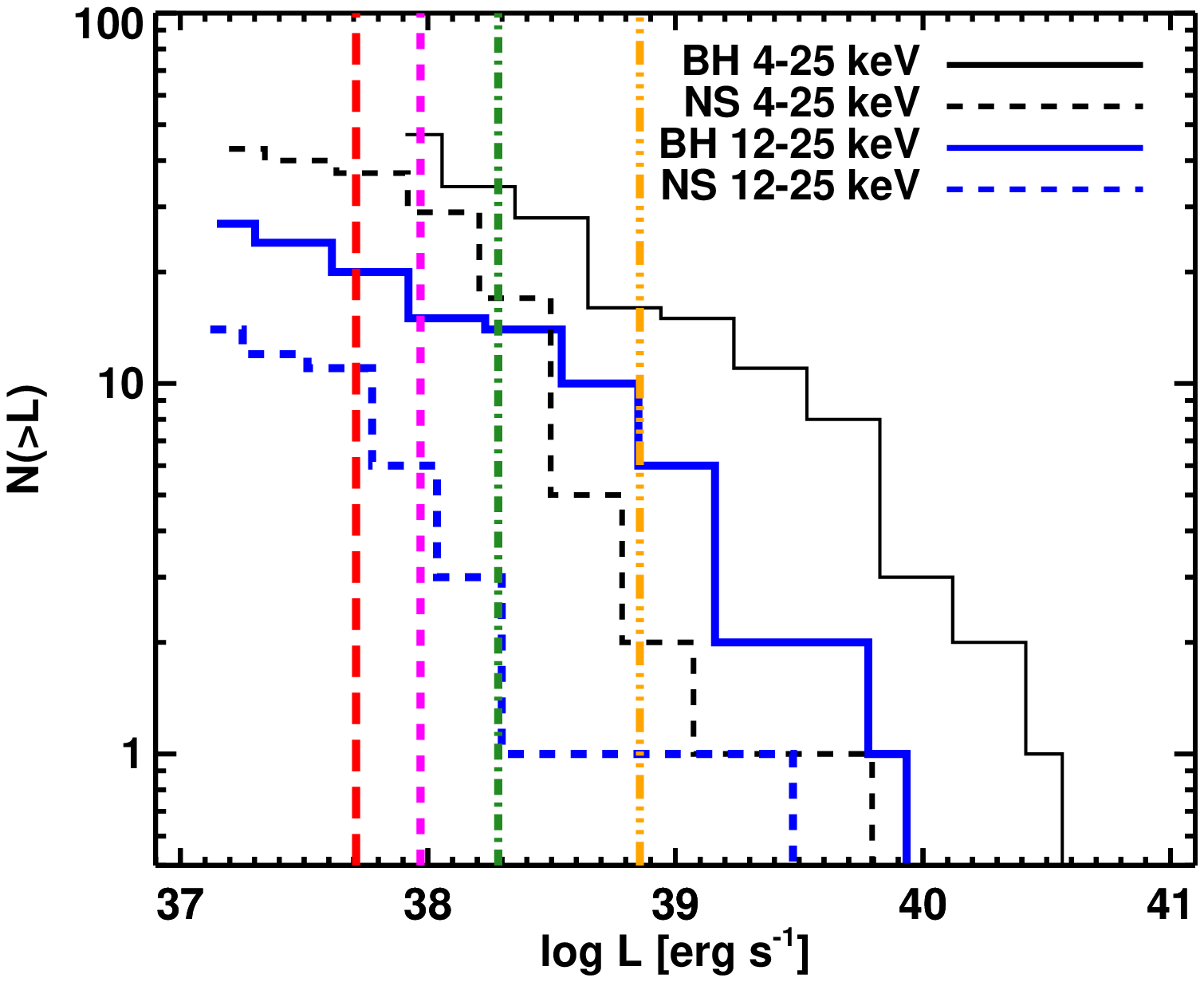}
\caption{\nustar\ XLFs for candidate BH and NS sources from the sample in the \full and \hard keV energy ranges. Vertical lines represent the Eddington limits for a 1.4 \msun\ NS and a 10 \msun\ BH as in Figure \ref{fig:bhnsratio} (see Section \ref{sec:bhnsxlfs} for details). The \hard keV NS XLF suggests a drop at $\approx6\times10^{37}$ \es, attributable to the \hard keV Eddington limit for a 1.4 \msun\ NS.}	\label{fig:xlfsratio}
\end{figure}

In Figure \ref{fig:xlfsratio} we show the NS (dashed lines) and BH (solid lines) XLFs in the \full keV (black) and \hard keV (blue) energy bands. XLFs represent all detected sources from the sample in the given energy band (including sources from galaxies whose individual XLFs were not shown in Figure \ref{fig:galxlfs}). 
Conversions from the bolometric Eddington limits were calculated for BH using the best-fit \textsc{diskbb + cutoffpl} spectrum from IC 342 X-1 \citep{rana02-15} and for NS using the best-fit \textsc{powerlaw} spectrum from M82 X-2 \citep{brightman01-16}. For the BH model, the \hard keV flux is 27\% of the \full keV flux, while for the NS model it is 55\%. The \full and \hard keV NS XLFs have similar shapes (as do the BH XLFs), which can be confirmed by applying the energy band conversion to the \full keV XLFs. Both the \full and \hard keV NS XLFs suggest a drop beginning at $\approx10^{38}$ and $\approx6\times10^{37}$ \es, respectively, attributable to their 1.4 \msun\ NS Eddington limits. Increased sensitivity and completeness will be required to confirm these declines. The \full keV BH XLF has a more gradual decline past the \full keV Eddington limit for a 10 \msun\ BH, with an abrupt drop at $\approx7\times10^{39}$ \es, the bolometric Eddington limit for a 50 \msun\ BH. Given the detection of $\sim30$ \msun\ BH by the \ligo, it is possible that these sources are stellar-mass BH accreting at the Eddington limit as opposed to super-Eddington sources \citep{abbott02-16, abbott06-17, marchant08-17, abbott10-17}. The \hard keV NS and BH XLFs are essentially cutoff at their respective bolometric Eddington limits. 
How do we interpret the distribution of BH across our luminosity range? \citet{elbert01-18} recently predicted the BH number as a function of galaxy stellar mass using an empirical approach based on the relationship between galaxy stellar mass and stellar metallicity. They estimated that an $L^{\star}$ galaxy should host millions of $\sim30$ \msun\ BH, while dwarf satellite galaxies like Draco should host $\sim100$. 
They determined that most low-mass BH of $\sim10$ \msun\ reside in massive galaxies ($M_{\star}\simeq10^{11}$ \msun) while massive BH of $\sim50$ \msun\ are typically found in dwarf galaxies ($M_{\star}\simeq10^{9}$ \msun). This result may explain the prevalence of many luminous ($\simeq10^{40}$ \es) ULXs that have been found in dwarf galaxies such as Holmberg II/IX, which contribute to the bright-end of the BH XLF.

\begin{figure}[!t]
\includegraphics[width=1.0\columnwidth]{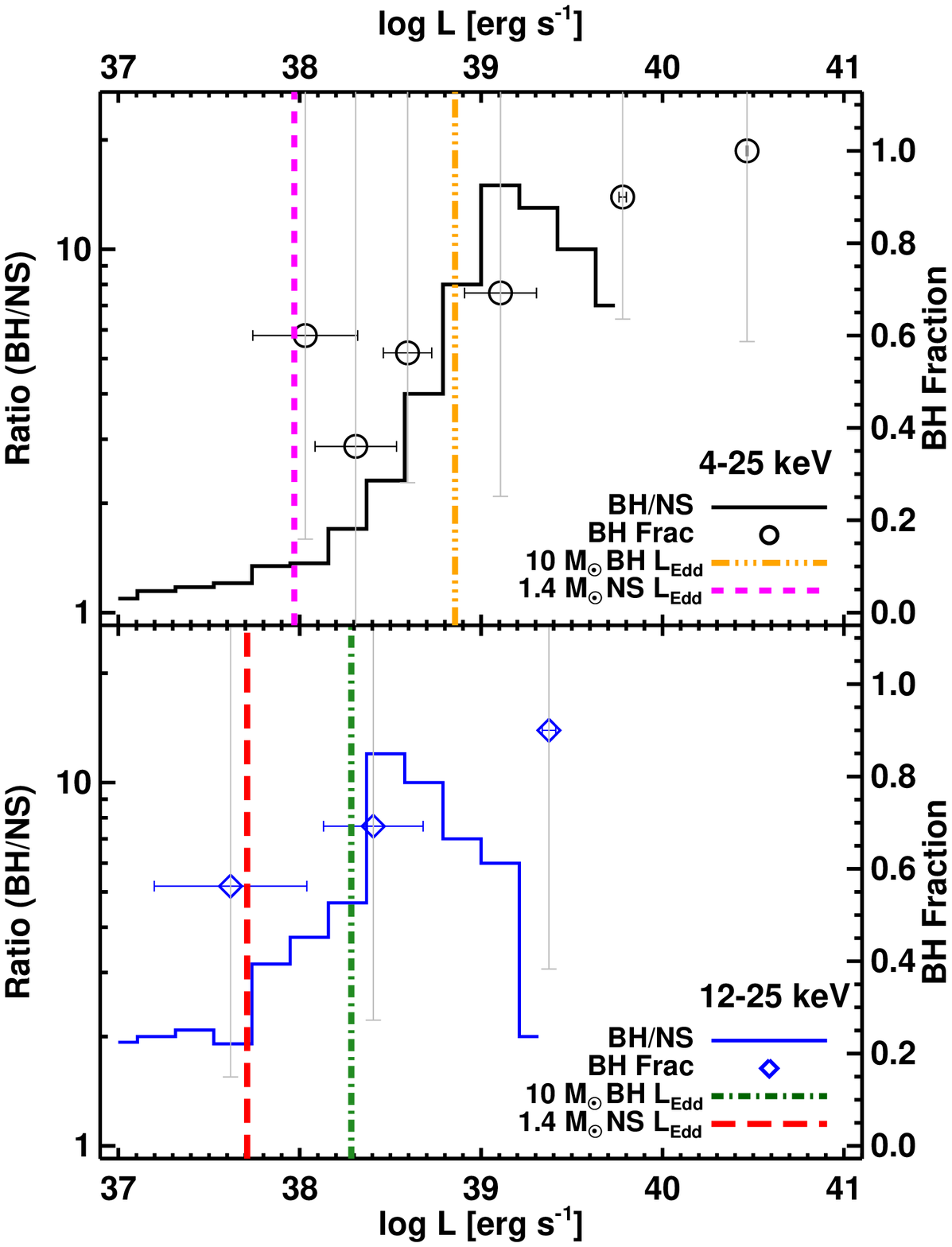}
\caption{Ratio of the cumulative number of BH to NS for the \full kev (top panel, black line) and \hard keV (bottom panel, blue line) sources. The lines are cutoff where no NS are detected/classified, even though BH are found at larger \lx. We also show the BH fraction $N_{\rm{BH}}$ / ($N_{\rm{BH}}$+$N_{\rm{NS}}$), in the \full (\hard) keV energy band as black circles (blue diamonds), each grouped to bins of 9 BH sources, except the \full keV bin at highest \lx\ that has 2 BH sources. BH fraction $1\sigma$ uncertainties were calculated using Poisson statistics \citep{gehrels04-86}. We found that the overall ratio of BH to NS was $\approx1$ for \full keV and $\approx2$ for \hard keV energy band. The \full keV BH fraction decreases at the bolometric Eddington limit for a 1.4 \msun\ NS (beyond the \full keV value), suggesting a copious NS population at this \lx\ (see Section \ref{sec:bhnsxlfs} for details).}\label{fig:bhnsratio}
\end{figure}

One of the main goals of this work was to determine the ratio of BH to NS. In Figure \ref{fig:bhnsratio} we used the cumulative XLFs to plot the ratio of BH to NS and the BH fraction\footnote{The BH fraction is not cumulative, unlike the cumulative BH/NS ratio.} $N_{\rm{BH}}$ / ($N_{\rm{BH}}$+$N_{\rm{NS}}$) in the \full keV (top panel, black) and \hard keV (bottom panel, blue) energy bands. Vertical lines are as in Figure \ref{fig:xlfsratio}. By taking the ratio of the BH and NS cumulative XLFs we were able to determine the X-ray luminosities at which each source population is prevalent relative to the other. The \full keV band has a ratio of $\approx1$ that begins to rise at the \full keV NS Eddington limit, as one would expect the number of NS to decrease. The same is true in the \hard keV energy band. It is apparent that the \full keV ratio of BH/NS peaks past the 10 \msun\ BH Eddington limit with a value of 15. However, the BH fraction rises past this point, indicating BH dominate but may decrease in total number. Whether this is a statistical fluctuation due to the small sample size or a real peak in total BH number, coincidence with the 10 \msun\ BH Eddington limit is intriguing in relation to the BH mass distribution. While the BH/NS ratio declines past this point, despite the existence of only one NS source $>10^{39}$ \es, a larger sample is needed to determine if stellar mass BH in late-type galaxies are not copious in this mass-luminosity regime. A confirmation of the utility for the \hard keV analysis are the similar ratios and shapes of both histograms.

The BH fraction enables us to study the relation between the BH and NS population at larger \lx\ compared to the BH/NS ratio. Beyond the Eddington limits for a 1.4 \msun\ NS in each energy band, we see an approximately monotonic increase in the BH fraction, as expected. The BH fraction reaches 75\% near the \full keV 10 \msun\ BH Eddington limit, quickly approaching unity at the highest \lx. The \full keV BH fraction decreases below 40\% at the bolometric Eddington limit for a 1.4 \msun\ NS, but given the lone data point to the left and uncertainties, this may not indicate a copious NS population at this \lx. To determine if NS cluster near their Eddington luminosities requires a larger sample with uniform completeness for \lx$<10^{38}$ \es.

From our previous analysis we quoted \numbh\ BH and \numns\ NS that were detected in \full keV energy band, giving a ratio of $\approx1$ across our luminosity range. We did not detect/identify many BH at faint luminosities compared with NS (\full keV), however, the opposite is true at high-\lx\ in both energy bands. A broader sample with uniform completeness will be required to eliminate ambiguities. 
The \hard keV ratio maintains a similar shape when scaling the \full keV luminosity by $\approx50$\%. When determining totals for BH and NS, we found the \hard keV band had a ratio of $\numbhh/\numnsh \approx2$, a factor of two larger than the \full keV band, suggesting that BH XRBs are harder than NS XRBs. When comparing relative percentages to the \full keV band, a lack of detected NS sources in the \hard keV band is apparent, possibly due to the softer spectra of Z/Atolls NS. Whether this ratio of unity holds down to the lower limit for actively accreting (`luminous') XRBs at $\approx10^{36}$ \es\ and into the quiescent XRB range will require deeper data.

\section{Conclusions and Future Directions}	\label{sec:con}

Using a \nustar-selected sample of 12 late-type and dwarf galaxies, we investigated the \full keV properties of the XRB population. With novel diagnostic methods that leverage the E $>10$ keV energy band, we were able to distinguish between compact object types and accretion states via hardness-intensity and color-color diagrams. Specifically, we were able to classify \numclass/\numsrc\ sources in the \full keV energy band: \numbh\ BH and \numns\ NS. This is a marked improvement from previous studies where identifying extragalactic XRB compact object types and spectral states has only been possible for a select few of the brightest systems.

We studied the relationship between BH and NS and the sSFR of a galaxy. A Spearman's Rank test on the BH fraction, $N_{\rm{BH}}$ / ($N_{\rm{BH}}+N_{\rm{NS}}$), versus sSFR gave a $p$-value of 0.072 and coefficient $r_{s}=0.56$, indicating weak monotonicity but no correlation. Including M31, which is dominated by NS and has low-sSFR, we obtained a $p$-value of 0.028 and coefficient $r_{s}=0.63$. The data suggests that BH dominate star-forming galaxies and NS dominate low-sSFR galaxies. However, due to the varying sensitivity and completeness of our sample, we require improved statistics to investigate this further. Similarly, while there were indications from the data, in agreement with theoretical expectations, that accreting pulsars dominate at high-sSFR and Z/Atoll sources were prevalent at low-sSFR, no statistically significant correlation was found. We found that most BH were identified with the hard accretion state, regardless of host galaxy sSFR, similar to the Galactic BH-HMXB Cygnus X-1. Subsequent analysis of the rich multiwavelength data sets using UV/optical/IR catalogs (in combination with the \nustar-\chandra/\xmmn\ data we analyzed) can help confirm the nature of these sources. 

We classified galaxies as BH, NS, and ULX-dominated if $>70$\% of their total \full or \hard keV X-ray point source emission came from one of these groups. We found that galaxies with sSFR $\gtrsim2\times10^{-9}$ yr$^{-1}$ were all ULX-dominated, which included all four dwarf galaxies in our sample as well as the starburst galaxies IC 342 and M82. Most galaxies were BH-dominated, whereas in the \full keV energy band, only M31 and NGC 4945 were NS-dominated. We confirmed the \lx-SFR correlation from previous studies by investigating the 8 normal (Milky Way-type) galaxies in the \nustar\ sample with SFR $0.3-12.5$ \sfr. The best-fitting parameters for the relation $\log L_{\text{X}}=\log A+B\log \text{SFR}$ can be found in Table \ref{tab:paramlxsfr}. The \full keV result was nearly identical to previous work in the $2-10$ keV range despite the use of different galaxy samples.

We constrained the correlation of X-ray luminosity with SFR and stellar mass using the relationship $L_{\rm{X}}=\alpha M_{\star}+\beta\text{SFR}$. We determined the best-fitting values (see Table \ref{tab:params}) for $\alpha$ and $\beta$ based on the 8 normal (Milky Way-type) galaxies in the \nustar\ sample. In particular, the four dwarf galaxies had increased \lx/SFR compared to normal galaxies, based on past scaling relations. This is not surprising as these dwarf galaxies were selected as ULX hosts, and as such are a biased sample. Studying an unbiased sample of dwarf galaxies would help determine a statistically significant \lx/SFR relation. With the introduction of new programs/observatories to identify faint dwarf galaxies in the optical (e.g.\ Dragonfly, \citealt{abraham01-14}; Dark Energy Survey, \citealt{the-des-collaboration08-16}), multiwavelength studies incorporating X-ray emission would improve our constraints on X-ray source populations in the low-mass regime.

We measured galaxy XLFs in the \full and \hard keV energy bands, including the first E $>10$ keV extragalactic XLF for an ensemble of galaxies. We determined that the combined XLF of all sample galaxies in each energy band followed that of the canonical HMXB XLF found by previous studies at E $<10$ keV. Using the classifications for BH and NS, we produced cumulative BH and NS XLFs in the \full and \hard keV energy bands. The \full and \hard keV NS XLFs each indicated a decline beginning at $\approx10^{38}$ and $\approx6\times10^{37}$ \es, respectively, attributable to the 1.4 \msun\ NS Eddington limit. Increased sensitivity and completeness in the \hard keV energy band is required to confirm the decline seen in the \hard keV NS XLF.

Using our classifications we investigated the characteristics of BH and NS at different \lx, with a focus on behavior near the Eddington limits. We calculated the overall BH to NS ratio, finding $N_{\rm{BH}}/N_{\rm{NS}}\approx1$ (\full keV) and $N_{\rm{BH}}/N_{\rm{NS}}\approx2$ (\hard keV), over a \full and \hard keV luminosity range for all detected sources of $\sim10^{37}-10^{40.5}$ and $\sim10^{37}-10^{40}$ \es, respectively. We found that the \full keV ratio of BH to NS increased from a value of 1 past the \full keV Eddington limit for a 1.4 \msun\ NS and reached a maximum value of 15 near the \full keV Eddington limit for a 10 \msun\ BH. However, while the total number of accreting stellar-mass BH may decrease beyond the 10 \msun\ BH $L_{\rm{Edd}}$, an improved statistical sample is required to determine its validity. To extend to larger \lx\ we investigated the BH fraction, $N_{\rm{BH}}$ / ($N_{\rm{BH}}$+$N_{\rm{NS}}$), finding approximately monotonic increase beyond the Eddington limits for a 1.4 \msun\ NS in both energy bands. We found evidence for a decrease in the BH fraction below 40\% beyond the \full keV Eddington limit for a 1.4 \msun\ NS (the data point was coincident with the bolometric Eddington limit for a 1.4 \msun\ NS). A larger sample with uniform completeness for \lx$<10^{38}$ \es\ is required to determine if NS cluster near their Eddington luminosities. 

This galaxy sample was biased towards late-type/spiral galaxies and contained no early-type galaxies, meaning that we did not offer a comprehensive view of older stellar populations. Future observations that focus on building a sample of elliptical galaxies would be of great interest. Such galaxies have inherently faint LMXB populations and thus require extended observing campaigns. However, the nearest candidate, Cen A, is problematic due to its AGN, and most giant elliptical galaxies are located at $d\gtrsim10$ Mpc, prohibiting resolved XRB studies with \nustar. Therefore, to improve our understanding of the XRB population in elliptical galaxies at E $>10$ keV requires a next generation hard X-ray telescope. Using our XRB classifications and XLFs enables comparison with binary population synthesis modeling \citep{fragos09-09, sorensen01-17} that predicts the NS and BH XRB populations in these galaxies. Expanding the range of sSFR coupled with increased sensitivity has the potential to profoundly impact the study of accreting compact objects.

\acknowledgements

We thank the referee for useful comments that improved the manuscript. VA acknowledges financial support from NASA grant NNX15AR30G. This research made use of \textsc{astropy}, a community-developed core \textsc{Python} package for Astronomy \citep{astropy-collaboration10-13}. We acknowledge the usage of the HyperLeda database (http://leda.univ-lyon1.fr). This research has made use of the NASA/IPAC Extragalactic Database (NED) which is operated by the Jet Propulsion Laboratory, California Institute of Technology, under contract with the National Aeronautics and Space Administration.
\facilities{\nustar, \emph{CXO} (ACIS), \emph{XMM} (PN, MOS)}
\software{astropy \citep{astropy-collaboration10-13}}

\bibliographystyle{aasjournal62}

\appendix

\section{Notes on Individual Galaxies}\label{sec:sampsum}

\subsection*{IC 342}

IC 342 is a nearly face-on spiral galaxy with intense star formation activity in its core \citep{becklin03-80}. A member of the IC 342/Maffei group of galaxies, it is located near the Galactic plane at $b\sim10\degr$, making it difficult to constrain X-ray emission $\lesssim1$ keV. IC 342 has been studied by all major X-ray observatories over the past four decades. 
\chandra\ high-resolution camera imaging (HRC-I) of the central 30\arcmin\ by 30\arcmin\ region by \citet{mak10-08} resulted in the detection of 23 sources. 
They found that one of the three historical ULXs detected by \einstein\ was actually coincident with the nuclear center and thus not a ULX. Multiple ACIS-S observations have been used to study the spectrum of ULX IC 342 X-1 and to create a point source catalog consisting of 61 sources \citep{liu01-11}. \citet{mak02-11} studied the long-term flux and spectral variability of a 35\arcmin\ by 30\arcmin\ region with \xmmn\ to a limiting luminosity of $10^{37}$ \es; 39 of the 61 detected sources showed long-term variability, 26 of which were classified as X-ray transients. Of the identified transients, 8 also showed spectral variability indicative of XRBs. \citet{rana02-15} recently used two epochs of \nustar/\xmmn\ observations to investigate the two ULXs in IC 342. They found luminosities of $\sim10^{40}$ and $7.4\times10^{39}$ \es\ for sources X-1 and X-2, respectively. Joint spectral fitting of each source ruled out the possibility of a BH binary in a low/hard accretion state. They concluded that further broadband spectral studies are needed to identify the origin of the spectral components.

\subsection*{M83}

M83 (NGC 5236) is a nearly face-on grand design spiral galaxy with a SFR of $\sim3$ \sfr. These characteristics have made it an ideal object for X-ray population studies. A deep (790 ks) \chandra\ ACIS survey of M83 by \citet{long06-14} detected 378 point sources within the D$_{25}$ ellipse and reached luminosities of $\sim10^{36}$ \es. Based on multiwavelength data they identified 87 supernova remnant (SNR) candidates, which dominated the population in the disk. \citet{long06-14} classified X-ray point sources using spectral and temporal analysis. Spectral fitting of the 29 brightest ($>2000$ counts) sources showed that most SNRs were associated with the spiral arms while the harder sources (likely XRBs) were not. Analysis of the cleaned XLF (foreground sources, AGN, and SNRs removed) indicated that most of the XRBs in the disk are LMXBs as opposed to HMXBs. 
The recent \nustar/\xmmn/\chandra\ survey by \citet{yukita06-16} detected 21 point sources and found that the hard X-ray emission E $>10$ keV was dominated by intermediate accretion state BH XRBs and NS LMXBs.

\subsection*{M82}

M82 (NGC 3034) is an example of an extreme starbursting galaxy with SFR of 12 \sfr. Part of the M81 group of galaxies, the starburst is likely a result of interaction with M81. 
Until the launch of \chandra, the discrete X-ray point source population of M82 was not well-studied due to the large number density of sources in the nucleus and the presence of X-ray emission from hot gas from the starburst. To date, most studies still focus on the brightest few point sources as opposed to the population.
M82 has more luminous XRBs (i.e. a flatter XLF; \citealt{kilgard07-02}) and its luminous source population appears to be HMXBs associated with young star clusters \citep{zezas00-04} exhibiting variability and spectral shapes consistent with BH HMXBs \citep{kilgard07, chiang06-11}. 
M82 has been particularly well-studied in X-rays due to the large population of 6 ULXs \citep{gladstone06-13}, the brightest of which reside in the nucleus, X-1 and X-2. M82 X-1 was long thought to be an intermediate-mass BH due to its super-Eddington luminosity, an idea that has recently been confirmed. \citet{pasham09-14} used \rxtet\ data to measure the quasi-periodic oscillations of the source and estimated the mass of the BH to be $429\pm105$ \msun. This important discovery bridged the divide between stellar mass BH in XRBs and supermassive BH at the centers of most galaxies. 
At nearly the same time as this discovery concerning M82 X-1, the ULX M82 X-2, which was long thought to be an intermediate-mass BH candidate, was discovered to be the first-ever confirmed ULX {\it pulsar}.
\citet{bachetti10-14} discovered pulsed emission spatially coincident with M82 X-2 using \nustar/\chandra/\swift\ data, confirming X-2 as a NS. M82 X-2 reaches 100 times the Eddington limit for a NS, with a peak luminosity \lx\ ($0.3-10$ keV) of $1.8\times10^{40}$ \es. This result challenged the theory of accretion on to magnetized NS and has led to studies on what fraction of ULXs are NS rather than BH \citep[e.g.][]{fragos03-15, shao04-15, wiktorowicz09-15, king05-16}. 

\subsection*{NGC 253}

NGC 253 is also an edge-on starburst galaxy with a similarly high SFR of $6$ \sfr. Although distance estimates vary, a census of ULXs in the nearby ($<5$ Mpc) Universe by \citet{gladstone06-13} reported that 8 ULXs reside in NGC 253. 
\citet{vogler02-99} used \rosat\ to detect 32 sources within the D$_{25}$ ellipse of NGC 253 to a luminosity of $7\times10^{36}$ \es.
They showed that most sources exhibit time variability and are likely XRBs. \citet{pietsch01-01} studied the 10 brightest sources in NGC 253 using \xmmn\ data and classified them using $0.5-10$ keV color diagnostics. They identified 3 sources with soft colors as likely LMXBs and noted most sources showed time variability, also indicative of LMXBs. The recent \nustar/\chandra\ study by \citet{lehmer07-13} determined that the ULX source dominating the entire galaxy over $0.5-30$ keV is distinct from the nuclear SMBH, which apparently was actively accreting a decade prior and had turned off in the 2012 observations. 
A comprehensive study of the \nustar\ point sources by \citet{wik12-14} detected 21 sources (\full keV) and found that most were BH XRBs in an intermediate accretion state.

\subsection*{M81}

M81 (NGC 3031) is a nearby grand design spiral galaxy with low SFR and a low-ionization nuclear emission region (LINER). The X-ray population of M81 has been studied by \einstein\ \citep{fabbiano02-88}, \rosat\ \citep{immler06-01}, and \chandra\ \citep{tennant03-01,swartz02-03}. The most comprehensive analysis was completed by \citet{sell07-11}, who used 220 ks of \chandra\ data from 16 observations to classify and investigate the variability of 265 sources detected above $\sim10^{37}$ \es. They found significant variability in $\sim36-60$\% of their sources but concluded that snapshot observations provided a consistent determination of the XLF of M81. Color diagnostics identified large numbers of many different source types such as LMXBs, HMXBs, and SNRs.

\subsection*{M31}

M31 is both the nearest large spiral galaxy and most similar to the Milky Way, and thus has been the target of the most detailed studies of any extragalactic X-ray point source population \citep[e.g.][]{trinchieri11-91,supper07-01,peacock10-10,barnard01-14,henze03-14}. Previous \chandra\ observations have mostly been focused on monitoring the bulge region for both the activity of the SMBH as well as XRB variability, until the recent  2015 \chandra\ Large Project to cover the Panchromatic Hubble Andromeda Treasury \citep[PHAT;][]{dalcanton06-12} survey area (PI: B. F. Williams). 
Notable results included the finding that the XLF of the bulge was flatter than the disk \citep{kong10-02, williams07-04, vulic10-16}, at odds with studies of other spiral galaxies \citep{colbert02-04, binder10-12}. \citet{stiele10-11} used \xmmn\ data covering an area greater than the D$_{25}$ ellipse to compile a catalog of 1897 sources above $\sim10^{35}$ \es. Source classification techniques included using X-ray hardness ratios, spatial extent of the sources, long-term X-ray variability, and cross-correlation with X-ray, optical, infrared, and radio catalogs. Despite their robust analysis, 65\% of their sources remained unclassified. 
This included only having 2 HMXB candidates to add to the 18 candidates found by \citet{shaw-greening03-09}, which was unusual given that there are $\sim100$ HMXBs in the Milky Way. This prompted a \chandra\ legacy survey of the M31 disk (PI: B. F. Williams) to complement the PHAT survey \citep{dalcanton06-12}. A \nustar\ legacy project was begun in early 2014 to cover part of the Chandra-PHAT area (P.I. A. E. Hornschemeier) and \nustar\ GO observations of the bulge have occurred in Cycles $1-3$ (P.I. M. Yukita). Details about the M31 source population are given in papers on the globular cluster LMXB population \citep{maccarone06-16}, which comprises most of the bright X-ray sources in M31, and the bright pulsar candidate dominating the entire galaxy at E $>25$ keV \citep{yukita03-17}.

\subsection*{NGC 5204}

NGC 5204 is a Magellanic spiral galaxy that is part of the M101 group of galaxies and has a large sSFR comparable to that of M82. The ULX X-1 originally discovered by \einstein\ \citep{fabbiano-82} has been the motivation for most X-ray/multiwavelength observations of NGC 5204. \citet{roberts10-06} used a 2 month \chandra\ monitoring campaign to study the variability of X-1 and found that its spectrum became harder (e.\ g.\ heating of the corona) as its flux increased. They found no evidence supporting the presence of an intermediate-mass BH in X-1. \citet{mukherjee07-15} used two epochs of \nustar/\xmmn\ coverage of X-1 to study its $0.3-20$ keV properties. No significant spectral variations were observed for the $5\times10^{39}$ \es\ source and the broadband spectrum was consistent with super-Eddington accretion on to a stellar-mass BH. 

\subsection*{NGC 1313}

NGC 1313 is an isolated peculiar spiral galaxy with starburst activity and a similar sSFR to NGC 5204.
It has a specific frequency of young massive star clusters similar to that of M83 \citep{larsen05-99} and a concentration of stars at a stellar age of $\approx200$ Myr. \citet{trudolyubov06-08} discovered a transient X-ray pulsar in \xmmn\ data with a period of 766 s that reached \lx\  ($0.3-7$ keV) $\approx2\times10^{39}$ \es. Based on the X-ray properties it was classified as a \be\ X-ray pulsar candidate. \citet{bachetti12-13} recently investigated the $0.3-30$ keV spectra of the ULXs X-1 and X-2 with \nustar/\xmmn\ observations. While X-2 was not detected for E $>10$ keV, X-1 showed a clear spectral cutoff that ruled out a BH in a low/hard accretion state. The characteristics of a large spectral variation found in X-2 was indicative of a BH in the hard state.

\subsection*{NGC 4945}

NGC 4945 is an edge-on barred spiral starburst galaxy with a Type II Seyfert nucleus. The galaxy is the brightest extragalactic hard X-ray source ($\sim50-100$ keV) and hosts one of the nearest AGN. The nuclear region contains an obscured starburst region with a 10\arcsec\ ring morphology \citep{schurch09-02}. Most X-ray investigations have focused on the AGN, although multiple \chandra\ studies \citep{colbert02-04, kaaret08-08, liu01-11} have detected up to $\approx50$ X-ray sources in NGC 4945 to sensitivities of $\sim10^{37}$ \es. In particular, \chandra\ \citep{swartz10-04} and \xmmn\ \citep{berghea11-08} studies found 2 ULXs that were used as part of a review of nearby galaxy ULX populations. \citet{colbert02-04} calculated a cumulative XLF slope of $\gamma=0.7$ using 22 X-ray point sources, finding agreement with other spiral galaxies in their sample. SN 2011ja occurred in NGC 4945 and was studied by \citet{chakraborti09-13} using \chandra. X-ray observations allowed the authors to probe the history of variable mass loss from the progenitor, suggesting that SN may interact with circumstellar material ejected by non-steady winds (varying densities). \citet{puccetti09-14} used \nustar\ observations in combination with other archival X-ray data of NGC 4945 to investigate the spectral properties and variability of the galaxy. They found strong spectral variability above E $>10$ keV and that most of the high-energy flux was transmitted rather than Compton-scattered.

\subsection*{Circinus}

Circinus is a spiral galaxy with similar sSFR to the Milky Way but located 4\degr\ below the Galactic plane, thus having a large \nh. Circinus is an active galaxy with a Type II Seyfert nucleus and complex structure. It has been observed many times by various X-ray observatories \citep[e.g.][]{smith08-01, bianchi12-02, yang01-09}. 
\citet{bauer07-01} completed the first point source population study with \chandra, detecting 16 point sources to a $0.5-10$ keV sensitivity of $10^{37}$ \es, with $25\%$ of the sources being variable. \citet{walton12-13} studied Circinus ULX5 (there are 4 other known/candidate ULXs in Circinus, e.g.\ \citealt{swartz10-04, ptak09-06}), a variable source in the outskirts of the galaxy beyond D$_{25}$, using coordinated \nustar-\xmmn\ observations and archival X-ray data from other observatories. They determined a $0.3-30$ keV luminosity of $1.6\times10^{40}$ \es\ and BH mass of 90 \msun, found variability on long time scales of at least a factor of $\sim5$, and spectral variability similar to luminous Galactic BH XRBs. \citet{esposito09-15} used archival \chandra-\xmmn\ observations and discovered 2 pulsators that were identified as likely foreground cataclysmic variables. The ULX candidate CG X-1 had properties consistent with a Wolf-Rayet BH XRB, the rare class of sources for which only 4 confirmed and 3 candidates exist.

\subsection*{Holmberg II}

Holmberg II is a dwarf irregular galaxy that is part of the M81 group and has properties very similar to the Small Magellanic Cloud. 
\citet{kerp06-02} detected 31 X-ray sources located within the H {\sc i} column density distribution of Holmberg II to a sensitivity of $10^{37}$ \es\ using \rosat\ PSPC data. Many studies of Holmberg II have focused on the unique ULX source Holmberg II X-1, located inside the ``Foot nebula'', from radio \citep{miller04-05} to the optical \citep{abolmasov03-07} and IR \citep{berghea01-102, heida06-16}. Many interpretations for the nature of this ULX have been put forth, although there has been general consensus for a $\approx100$ \msun\ BH \citep[e.g.][]{goad01-06, berghea01-10}. \citet{walton06-15} studied the $0.3-25$ keV emission from the ULX Holmberg II X-1 with \nustar-\xmmn-\suzaku\ observations, finding \lx$=8.1\times10^{39}$ \es, which is typical for this source. They implied that the source was accreting near or above its Eddington limit and found 90\% of the flux was emitted at E $<10$ keV. \citet{egorov05-17} analyzed the structure and kinematics of ionized gas around X-1 using optical emission lines, finding evidence that the ULX may have escaped its parent star cluster.

\subsection*{Holmberg IX}

Holmberg IX is a dwarf irregular galaxy that is also part of the M81 group, located near the outskirts of M81's D$_{25}$ ellipse. It is the nearest young galaxy, having stellar populations with ages $\lesssim200$ Myr and dominated by blue and red supergiants. Thought to be formed by the recent tidal interaction between M81 and another M81 group galaxy, Holmberg IX hosts one of the best-studied ULXs (Ho IX X-1, also known as M81 X-9), which is persistently detected at \lx\ ($0.3-10$ keV) $>10^{40}$ \es\ \citep[e.g.][]{walton09-14}. First discovered by the \einstein\ observatory \citep{fabbiano02-88}, X-1 has been well studied by all X-ray observatories \citep[e.g.][]{la-parola07-01, walton09-14}. \xmmn\ observations initially revealed a cool accretion disk ($kT\sim0.1-0.2$ keV), suggesting an intermediate-mass BH \citep[e.g.][]{miller06-04, miller10-04}. However, recent work has indicated a spectrum consistent with a 100 \msun\ BH accreting at the Eddington limit or a 10 \msun\ BH above $L_{\rm{Edd}}$ \citep{kong10-10}. \nustar\ was critical in confirming the spectral cutoff and disfavoring an intermediate-mass BH \citep{walton09-14, walton04-17}.

\begin{longrotatetable}
\begin{deluxetable*}{c c  c  c  c  c  c  c  c  c  c  c  c  c  c  c  c  c  c  c  c  c  c  c  c  c  c}
{\setlength\tabcolsep{4pt}
\tablecaption{\nustar\ Point Source Properties\label{tab:rates}}
\tabletypesize{\tiny}
\tablehead{
\colhead{Galaxy} & \colhead{ID}  & \colhead{R.A.} & \colhead{Decl.} &	\multicolumn{12}{c}{\nustar\ Count Rates}	&	\multicolumn{6}{c}{\nustar\ colors}	&	\multicolumn{3}{c}{\nustar\ \lx}	&	&	\\
	&	&	&	& \colhead{($4-6$ keV)}	&	\colhead{$\sigma_{up}$}	&	\colhead{$\sigma_{down}$}	& \colhead{($6-12$ keV)}		&	\colhead{$\sigma_{up}$}	&	\colhead{$\sigma_{down}$}	& \colhead{(\hard keV)}	&	\colhead{$\sigma_{up}$}	&	\colhead{$\sigma_{down}$}	& \colhead{(\full keV)}	&	\colhead{$\sigma_{up}$}	&	\colhead{$\sigma_{down}$}	& \colhead{HR1}	&	\colhead{$\sigma_{up}$}	&	\colhead{$\sigma_{down}$}	& \colhead{HR2}	&	\colhead{$\sigma_{up}$}	&	\colhead{$\sigma_{down}$}	& \colhead{(\full keV)}	&	\colhead{$\sigma_{up}$}	&	\colhead{$\sigma_{down}$}	& \colhead{Type}	& \colhead{State}	\\
\cmidrule(lr){5-16}	\cmidrule(lr){17-22}	\cmidrule(lr){23-25}	&	&	\multicolumn{2}{c}{(J2000.0)}	&	\multicolumn{12}{c}{($10^{-4}$ counts s$^{-1}$)}	&	&	&	&	&	&	&	\multicolumn{3}{c}{(10$^{38}$ \es)}	&	&
}
\startdata
        Circinus & 1 & 213.162556 & -65.392239 & 445.43 & 4.68 & 4.57 & 401.37 & 4.58 & 4.45 & 39.48 & 1.92 & 1.85 & 1078.08 & 8.11 & 8.11 & -0.04 & 0.01 & 0.01 & -0.80 & 0.01 & 0.01 & 113.98 & 0.86 & 0.86 & BH & ULX
 \\ Circinus & 2 & 213.291263 & -65.345541 & 114.42 & 9.28 & 9.14 & 324.15 & 14.37 & 14.14 & 116.70 & 13.62 & 13.51 & 589.66 & 24.31 & 24.22 & 0.53 & 0.04 & 0.04 & -0.49 & 0.05 & 0.05 & 62.34 & 2.57 & 2.56 & NS & AP
 \\ Circinus & 3 & 213.079532 & -65.433248 & 4.78 & 1.58 & 1.36 & 9.38 & 1.93 & 1.73 & 4.78 & 1.82 & 1.56 & 16.18 & 2.58 & 2.41 & 0.35 & 0.17 & 0.16 & -0.42 & 0.18 & 0.21 & 1.71 & 0.27 & 0.25 & NS & AP
 \\ Circinus & 4 & 213.253387 & -65.429618 & 4.89 & 1.49 & 1.30 & 4.15 & 1.69 & 1.47 & 1.91 & 0.00 & 0.00 & 11.92 & 2.39 & 1.97 & -0.14 & 0.17 & 0.18 & -0.56 & 0.00 & 0.00 & 1.26 & 0.25 & 0.21 & BH & -
 \\ IC342 & 1 & 56.479998 & 68.081921 & 365.84 & 3.63 & 3.62 & 458.99 & 4.05 & 4.03 & 87.25 & 1.97 & 1.94 & 1065.63 & 6.82 & 6.85 & 0.11 & 0.01 & 0.01 & -0.68 & 0.01 & 0.01 & 73.40 & 0.47 & 0.47 & BH & ULX
 \\ IC342 & 2 & 56.564447 & 68.186753 & 258.74 & 3.54 & 3.51 & 401.89 & 4.43 & 4.32 & 100.29 & 2.70 & 2.66 & 883.17 & 7.23 & 7.13 & 0.21 & 0.01 & 0.01 & -0.59 & 0.01 & 0.01 & 60.83 & 0.50 & 0.49 & BH & ULX
 \\ IC342 & 3 & 56.416343 & 68.052534 & 23.07 & 1.23 & 1.21 & 21.81 & 1.32 & 1.25 & 3.50 & 0.89 & 0.79 & 56.64 & 2.28 & 2.19 & -0.03 & 0.04 & 0.04 & -0.71 & 0.06 & 0.06 & 3.90 & 0.16 & 0.15 & BH & I
 \\ IC342 & 4 & 56.738231 & 68.105118 & 23.86 & 1.65 & 1.52 & 14.15 & 1.50 & 1.42 & 0.84 & 0.00 & 0.00 & 46.01 & 2.66 & 2.53 & -0.27 & 0.05 & 0.05 & -0.89 & 0.00 & 0.00 & 3.17 & 0.18 & 0.17 & BH & -
 \\ IC342 & 5 & 56.683009 & 68.102598 & 18.58 & 1.51 & 1.42 & 16.73 & 1.53 & 1.43 & 1.00 & 0.00 & 0.00 & 44.89 & 2.74 & 2.62 & -0.08 & 0.06 & 0.06 & -0.78 & 0.00 & 0.00 & 3.09 & 0.19 & 0.18 & - & -
 \\ IC342 & 6 & 56.498453 & 68.093783 & 17.59 & 1.92 & 1.84 & 16.24 & 1.95 & 1.90 & 3.24 & 1.04 & 0.99 & 41.58 & 3.46 & 3.40 & -0.03 & 0.09 & 0.08 & -0.69 & 0.10 & 0.10 & 2.86 & 0.24 & 0.23 & BH & I
 \\ IC342 & 7 & 56.701628 & 68.096126 & 12.15 & 1.76 & 1.65 & 16.29 & 1.86 & 1.76 & 4.11 & 1.22 & 1.10 & 38.21 & 3.29 & 3.12 & 0.15 & 0.09 & 0.09 & -0.60 & 0.10 & 0.10 & 2.63 & 0.23 & 0.21 & NS & ZA
 \\ IC342 & 8 & 56.197127 & 68.144554 & 8.20 & 1.56 & 1.42 & 8.34 & 1.79 & 1.65 & 1.74 & 0.00 & 0.00 & 30.79 & 2.82 & 2.67 & 0.03 & 0.09 & 0.09 & -0.51 & 0.14 & 0.16 & 2.12 & 0.19 & 0.18 & BH & I
 \\ IC342 & 9 & 56.482757 & 67.990453 & 5.09 & 1.83 & 1.57 & 10.70 & 2.45 & 2.20 & 2.40 & 0.00 & 0.00 & 25.66 & 3.52 & 3.30 & 0.29 & 0.13 & 0.13 & -0.33 & 0.00 & 0.00 & 1.77 & 0.24 & 0.23 & - & -
 \\ IC342 & 10 & 56.527707 & 68.174742 & 8.36 & 1.34 & 1.28 & 10.79 & 1.54 & 1.45 & 4.75 & 1.19 & 1.09 & 22.48 & 2.53 & 2.39 & 0.39 & 0.14 & 0.13 & -0.30 & 0.12 & 0.11 & 1.55 & 0.17 & 0.16 & NS & AP
 \\ IC342 & 11 & 56.715792 & 68.146976 & 2.97 & 1.00 & 0.89 & 6.97 & 1.24 & 1.14 & 3.69 & 1.36 & 1.22 & 16.06 & 2.27 & 2.15 & 0.49 & 0.17 & 0.15 & -0.29 & 0.15 & 0.17 & 1.11 & 0.16 & 0.15 & NS & AP
 \\ IC342 & 12 & 56.720147 & 68.094210 & 4.56 & 1.71 & 1.57 & 1.67 & 0.00 & 0.00 & 1.11 & 0.00 & 0.00 & 10.48 & 2.96 & 2.70 & -0.18 & 0.29 & 0.30 & -0.47 & 0.00 & 0.00 & 0.72 & 0.20 & 0.19 & - & -
 \\ IC342 & 13 & 56.766922 & 68.151472 & 4.62 & 1.21 & 1.06 & 1.21 & 0.00 & 0.00 & 1.40 & 0.00 & 0.00 & 9.64 & 2.48 & 2.09 & -0.31 & 0.00 & 0.00 & 0.09 & 0.00 & 0.00 & 0.66 & 0.17 & 0.14 & - & -
 \\ IC342 & 14 & 56.720092 & 68.084930 & 2.90 & 1.23 & 1.08 & 1.27 & 0.00 & 0.00 & 1.11 & 0.00 & 0.00 & 8.25 & 2.36 & 2.21 & -0.10 & 0.31 & 0.30 & -0.02 & 0.00 & 0.00 & 0.57 & 0.16 & 0.15 & - & -
 \\ IC342 & 15 & 56.672095 & 68.145262 & 2.91 & 0.88 & 0.79 & 0.93 & 0.00 & 0.00 & 0.99 & 0.00 & 0.00 & 7.75 & 1.81 & 1.67 & -0.14 & 0.22 & 0.24 & 0.03 & 0.00 & 0.00 & 0.53 & 0.12 & 0.12 & - & -
 \\ IC342 & 16 & 56.635542 & 68.065637 & 2.42 & 0.76 & 0.66 & 2.66 & 0.84 & 0.74 & 0.84 & 0.00 & 0.00 & 6.60 & 1.40 & 1.40 & 0.04 & 0.20 & 0.20 & 0.41 & 0.00 & 0.00 & 0.12 & 0.00 & 0.00 & - & -
  \\ NGC4945 & 1 & 196.408793 & -49.429173 & 31.83 & 7.65 & 7.77 & 29.81 & 6.98 & 6.58 & 3.63 & 0.00 & 0.00 & 73.98 & 13.09 & 12.34 & -0.02 & 0.19 & 0.16 & -0.81 & 0.00 & 0.00 & 6.14 & 1.09 & 1.02 & NS & ZA
 \\ NGC4945 & 2 & 196.355946 & -49.473268 & 9.07 & 0.00 & 0.00 & 27.48 & 7.39 & 7.34 & 7.68 & 0.00 & 0.00 & 67.05 & 13.24 & 12.02 & 0.01 & 0.19 & 0.18 & -0.61 & 0.00 & 0.00 & 5.56 & 1.10 & 1.00 & - & -
 \\ NGC4945 & 3 & 196.296232 & -49.524091 & 12.99 & 1.60 & 1.49 & 21.33 & 2.05 & 1.95 & 1.78 & 0.00 & 0.00 & 46.40 & 3.45 & 3.29 & 0.24 & 0.07 & 0.07 & -0.73 & 0.00 & 0.00 & 3.85 & 0.29 & 0.27 & NS & ZA
 \\ NGC4945 & 4 & 196.387047 & -49.459326 & 41.37 & 2.98 & 2.84 & 29.17 & 3.60 & 3.45 & 5.18 & 0.00 & 0.00 & 39.96 & 4.58 & 4.49 & -0.44 & 0.00 & 0.00 & -0.83 & 0.00 & 0.00 & 3.31 & 0.38 & 0.37 & BH & S
 \\ NGC4945 & 5 & 196.338298 & -49.461319 & 5.27 & 0.00 & 0.00 & 5.40 & 0.00 & 0.00 & 1.66 & 0.00 & 0.00 & 31.45 & 4.18 & 4.01 & -0.06 & 0.13 & 0.15 & -0.92 & 0.00 & 0.00 & 2.61 & 0.35 & 0.33 & NS & ZA
 \\ NGC4945 & 6 & 196.327271 & -49.473171 & 12.04 & 2.23 & 2.10 & 16.31 & 2.61 & 2.53 & 1.62 & 0.00 & 0.00 & 30.30 & 3.79 & 3.64 & 0.04 & 0.12 & 0.13 & -0.93 & 0.00 & 0.00 & 2.51 & 0.31 & 0.30 & NS & ZA
 \\ NGC4945 & 7 & 196.404702 & -49.426126 & 5.81 & 0.00 & 0.00 & 5.33 & 0.00 & 0.00 & 8.77 & 2.46 & 2.53 & 29.79 & 9.82 & 8.93 & 0.31 & 0.93 & 0.42 & 0.01 & 0.00 & 0.00 & 2.47 & 0.81 & 0.74 & - & -
 \\ NGC4945 & 8 & 196.412435 & -49.424898 & 3.83 & 0.00 & 0.00 & 12.22 & 3.81 & 3.64 & 2.85 & 0.00 & 0.00 & 23.61 & 7.03 & 6.67 & 0.34 & 0.49 & 0.29 & -0.42 & 0.00 & 0.00 & 1.96 & 0.58 & 0.55 & NS & AP
 \\ NGC4945 & 9 & 196.397906 & -49.486358 & 12.65 & 1.65 & 1.53 & 9.77 & 1.96 & 1.84 & 2.68 & 0.00 & 0.00 & 15.00 & 2.73 & 2.55 & -0.22 & 0.00 & 0.00 & -0.66 & 0.00 & 0.00 & 1.24 & 0.23 & 0.21 & BH & S
 \\ NGC4945 & 10 & 196.343820 & -49.493043 & 2.25 & 0.00 & 0.00 & 6.50 & 2.53 & 2.43 & 1.83 & 0.00 & 0.00 & 13.70 & 3.44 & 3.31 & 0.08 & 0.27 & 0.29 & -0.85 & 0.00 & 0.00 & 1.14 & 0.29 & 0.27 & NS & ZA
 \\ NGC4945 & 11 & 196.438717 & -49.489241 & 1.10 & 0.00 & 0.00 & 4.56 & 1.46 & 1.35 & 1.68 & 0.00 & 0.00 & 8.24 & 2.64 & 2.11 & 0.24 & 0.00 & 0.00 & -0.45 & 0.00 & 0.00 & 0.68 & 0.22 & 0.18 & - & -
 \\ NGC4945 & 12 & 196.375811 & -49.413063 & 1.25 & 0.00 & 0.00 & 4.07 & 1.52 & 1.39 & 1.74 & 0.00 & 0.00 & 7.82 & 2.88 & 2.35 & 0.17 & 0.40 & 0.36 & -0.42 & 0.00 & 0.00 & 0.65 & 0.24 & 0.19 & - & -
 \\ HolmbergII & 1 & 124.869805 & 70.705453 & 288.82 & 3.49 & 3.51 & 305.27 & 3.66 & 3.59 & 42.73 & 1.68 & 1.64 & 752.54 & 6.51 & 6.37 & 0.02 & 0.01 & 0.01 & -0.75 & 0.01 & 0.01 & 48.23 & 0.42 & 0.41 & BH & ULX
 \\ HolmbergII & 2 & 124.788852 & 70.657370 & 6.99 & 0.95 & 0.87 & 11.90 & 1.30 & 1.19 & 2.61 & 1.07 & 0.95 & 29.96 & 2.20 & 2.12 & 0.25 & 0.07 & 0.07 & -0.57 & 0.10 & 0.10 & 1.92 & 0.14 & 0.14 & - & -
 \\ HolmbergII & 3 & 124.790615 & 70.777998 & 5.59 & 0.00 & 0.00 & 6.95 & 1.37 & 1.25 & 4.30 & 1.51 & 1.37 & 24.31 & 2.59 & 2.47 & 0.13 & 0.11 & 0.11 & -0.25 & 0.14 & 0.16 & 1.56 & 0.17 & 0.16 & BH & H
 \\ HolmbergII & 4 & 124.666558 & 70.767633 & 1.69 & 0.00 & 0.00 & 5.88 & 2.02 & 1.79 & 2.23 & 0.00 & 0.00 & 21.81 & 3.98 & 3.69 & 0.20 & 0.21 & 0.20 & -0.17 & 0.21 & 0.26 & 1.40 & 0.26 & 0.24 & BH & H
 \\ HolmbergII & 5 & 124.616683 & 70.714178 & 3.05 & 1.43 & 1.23 & 1.71 & 0.00 & 0.00 & 0.90 & 0.00 & 0.00 & 11.56 & 2.96 & 2.57 & -0.05 & 0.24 & 0.27 & -0.56 & 0.00 & 0.00 & 0.74 & 0.19 & 0.16 & - & -
 \\ HolmbergII & 6 & 124.967121 & 70.711546 & 2.87 & 0.97 & 0.86 & 4.22 & 1.11 & 1.03 & 2.10 & 0.93 & 0.79 & 11.00 & 1.94 & 1.84 & 0.33 & 0.22 & 0.20 & -0.31 & 0.19 & 0.20 & 0.70 & 0.12 & 0.12 & - & -
 \\ M81 & 1 & 148.886846 & 69.009754 & 95.39 & 2.71 & 2.59 & 107.25 & 2.81 & 2.69 & 22.62 & 1.50 & 1.43 & 262.46 & 4.87 & 4.75 & 0.07 & 0.02 & 0.02 & -0.64 & 0.02 & 0.02 & 20.50 & 0.38 & 0.37 & BH & ULX
 \\ M81 & 2 & 148.908881 & 69.067523 & 27.00 & 7.21 & 7.03 & 34.60 & 8.35 & 8.30 & 4.79 & 0.00 & 0.00 & 84.01 & 14.08 & 14.06 & 0.10 & 0.19 & 0.18 & -0.37 & 0.00 & 0.00 & 6.56 & 1.10 & 1.10 & - & -
 \\ M81 & 3 & 148.956128 & 69.092627 & 22.42 & 2.41 & 2.31 & 26.49 & 2.00 & 1.92 & 1.19 & 0.00 & 0.00 & 52.47 & 3.86 & 3.72 & 0.21 & 0.08 & 0.07 & -0.76 & 0.00 & 0.00 & 4.10 & 0.30 & 0.29 & NS & ZA
 \\ M81 & 4 & 148.792621 & 69.084213 & 20.26 & 1.74 & 1.61 & 18.20 & 1.79 & 1.71 & 1.12 & 0.00 & 0.00 & 43.59 & 3.05 & 2.96 & 0.05 & 0.07 & 0.07 & -0.69 & 0.00 & 0.00 & 3.40 & 0.24 & 0.23 & NS & ZA
 \\ M81 & 5 & 148.753974 & 69.124464 & 7.61 & 1.85 & 1.69 & 12.02 & 2.27 & 2.13 & 5.74 & 1.94 & 1.74 & 40.23 & 4.25 & 4.09 & 0.19 & 0.11 & 0.11 & -0.40 & 0.13 & 0.13 & 3.14 & 0.33 & 0.32 & BH & H
 \\ M81 & 6 & 148.749554 & 69.129380 & 9.11 & 2.04 & 1.85 & 11.39 & 2.41 & 2.19 & 1.96 & 0.00 & 0.00 & 29.85 & 4.36 & 4.11 & 0.09 & 0.14 & 0.14 & -0.45 & 0.00 & 0.00 & 2.33 & 0.34 & 0.32 & - & -
 \\ M81 & 7 & 149.010120 & 68.993406 & 2.31 & 0.99 & 0.91 & 3.50 & 1.23 & 1.12 & 2.80 & 1.31 & 1.17 & 17.83 & 2.36 & 2.23 & 0.20 & 0.14 & 0.14 & -0.21 & 0.16 & 0.18 & 1.39 & 0.18 & 0.17 & BH & H
 \\ M81 & 8 & 148.955677 & 69.136729 & 3.34 & 1.16 & 1.04 & 3.91 & 1.34 & 1.18 & 0.80 & 0.00 & 0.00 & 17.68 & 2.65 & 2.19 & 0.06 & 0.13 & 0.13 & -0.74 & 0.00 & 0.00 & 1.38 & 0.21 & 0.17 & NS & ZA
 \\ M81 & 9 & 148.956347 & 68.977013 & 2.07 & 0.97 & 0.86 & 1.12 & 0.00 & 0.00 & 1.18 & 0.00 & 0.00 & 11.12 & 2.20 & 2.05 & 0.07 & 0.20 & 0.20 & -0.13 & 0.00 & 0.00 & 0.87 & 0.17 & 0.16 & - & -
 \\ M81 & 10 & 148.740078 & 69.045188 & 3.19 & 0.95 & 0.86 & 1.03 & 0.00 & 0.00 & 0.45 & 0.00 & 0.00 & 6.83 & 1.68 & 1.54 & -0.01 & 0.25 & 0.26 & -0.66 & 0.00 & 0.00 & 0.53 & 0.13 & 0.12 & - & -
 \\ HolmbergIX & 1 & 149.471230 & 69.063248 & 881.68 & 3.90 & 3.90 & 1104.69 & 4.28 & 4.38 & 209.73 & 2.04 & 2.06 & 2572.42 & 7.58 & 7.53 & 0.11 & 0.00 & 0.00 & -0.68 & 0.00 & 0.00 & 219.14 & 0.65 & 0.64 & BH & ULX
 \\ HolmbergIX & 2 & 149.700595 & 69.092578 & 7.08 & 0.84 & 0.77 & 12.64 & 1.02 & 1.00 & 2.55 & 0.91 & 0.86 & 35.32 & 1.92 & 1.86 & 0.20 & 0.05 & 0.05 & -0.60 & 0.07 & 0.08 & 3.01 & 0.16 & 0.16 & NS & ZA
 \\ HolmbergIX & 3 & 149.246174 & 69.072843 & 5.47 & 0.85 & 0.82 & 7.15 & 0.98 & 0.91 & 2.44 & 0.00 & 0.00 & 28.45 & 1.81 & 1.77 & 0.09 & 0.06 & 0.06 & -0.43 & 0.09 & 0.09 & 2.42 & 0.15 & 0.15 & BH & H
 \\ HolmbergIX & 4 & 149.401430 & 69.001676 & 2.89 & 0.63 & 0.59 & 4.37 & 0.76 & 0.72 & 2.64 & 0.00 & 0.00 & 13.10 & 1.29 & 1.25 & 0.17 & 0.10 & 0.10 & -0.43 & 0.13 & 0.14 & 1.12 & 0.11 & 0.11 & BH & H
 \\ HolmbergIX & 5 & 149.701622 & 69.013395 & 0.67 & 0.00 & 0.00 & 0.83 & 0.00 & 0.00 & 0.99 & 0.00 & 0.00 & 11.18 & 1.77 & 1.66 & -0.01 & 0.17 & 0.18 & -0.09 & 0.22 & 0.25 & 0.95 & 0.15 & 0.14 & BH & H
 \\ HolmbergIX & 6 & 149.264037 & 69.056331 & 1.65 & 0.65 & 0.60 & 1.97 & 0.79 & 0.70 & 0.70 & 0.00 & 0.00 & 9.58 & 1.40 & 1.34 & 0.14 & 0.15 & 0.15 & -0.23 & 0.00 & 0.00 & 0.82 & 0.12 & 0.11 & - & -
 \\ HolmbergIX & 7 & 149.403636 & 69.124751 & 0.62 & 0.00 & 0.00 & 2.81 & 0.73 & 0.68 & 0.69 & 0.00 & 0.00 & 9.54 & 1.31 & 1.24 & 0.30 & 0.15 & 0.14 & -0.36 & 0.16 & 0.18 & 0.81 & 0.11 & 0.11 & BH & H
 \\ HolmbergIX & 8 & 149.527654 & 69.076959 & 5.57 & 1.10 & 1.00 & 2.89 & 1.07 & 1.02 & 0.65 & 0.00 & 0.00 & 5.74 & 1.74 & 1.31 & 0.69 & 0.00 & 0.00 & -0.22 & 0.00 & 0.00 & 0.49 & 0.15 & 0.11 & NS & AP
 \\ NGC5204 & 1 & 202.410891 & 58.418104 & 90.90 & 2.81 & 2.58 & 97.78 & 2.47 & 2.40 & 14.38 & 1.31 & 1.25 & 246.82 & 4.05 & 3.92 & 0.03 & 0.02 & 0.02 & -0.73 & 0.02 & 0.02 & 35.23 & 0.58 & 0.56 & BH & ULX
 \\ NGC5204 & 2 & 202.365137 & 58.426330 & 5.67 & 1.45 & 1.37 & 0.95 & 0.00 & 0.00 & 0.86 & 0.00 & 0.00 & 10.78 & 2.64 & 2.30 & -0.25 & 0.23 & 0.25 & -0.51 & 0.00 & 0.00 & 1.54 & 0.38 & 0.33 & BH & S
 \\ NGC5204 & 3 & 202.357118 & 58.424131 & 2.46 & 0.00 & 0.00 & 5.01 & 1.10 & 0.99 & 2.27 & 0.00 & 0.00 & 9.85 & 2.56 & 2.40 & 0.20 & 0.43 & 0.33 & 0.06 & 0.00 & 0.00 & 1.41 & 0.37 & 0.34 & - & -
 \\ NGC1313 & 1 & 49.583262 & -66.486433 & 251.07 & 2.56 & 2.52 & 309.54 & 2.83 & 2.81 & 59.09 & 1.45 & 1.41 & 728.94 & 4.97 & 4.88 & 0.10 & 0.01 & 0.01 & -0.68 & 0.01 & 0.01 & 78.92 & 0.54 & 0.53 & BH & ULX
 \\ NGC1313 & 2 & 49.592608 & -66.600926 & 98.98 & 2.27 & 2.16 & 69.05 & 2.07 & 2.00 & 9.66 & 1.39 & 1.29 & 215.88 & 3.98 & 3.89 & -0.17 & 0.02 & 0.02 & -0.73 & 0.03 & 0.03 & 23.37 & 0.43 & 0.42 & BH & ULX
 \\ NGC1313 & 3 & 49.576007 & -66.500731 & 23.27 & 1.69 & 1.62 & 28.30 & 1.77 & 1.71 & 3.47 & 0.89 & 0.87 & 64.43 & 3.09 & 3.03 & 0.11 & 0.05 & 0.05 & -0.76 & 0.05 & 0.05 & 6.98 & 0.33 & 0.33 & NS & ZA
 \\ NGC1313 & 4 & 49.410931 & -66.551014 & 17.97 & 1.22 & 1.16 & 14.25 & 1.26 & 1.19 & 1.09 & 0.00 & 0.00 & 44.62 & 2.40 & 2.30 & -0.12 & 0.05 & 0.05 & -0.65 & 0.00 & 0.00 & 4.83 & 0.26 & 0.25 & BH & I
 \\ NGC1313 & 5 & 49.453570 & -66.512900 & 4.43 & 0.73 & 0.68 & 5.58 & 0.85 & 0.79 & 2.46 & 0.85 & 0.79 & 17.91 & 1.59 & 1.55 & 0.06 & 0.09 & 0.09 & -0.31 & 0.12 & 0.13 & 1.94 & 0.17 & 0.17 & BH & H
 \\ NGC1313 & 6 & 49.849423 & -66.584843 & 3.20 & 0.97 & 0.87 & 5.53 & 1.29 & 1.18 & 1.42 & 0.00 & 0.00 & 16.70 & 2.45 & 2.31 & 0.17 & 0.13 & 0.14 & -0.28 & 0.00 & 0.00 & 1.81 & 0.27 & 0.25 & - & -
 \\ NGC1313 & 7 & 49.887555 & -66.532604 & 2.97 & 0.99 & 0.88 & 4.77 & 1.19 & 1.16 & 1.26 & 0.00 & 0.00 & 14.77 & 2.31 & 2.19 & 0.11 & 0.15 & 0.15 & -0.30 & 0.00 & 0.00 & 1.60 & 0.25 & 0.24 & - & -
 \\ NGC1313 & 8 & 49.857416 & -66.496388 & 0.96 & 0.00 & 0.00 & 5.38 & 1.26 & 1.16 & 1.22 & 0.00 & 0.00 & 13.16 & 2.38 & 2.24 & 0.29 & 0.19 & 0.17 & -0.41 & 0.00 & 0.00 & 1.42 & 0.26 & 0.24 & NS & AP
 \\ NGC1313 & 9 & 49.588208 & -66.509542 & 5.67 & 1.05 & 1.01 & 2.70 & 1.05 & 1.00 & 0.72 & 0.00 & 0.00 & 11.26 & 1.91 & 1.84 & -0.29 & 0.16 & 0.18 & -0.40 & 0.00 & 0.00 & 1.22 & 0.21 & 0.20 & BH & S
 \\ NGC1313 & 10 & 49.684068 & -66.428079 & 2.09 & 0.61 & 0.57 & 1.81 & 0.75 & 0.70 & 0.85 & 0.00 & 0.00 & 10.32 & 1.48 & 1.41 & -0.06 & 0.15 & 0.17 & -0.05 & 0.20 & 0.23 & 1.12 & 0.16 & 0.15 & BH & H
 \\ NGC1313 & 11 & 49.591024 & -66.434908 & 2.02 & 0.58 & 0.53 & 2.40 & 0.71 & 0.66 & 0.57 & 0.00 & 0.00 & 7.98 & 1.32 & 1.21 & 0.07 & 0.15 & 0.15 & -0.47 & 0.00 & 0.00 & 0.86 & 0.14 & 0.13 & - & -
 \\ M83 & 1 & 204.271280 & -29.868466 & 58.71 & 2.12 & 2.05 & 41.51 & 1.92 & 1.84 & 3.01 & 0.96 & 0.87 & 129.29 & 3.56 & 3.45 & -0.15 & 0.03 & 0.03 & -0.85 & 0.04 & 0.04 & 16.83 & 0.46 & 0.45 & BH & ULX
 \\ M83 & 2 & 204.121141 & -29.856414 & 27.35 & 2.40 & 2.22 & 32.63 & 2.70 & 2.52 & 11.72 & 2.39 & 2.16 & 91.78 & 4.09 & 3.94 & 0.08 & 0.04 & 0.04 & -0.51 & 0.06 & 0.06 & 11.95 & 0.53 & 0.51 & BH & I
 \\ M83 & 3 & 204.332434 & -29.896720 & 12.73 & 1.93 & 1.80 & 7.44 & 1.79 & 1.63 & 3.24 & 0.00 & 0.00 & 54.55 & 3.39 & 3.24 & -0.16 & 0.06 & 0.06 & -0.35 & 0.10 & 0.10 & 7.10 & 0.44 & 0.42 & - & -
 \\ M83 & 4 & 204.247658 & -29.832904 & 14.74 & 1.28 & 1.20 & 13.02 & 1.32 & 1.25 & 0.95 & 0.00 & 0.00 & 42.86 & 2.41 & 2.32 & -0.04 & 0.05 & 0.05 & -0.70 & 0.08 & 0.09 & 5.58 & 0.31 & 0.30 & BH & I
 \\ M83 & 5 & 204.302098 & -29.864971 & 8.58 & 1.23 & 1.17 & 14.35 & 1.42 & 1.38 & 5.99 & 1.26 & 1.18 & 41.31 & 2.54 & 2.46 & 0.24 & 0.07 & 0.06 & -0.41 & 0.07 & 0.07 & 5.38 & 0.33 & 0.32 & BH & H
 \\ M83 & 6 & 204.253488 & -29.865445 & 4.77 & 0.00 & 0.00 & 6.58 & 0.00 & 0.00 & 1.06 & 0.00 & 0.00 & 39.10 & 7.63 & 8.20 & 0.37 & 0.31 & 0.22 & -0.73 & 0.00 & 0.00 & 5.09 & 0.99 & 1.07 & NS & -
 \\ M83 & 7 & 204.318179 & -29.827451 & 7.16 & 1.18 & 1.10 & 8.80 & 1.41 & 1.32 & 3.28 & 1.36 & 1.24 & 36.44 & 2.67 & 2.55 & 0.07 & 0.07 & 0.07 & -0.42 & 0.10 & 0.11 & 4.74 & 0.35 & 0.33 & BH & H
 \\ M83 & 8 & 204.279501 & -29.850263 & 14.51 & 1.34 & 1.27 & 8.30 & 1.22 & 1.13 & 0.85 & 0.00 & 0.00 & 33.35 & 2.37 & 2.27 & -0.21 & 0.07 & 0.06 & -0.67 & 0.00 & 0.00 & 4.34 & 0.31 & 0.30 & BH & I
 \\ M83 & 9 & 204.258357 & -29.921444 & 6.90 & 0.96 & 0.88 & 10.93 & 1.15 & 1.11 & 2.18 & 0.94 & 0.85 & 29.78 & 2.02 & 1.94 & 0.19 & 0.07 & 0.07 & -0.64 & 0.09 & 0.09 & 3.88 & 0.26 & 0.25 & NS & ZA
 \\ M83 & 10 & 204.311241 & -29.908424 & 6.87 & 1.23 & 1.15 & 6.27 & 1.35 & 1.26 & 5.33 & 1.37 & 1.25 & 28.39 & 2.61 & 2.48 & 0.01 & 0.10 & 0.10 & -0.15 & 0.12 & 0.13 & 3.70 & 0.34 & 0.32 & BH & H
 \\ M83 & 11 & 204.181380 & -29.851747 & 10.82 & 1.45 & 1.34 & 9.75 & 1.53 & 1.40 & 0.80 & 0.00 & 0.00 & 27.12 & 2.26 & 2.07 & 0.03 & 0.08 & 0.08 & -0.88 & 0.00 & 0.00 & 3.53 & 0.29 & 0.27 & NS & ZA
 \\ M83 & 12 & 204.320495 & -29.893943 & 7.29 & 1.33 & 1.23 & 8.41 & 1.50 & 1.38 & 5.73 & 1.42 & 1.30 & 23.81 & 2.85 & 2.72 & 0.17 & 0.14 & 0.13 & -0.24 & 0.14 & 0.15 & 3.10 & 0.37 & 0.35 & BH & H
 \\ M83 & 13 & 204.249872 & -29.863881 & 3.02 & 0.00 & 0.00 & 9.69 & 3.58 & 3.42 & 1.55 & 0.00 & 0.00 & 20.65 & 5.77 & 5.58 & -0.13 & 0.31 & 0.30 & -0.41 & 0.00 & 0.00 & 2.69 & 0.75 & 0.73 & - & -
 \\ M83 & 14 & 204.236728 & -29.820689 & 3.68 & 1.00 & 0.92 & 4.50 & 1.10 & 1.02 & 0.97 & 0.00 & 0.00 & 17.07 & 2.14 & 2.02 & 0.09 & 0.12 & 0.11 & -0.55 & 0.00 & 0.00 & 2.22 & 0.28 & 0.26 & - & -
 \\ M83 & 15 & 204.238592 & -29.894139 & 4.69 & 0.96 & 0.88 & 4.88 & 1.06 & 0.96 & 0.59 & 0.00 & 0.00 & 15.60 & 1.96 & 1.72 & 0.03 & 0.11 & 0.11 & -0.78 & 0.00 & 0.00 & 2.03 & 0.26 & 0.22 & NS & ZA
 \\ M83 & 16 & 204.256672 & -29.795013 & 0.93 & 0.00 & 0.00 & 3.93 & 1.24 & 1.15 & 1.18 & 0.00 & 0.00 & 14.09 & 2.30 & 2.16 & 0.25 & 0.15 & 0.15 & -0.43 & 0.00 & 0.00 & 1.83 & 0.30 & 0.28 & - & -
 \\ M83 & 17 & 204.243149 & -29.851161 & 0.96 & 0.00 & 0.00 & 4.42 & 1.13 & 1.04 & 0.85 & 0.00 & 0.00 & 12.20 & 1.96 & 1.88 & 0.41 & 0.20 & 0.17 & -0.42 & 0.00 & 0.00 & 1.59 & 0.26 & 0.24 & NS & AP
 \\ M83 & 18 & 204.266172 & -29.825121 & 0.85 & 0.00 & 0.00 & 3.42 & 1.04 & 0.95 & 0.57 & 0.00 & 0.00 & 11.90 & 1.88 & 1.70 & 0.28 & 0.15 & 0.14 & -0.65 & 0.00 & 0.00 & 1.55 & 0.24 & 0.22 & NS & ZA
 \\ M83 & 19 & 204.267739 & -29.900895 & 2.86 & 0.97 & 0.87 & 0.94 & 0.00 & 0.00 & 2.16 & 0.95 & 0.86 & 11.11 & 1.90 & 1.78 & -0.23 & 0.19 & 0.20 & -0.01 & 0.27 & 0.27 & 1.45 & 0.25 & 0.23 & - & -
 \\ M83 & 20 & 204.231018 & -29.919042 & 0.69 & 0.00 & 0.00 & 2.00 & 0.87 & 0.79 & 0.89 & 0.00 & 0.00 & 9.11 & 1.67 & 1.56 & 0.29 & 0.19 & 0.18 & -0.22 & 0.00 & 0.00 & 1.19 & 0.22 & 0.20 & - & -
 \\ M83 & 21 & 204.260241 & -29.888722 & 1.86 & 0.00 & 0.00 & 1.60 & 0.00 & 0.00 & 2.66 & 0.92 & 0.82 & 7.28 & 3.19 & 2.36 & 0.32 & 0.00 & 0.00 & 0.04 & 0.00 & 0.00 & 0.95 & 0.42 & 0.31 & NS & AP
 \\ NGC253 & 1 & 11.887405 & -25.296824 & 183.31 & 3.90 & 3.77 & 134.68 & 5.76 & 5.73 & 9.33 & 1.70 & 2.16 & 397.00 & 8.50 & 8.64 & -0.15 & 0.02 & 0.02 & -0.87 & 0.02 & 0.03 & 30.16 & 0.65 & 0.66 & BH & ULX
 \\ NGC253 & 2 & 11.887485 & -25.288801 & 87.88 & 36.78 & 37.24 & 161.89 & 19.58 & 34.17 & 14.76 & 0.00 & 0.00 & 285.48 & 51.43 & 60.23 & 0.32 & 0.38 & 0.20 & -0.75 & 0.00 & 0.00 & 21.69 & 3.91 & 4.58 & BH & ULX
 \\ NGC253 & 3 & 11.889146 & -25.289359 & 12.51 & 0.00 & 0.00 & 82.16 & 12.76 & 12.24 & 5.62 & 0.00 & 0.00 & 134.19 & 22.51 & 24.49 & 0.53 & 0.44 & 0.19 & -0.65 & 0.00 & 0.00 & 10.19 & 1.71 & 1.86 & NS & AP
 \\ NGC253 & 4 & 11.928089 & -25.250713 & 28.88 & 1.74 & 1.64 & 19.95 & 1.55 & 1.47 & 0.76 & 0.00 & 0.00 & 64.61 & 2.91 & 2.85 & -0.18 & 0.04 & 0.04 & -0.82 & 0.06 & 0.06 & 4.91 & 0.22 & 0.22 & BH & I
 \\ NGC253 & 5 & 11.896703 & -25.253366 & 29.58 & 1.10 & 1.07 & 14.26 & 0.93 & 0.89 & 1.62 & 0.60 & 0.56 & 55.68 & 1.83 & 1.77 & -0.33 & 0.03 & 0.03 & -0.76 & 0.06 & 0.06 & 4.23 & 0.14 & 0.13 & BH & S
 \\ NGC253 & 6 & 11.844211 & -25.347459 & 20.71 & 0.97 & 0.94 & 20.87 & 1.08 & 1.03 & 2.52 & 0.82 & 0.75 & 55.46 & 1.91 & 1.83 & 0.00 & 0.03 & 0.03 & -0.75 & 0.05 & 0.06 & 4.21 & 0.15 & 0.14 & NS & ZA
 \\ NGC253 & 7 & 11.878991 & -25.307364 & 8.88 & 1.96 & 1.88 & 13.60 & 2.06 & 1.97 & 0.91 & 0.00 & 0.00 & 30.98 & 3.55 & 3.45 & 0.18 & 0.13 & 0.12 & -0.68 & 0.10 & 0.11 & 2.35 & 0.27 & 0.26 & NS & ZA
 \\ NGC253 & 8 & 11.827027 & -25.320633 & 10.14 & 0.80 & 0.75 & 9.59 & 0.87 & 0.83 & 0.72 & 0.00 & 0.00 & 25.86 & 1.55 & 1.48 & -0.04 & 0.05 & 0.05 & -0.79 & 0.00 & 0.00 & 1.96 & 0.12 & 0.11 & NS & ZA
 \\ NGC253 & 9 & 11.883434 & -25.289374 & 7.95 & 0.00 & 0.00 & 5.38 & 0.00 & 0.00 & 6.88 & 2.15 & 2.36 & 13.26 & 0.00 & 0.00 & 0.65 & 0.00 & 0.00 & -0.33 & 0.75 & 0.22 & 1.01 & 0.00 & 0.00 & - & -
 \\ NGC253 & 10 & 11.892656 & -25.284370 & 7.57 & 2.87 & 2.87 & 8.24 & 2.88 & 2.82 & 3.35 & 1.33 & 1.26 & 22.32 & 5.12 & 5.17 & 0.03 & 0.33 & 0.28 & -0.40 & 0.29 & 0.22 & 1.70 & 0.39 & 0.39 & BH & H
 \\ NGC253 & 11 & 11.864731 & -25.283095 & 3.81 & 0.92 & 0.87 & 9.81 & 1.11 & 1.07 & 6.87 & 0.83 & 0.79 & 18.74 & 1.77 & 1.71 & 0.70 & 0.00 & 0.00 & -0.12 & 0.08 & 0.08 & 1.42 & 0.13 & 0.13 & NS & AP
 \\ NGC253 & 12 & 11.878787 & -25.295633 & 2.44 & 0.00 & 0.00 & 9.89 & 2.78 & 2.75 & 1.19 & 0.00 & 0.00 & 18.09 & 4.58 & 4.43 & 0.39 & 0.42 & 0.27 & -0.49 & 0.00 & 0.00 & 1.37 & 0.35 & 0.34 & NS & AP
 \\ NGC253 & 13 & 11.919357 & -25.236955 & 5.68 & 0.78 & 0.72 & 6.21 & 0.83 & 0.80 & 0.65 & 0.00 & 0.00 & 17.00 & 1.49 & 1.44 & 0.03 & 0.09 & 0.08 & -0.54 & 0.00 & 0.00 & 1.29 & 0.11 & 0.11 & - & -
 \\ NGC253 & 14 & 11.854943 & -25.329247 & 6.34 & 0.75 & 0.72 & 6.58 & 0.82 & 0.77 & 0.66 & 0.00 & 0.00 & 16.94 & 1.48 & 1.43 & 0.03 & 0.08 & 0.08 & -0.59 & 0.00 & 0.00 & 1.29 & 0.11 & 0.11 & - & -
 \\ NGC253 & 15 & 11.866615 & -25.305688 & 5.61 & 0.98 & 0.92 & 7.11 & 1.04 & 1.00 & 0.69 & 0.00 & 0.00 & 15.82 & 1.78 & 1.73 & 0.21 & 0.12 & 0.11 & -0.54 & 0.00 & 0.00 & 1.20 & 0.14 & 0.13 & - & -
 \\ NGC253 & 16 & 11.878149 & -25.312474 & 4.43 & 1.47 & 1.44 & 4.73 & 1.53 & 1.52 & 0.90 & 0.00 & 0.00 & 12.86 & 2.65 & 2.60 & -0.07 & 0.21 & 0.22 & -0.51 & 0.00 & 0.00 & 0.98 & 0.20 & 0.20 & - & -
 \\ NGC253 & 17 & 11.901412 & -25.277326 & 5.29 & 1.13 & 1.07 & 5.72 & 1.15 & 1.10 & 0.62 & 0.00 & 0.00 & 12.82 & 2.01 & 1.91 & 0.07 & 0.16 & 0.15 & -0.74 & 0.00 & 0.00 & 0.97 & 0.15 & 0.15 & NS & ZA
 \\ NGC253 & 18 & 11.929490 & -25.256850 & 4.57 & 1.22 & 1.18 & 6.19 & 1.19 & 1.15 & 0.51 & 0.00 & 0.00 & 12.23 & 2.04 & 1.95 & 0.17 & 0.19 & 0.16 & -0.84 & 0.00 & 0.00 & 0.93 & 0.15 & 0.15 & NS & ZA
 \\ NGC253 & 19 & 11.936255 & -25.249076 & 3.19 & 1.10 & 1.07 & 4.64 & 1.10 & 1.06 & 0.80 & 0.00 & 0.00 & 10.88 & 2.02 & 1.95 & 0.13 & 0.24 & 0.20 & -0.39 & 0.21 & 0.22 & 0.83 & 0.15 & 0.15 & BH & H
 \\ NGC253 & 20 & 11.878623 & -25.247475 & 0.94 & 0.00 & 0.00 & 3.46 & 0.99 & 0.94 & 1.68 & 0.65 & 0.59 & 8.11 & 1.74 & 1.68 & 0.35 & 0.38 & 0.27 & -0.26 & 0.22 & 0.23 & 0.62 & 0.13 & 0.13 & - & -
 \\ NGC253 & 21 & 11.881610 & -25.251775 & 3.14 & 1.09 & 1.02 & 4.26 & 1.06 & 1.01 & 0.63 & 0.00 & 0.00 & 8.11 & 1.85 & 1.69 & 0.21 & 0.27 & 0.21 & -0.73 & 0.00 & 0.00 & 0.62 & 0.14 & 0.13 & NS & ZA
 \\ NGC253 & 22 & 11.868956 & -25.323236 & 3.59 & 0.78 & 0.74 & 2.80 & 0.81 & 0.76 & 0.59 & 0.00 & 0.00 & 7.87 & 1.49 & 1.35 & -0.08 & 0.17 & 0.18 & -0.52 & 0.00 & 0.00 & 0.60 & 0.11 & 0.10 & - & -
 \\ NGC253 & 23 & 11.904814 & -25.333960 & 0.39 & 0.00 & 0.00 & 2.08 & 0.62 & 0.57 & 1.78 & 0.66 & 0.60 & 4.96 & 1.06 & 1.08 & 0.71 & 0.00 & 0.00 & -0.02 & 0.20 & 0.21 & 0.38 & 0.08 & 0.08 & NS & AP
 \\ NGC253 & 24 & 11.855788 & -25.278774 & 0.71 & 0.00 & 0.00 & 0.79 & 0.00 & 0.00 & 0.65 & 0.00 & 0.00 & 4.57 & 1.42 & 1.32 & 0.07 & 0.36 & 0.33 & -0.14 & 0.00 & 0.00 & 0.35 & 0.11 & 0.10 & - & -
 \\ NGC253 & 25 & 11.837290 & -25.296446 & 1.87 & 0.58 & 0.54 & 1.81 & 0.66 & 0.60 & 0.40 & 0.00 & 0.00 & 4.29 & 1.15 & 0.95 & 0.03 & 0.24 & 0.26 & -0.54 & 0.00 & 0.00 & 0.33 & 0.09 & 0.07 & - & -
 \\ M82 & 1 & 148.959041 & 69.679695 & 1558.52 & 8.00 & 7.95 & 2201.69 & 11.12 & 10.79 & 475.27 & 3.98 & 4.30 & 4894.26 & 15.82 & 17.09 & 0.17 & 0.00 & 0.00 & -0.65 & 0.00 & 0.00 & 365.54 & 1.18 & 1.28 & BH & ULX
 \\ M82 & 2 & 148.947688 & 69.683262 & 298.98 & 7.28 & 7.26 & 394.00 & 8.55 & 8.42 & 74.54 & 3.26 & 3.37 & 888.56 & 13.42 & 13.15 & 0.15 & 0.02 & 0.02 & -0.68 & 0.01 & 0.01 & 66.36 & 1.00 & 0.98 & BH & ULX
 \\ M82 & 3 & 148.972467 & 69.683930 & 160.41 & 5.58 & 8.85 & 192.71 & 8.58 & 12.24 & 37.29 & 2.34 & 2.73 & 473.13 & 9.05 & 9.12 & 0.10 & 0.02 & 0.02 & -0.68 & 0.02 & 0.02 & 35.34 & 0.68 & 0.68 & BH & ULX
 \\ M82 & 4 & 148.943809 & 69.678010 & 45.53 & 5.05 & 5.02 & 7.86 & 0.00 & 0.00 & 1.95 & 0.00 & 0.00 & 65.37 & 11.10 & 10.06 & -0.45 & 0.00 & 0.00 & -0.48 & 0.00 & 0.00 & 4.88 & 0.83 & 0.75 & BH & S
 \\ M82 & 5 & 148.948396 & 69.688254 & 4.00 & 0.00 & 0.00 & 24.39 & 4.38 & 4.44 & 11.65 & 2.03 & 1.95 & 46.33 & 7.28 & 6.76 & 0.54 & 0.36 & 0.20 & -0.35 & 0.13 & 0.10 & 3.46 & 0.54 & 0.50 & NS & AP
 \\ M82 & 6 & 148.936554 & 69.679625 & 6.40 & 0.00 & 0.00 & 23.20 & 6.21 & 6.32 & 1.87 & 0.00 & 0.00 & 21.36 & 6.98 & 6.80 & 0.68 & 0.00 & 0.00 & -0.84 & 0.00 & 0.00 & 1.60 & 0.52 & 0.51 & NS & AP
 \\ M82 & 7 & 148.908934 & 69.674927 & 4.49 & 1.34 & 1.30 & 9.87 & 1.68 & 1.75 & 0.73 & 0.00 & 0.00 & 11.62 & 1.79 & 1.74 & 0.92 & 0.00 & 0.00 & -0.65 & 0.00 & 0.00 & 0.87 & 0.13 & 0.13 & NS & AP
 \\ M82 & 8 & 148.863555 & 69.656646 & 4.77 & 0.54 & 0.52 & 5.00 & 0.57 & 0.55 & 0.39 & 0.00 & 0.00 & 5.90 & 0.91 & 0.89 & 0.65 & 0.24 & 0.16 & -0.66 & 0.00 & 0.00 & 0.44 & 0.07 & 0.07 & NS & AP
 \\ M82 & 9 & 149.128916 & 69.705975 & 1.38 & 0.45 & 0.42 & 3.08 & 0.55 & 0.52 & 1.62 & 0.56 & 0.52 & 5.18 & 0.95 & 0.93 & 0.43 & 0.00 & 0.00 & -0.38 & 0.18 & 0.23 & 0.39 & 0.07 & 0.07 & NS & AP
 \\ M82 & 10 & 149.082374 & 69.696161 & 2.08 & 0.47 & 0.45 & 2.79 & 0.53 & 0.51 & 0.44 & 0.00 & 0.00 & 2.54 & 0.66 & 0.54 & 0.88 & 0.00 & 0.00 & -0.75 & 0.00 & 0.00 & 0.19 & 0.05 & 0.04 & NS & AP
 \\ M82 & 11 & 149.103354 & 69.715913 & 0.43 & 0.00 & 0.00 & 1.74 & 0.50 & 0.48 & 0.51 & 0.00 & 0.00 & 2.15 & 0.83 & 0.68 & 0.66 & 0.00 & 0.00 & -0.14 & 0.00 & 0.00 & 0.16 & 0.06 & 0.05 & NS & AP
 \\ M82 & 12 & 148.920935 & 69.657896 & 0.74 & 0.00 & 0.00 & 4.48 & 0.85 & 0.83 & 0.47 & 0.00 & 0.00 & 2.11 & 0.95 & 0.98 & 0.95 & 0.00 & 0.00 & 0.03 & 0.00 & 0.00 & 0.16 & 0.07 & 0.07 & NS & AP \\ 
\enddata
\tablecomments{Point source properties for galaxies from Table \ref{tab:gals}. Sources are grouped by galaxy and sorted/numbered by decreasing \full keV count rate. Count rates shown are the soft ($S$, $4-6$ keV), medium ($M$, $6-12$ keV), hard ($H$, \hard keV), and full ($F$, \full keV) energy bands. These count rates were derived from individual PSF fitting of each energy band (Section \ref{sec:psffit}). Where the $\sigma_{up}$ and $\sigma_{down}$ values are 0.00 the \nustar\ count rate represents the upper limit on the 90\% confidence interval. Section \ref{sec:simul} defines the derivation of hardness ratios $HR1=(M-S)/(M+S)$ and $HR2=(H-M)/(H+M)$. Luminosities were calculated using the $N_{\rm{H}}$ values and distances from Table \ref{tab:gals} assuming an absorbed power-law with spectral index $\Gamma=1.7$. The hardness ratios and luminosities were determined using the results of simultaneous PSF fitting described in Section \ref{sec:simul}, and thus can vary from values derived using count rates from individual PSF fitting in each energy band. State abbreviations are as follows: AP (accreting pulsar), S (BH soft state), I (BH intermediate state), H (BH hard state), ZA (Z/Atoll NS), ULX (ultraluminous source). Compact object type and accretion state classifications are described in Section \ref{sec:diag}.}
}

\end{deluxetable*}
\end{longrotatetable}

\end{document}